\documentclass[12pt,preprint]{aastex}

\shorttitle{Sub-millimeter Survey of HAEBE Stars}

\shortauthors{Sandell et al.}					
\slugcomment{Accepted by ApJ}

\begin{document}

\def\arcsec{{$^{\prime\prime}$}}
\def\mum{{$\mu$m}}
\def\ptsec{$''\mskip-7.6mu.\,$}
\def\ptmin{$'\mskip-7.6mu.\,$}
\def\psec{$^s\mskip-7.6mu.\,$}
\def\Msun{\,{\rm M$_{\odot}$}}
\def\Lsun{\,{\rm L$_{\odot}$}}
\def\ltsim{$\stackrel{<}{\sim}$}
\def\gtsim{$\stackrel{>}{\sim}$}
\def\eg{{\it e.g.}}
\def\ie{{\it i.e.}}
\def\cf{{\it cf.}}
\def\vs{{\it vs.}}
\def\etal{{\it et al.}}

\title{A sub-millimeter Mapping Survey of Herbig AeBe Stars}

\author{G\"oran Sandell\altaffilmark{1},  David A. Weintraub\altaffilmark{2}, \and Murad Hamidouche\altaffilmark{1} } 

\altaffiltext{1}{ SOFIA-USRA, NASA Ames Research Center, Mail Stop N211-3, Building N211/Rm. 249,
Moffett Field, CA 94035, U.S.A.}

\email{Goran.H.Sandell@nasa.gov}

\altaffiltext{2}{Department of Physics \& Astronomy, Vanderbilt University, P.O.\ Box 1807 Station B, Nashville, TN 37235, U.S.A.}
\email{mhamidouche@sofia.usra.edu} 
\email{david.a.weintraub@vanderbilt.edu}

\begin{abstract}

We have acquired sub-millimeter observations of 33 fields containing 37
Herbig Ae/Be (HAEBE) stars or potential HAEBE stars, including SCUBA
maps of all but two of these stars. Nine target stars show extended dust
emission. The other 18 are unresolved, suggesting that the dust
envelopes or disks around these stars are less than a few arcseconds in
angular size. In several cases we find that the strongest sub-millimeter
emission originates from younger, heavily embedded sources rather than
from the HAEBE star, which means that previous models must be viewed
with caution. ÊThese new data, in combination with far-infrared flux
measurements available in the literature, yield SEDs from far-infrared
to millimeter wavelengths for all the observed objects. ÊIsothermal fits
to these SEDs demonstrate excellent fits, in most cases, to the flux
densities longward of 100 $\mu$m. We find that a smaller proportion of
B-type stars than A and F-type stars are surrounded by circumstellar
disks, suggesting that disks around B stars dissipate on shorter time
scales than those around later spectral types. ÊOur models also reveal
that the mass of the circumstellar material and the value of $\beta$ are
correlated, with low masses corresponding to low values of $\beta$. Since
low values of $\beta$ imply large grain sizes, our results suggest that a
large fraction of the mass in low-beta sources is locked up in very
large grains.ÊSeveral of the isolated HAEBE stars have disks with very
flat sub-millimeter SEDs. These disks may be on the verge of forming
planetary systems.

\end{abstract}


\section{Introduction}\label{intro}

Herbig Ae/Be (HAEBE) stars are very young, intermediate mass (2 -- 8
\Msun) pre-main-sequence stars
\citep{Herbig60,Strom72,Finkenzeller84,Hillenbrand92,The94}. Since
\citeauthor{Herbig60} first established criteria for membership in this
class, our understanding and consequently our classification rules for
HAEBE stars have changed somewhat to include some stars of slightly
later spectral class and others that do not illuminate reflection
nebulae, including ''isolated'' HAEBE stars. At visible wavelengths,
spectra of HAEBE stars often include broad emission lines that show
rapid variability while at longer wavelengths HAEBE stars often are
associated with infrared and far-infrared excesses and millimeter
wavelength line emission.  These properties are very strongly associated
with circumstellar gas and dust at a wide range of temperatures.  Most
workers in the field now accept that the bulk of the circumstellar gas
and dust around many, if not most of these stars, is found in fairly
massive ($\sim$0.01 M$_\odot$) circumstellar disks
\citep{Hamidouche06,Eisner03,Meeus01,Natta01,Natta00a,Hillenbrand92}.
HAEBE stars are the progenitors of Vega like stars. Probing the
circumstellar disks of HAEBE stars is important to investigate the
likelihood of planet formation around intermediate mass stars. 
Accretion disks around the more massive Herbig Be stars also are
important in understanding the formation of massive stars. In addition,
it is possible that accretion plays a role in generating X-ray emission
in HAEBE stars \citep{Hamidouche08}.

\citet{Hillenbrand92} suggested a classification scheme for HAEBE
stars akin to that used for T Tauri stars. Using the slope of a star's
spectral energy distribution in the near to mid-infrared, stars with
large infrared excesses and spectral slopes with $\lambda$F$_\lambda$
$\sim$ $\lambda^{-4/3}$ were assigned to Group I; these stars
presumably have geometrically flat, optically thick circumstellar disks
and may have optically thin inner disks. Group II stars have flat or
rising spectra and were interpreted to have disk systems surrounded by
gas and dust envelopes. Finally, Group III HAEBE stars have very small
infrared excesses and appear to have only gaseous emission from any
residual circumstellar disk or envelope. 

The spectral energy distribution, which is composed of high
spatial resolution optical and near-infrared observations combined with
much lower spatial resolution {\it Infrared Space Observatory} (ISO),
{\it Infrared Astronomy Satellite} (IRAS), and ground-based
sub-millimeter continuum data may, however, be incorrect, if the
far-infrared and sub-millimeter emission seen toward these stars is
dominated by emission from extended reflection nebulae or nearby
embedded protostars. If this is the case, which it can be, especially
for B stars,  see e.g.,\citet{Henning98,Henning94,Sandell94a,Aspin94}
and the results from this paper,  then the disk and envelope models for
these HAEBE stars must be re-examined based on higher resolution
observations. 

In this paper, we present high spatial resolution 850 $\mu$m and 450
$\mu$m maps of HAEBE stars that enable us to examine 
what fraction of the long wavelength emission that has always been
assumed to be part of the spectral energy distributions for these stars
is associated with the stars and whether some of the emission originates
from the surrounding cloud and/or invisible nearby young objects.

\section{Observations}\label{obs}

Almost all of the sub-millimeter observations reported in this paper
have been obtained with SCUBA on JCMT\footnote{The JCMT is operated by
the Joint Astronomy Centre, on behalf of the UK Particle Physics and
Astronomy Research Council, the Netherlands Organization for Scientific
Research, and the Canadian National Research Council.}, Mauna Kea,
Hawaii. Some of the observations were carried out by us in August 2001;
others have been retrieved from the JCMT archive at the CADC\footnote{Guest
User, Canadian Astronomy Data Center, which is operated by the Dominion
Astrophysical Observatory for the National Research Council of Canada's
Herzberg Institute of Astrophysics}. We also present a few observations 
obtained using the single element bolometer UKT14 on JCMT \citep{Sandell94b}.

SCUBA \citep{Holland99} has 37 bolometers in the long and 91 in the
short wavelength array separated by approximately two beam widths in a
hexagonal pattern. The field of view of both arrays is $\sim$
2\ptmin3.  Both arrays can be used simultaneously by means of a
dichroic beamsplitter. The long wavelength array is optimized for 850
$\mu$m and the short wavelength array for 450 $\mu$m, but one can also
choose to observe at 750 $\mu$m and 350 $\mu$m for the long and short
wavelength array, respectively, with slightly reduced sensitivity. The
750/350 $\mu$m filter combination is no longer supported, but it was
used in 1997 and 1998 and we have retrieved and reduced some 750 $\mu$m
and 350 $\mu$m from this time period. The measured Half Power Beam Width (HPBW)
is 14\ptsec5 and 7\ptsec8 for 850 $\mu$m and 450 $\mu$m, respectively, 
for standard jiggle-map observing in stable night time conditions. For
the 750/350 $\mu$m filter combination, we measured HPBWs of 12\ptsec9
for 750 $\mu$m and 6\ptsec5 for 350 $\mu$m, respectively.

For sources with angular sizes less than or about equal to the field of
view, the preferred observing mode is to use the long and short
wavelength filter combination in jiggle-map mode. Larger emission
regions have to be observed in scan-map mode. In jiggle map mode, the
spacings between the bolometers are filled in by performing a jiggle
pattern with the chopping secondary, while the secondary is chopping
with a frequency of 7.8125 Hz. To completely sample both arrays, one
needs to sample 64 positions with the array, while only 16 jiggle
positions are needed if only one of the arrays is used.  The default
time for each jiggle step is 1 sec for standard observing. Each
64-point jiggle pattern is done in a set of four, 16-point jiggles, each
followed by a nod of the telescope into the other beam. In scan-map mode
the array is tilted relative to the scan direction to ensure full
sampling over the array with the telescope is scanning with a rate of
24\arcsec\ s$^{-1}$ while chopping with a frequency of 7.8125 Hz. Each
scan is overlapped by about half the array. The typical scan-map mode at
JCMT is a basket weaving technique first suggested by \citet{Emerson95} 
where one can scan the source in an arbitrary angle, but chop in two
orthogonal directions (fixed in equatorial coordinates) and restore the
dual beam map  in the Fourier space after converting the dual beam maps
to equatorial coordinates. The standard setup for SCUBA is to use six
maps, three of which are made while chopping in RA with three different
chop throws and three while chopping in Dec. also with three different
chop throws. The chop throws are chosen to so that the final map is
sensitive to most spatial frequencies. Scan-maps never go as deep as
jiggle-maps, because the restoration process results in baseline
uncertainties and sky noise and even small pointing errors cause
additional restoration noise \citep{Sandell01}. However, for compact
sources well above the noise level of the map, this is not a problem,
since one can always subtract or filter out the background emission.

The majority of the data presented in this paper have been obtained in
jiggle-map mode, typically  with a 120\arcsec\ chop in azimuth or at a
fixed position angle in order to avoid chopping onto extended emission.
A few compact isolated sources have been observed with shorter chop
throws, but not less than 80\arcsec.  Most HAEBE stars are known to
illuminate large reflection nebulae or are located in extended molecular
cloud complexes; these sources have been observed in scan-map mode. A
few stars --- HD\,200775 in NGC\,7023, and CoD$-$42$^\circ$
11721 --- have been observed in both scan-map and jiggle-map modes. For
these regions both jiggle and scan-maps agree within errors with each
other. The absolute pointing of SCUBA is typically checked on blazars or
calibrators near the target object before and after each map or set of
maps. This enables us, in the data reduction stage, to correct for
possible pointing drifts that may occur between the start and end of a
map. Some of the maps retrieved from the archive did not have such
frequent pointing observations. In these cases we have used the closest
pointing observation to correct the pointing for the target star and
then used the target star or another suitable point source in the map to
correct for possible pointing drifts during the observations. The
absolute pointing accuracy is expected to be better than 2\arcsec. The
sub-millimeter opacity was determined from sky dips at 850 $\mu$m and/or
derived from {\it CSO tau}, the sky dip meter located at CSO, which
provides opacity information every 10 minutes at $\lambda$ = 1.3~mm.
Two data sets were reduced from SCUBA in photometry mode. The reduction of
of these data sets is essentially similar, although we only obtain a total flux density, 
i.e., no information about whether the source is compact or extended.

All the data have been flux calibrated from maps of Mars, Uranus or Neptune
or from maps of the SCUBA secondary calibrator sources HL\,Tau,
OH231.8+4.2, IRC$+$10216, RAFGL\,618 and RAFGL\,2688
\citep{Sandell94,Sandell98,Jenness02} typically observed several times
during a night. If sufficient calibration information was not available
for a night, we have derived flux calibration from nearby nights with
similar weather conditions. A complete list of the HAEBE stars
observed in this survey is given in Table~1.

The data were reduced in a standard way using SURF
\citep{Jenness99,Sandell01} and STARLINK imaging software, i.e., we flat
fielded, extinction corrected, sky subtracted, despiked, and calibrated
the images in Jy~beam$^{-1}$. We then pointing-corrected each scan for
any drift in pointing between each pointing observation and added the
data together to determine the most likely sub-millimeter position at 850
$\mu$m. Once we had derived a source position this way, we then applied
small additional RA and Dec corrections to each scan (shift and add) to
sharpen the final image to this position. Both despiking and sky
subtraction can be done in several different ways depending on the
amount of data and what observing mode was used for collecting the data
\citep{Sandell01}. We generally followed the advice given in
\citet{Sandell01} or tried several different ways in order to find the
optimal way to ensure good data quality.

All our maps were converted to FITS files and exported to MIRIAD
\citep{Sault95} for further analysis. In order to correct for the error
lobe contribution, especially at 450 $\mu$m, we have deconvolved all the
maps using CLEAN and a circular model beam even though the beam is
generally somewhat  elliptical, i.e.,  broadened in the chop
direction.%\footnote{Beam maps of Uranus on Oct 3, 1997 were anomalously
%broadened in the chop direction. For HD\, 163296, observed at the same
%time, we therefore created an elliptical beam model, which reduced the
%clean errors, but in this case the beam profile was probably somewhat
%unstable, perhaps due to problems with the chopping secondary.}. 
However,
for large data sets sky rotation will circularize the average beam
profile anyway. Since our data set spans a variety of observing
conditions and several observing seasons the beam size and error lobe
contribution varies significantly from data set to data set. At 450
$\mu$m the HPBW can vary due to seeing conditions and changes in the
surface by as much as 1\arcsec. The error beam is even more sensitive to
thermal gradients over the telescope. Especially in the early evening
and during daytime, the 450 $\mu$m error beam can change 
due to these thermal gradients (Sandell, JCMT internal report).  We find
that the 450 $\mu$m HPBW varied by as much as 1\arcsec\ and occasionally
more due to all these factors.

For jigglemaps we measured HPBWs at 850 $\mu$m in the range 14\ptsec3 -
15\ptsec5, but most of our maps in good stable night time conditions
agree with the nominal value of 14\ptsec5. At 450 $\mu$m the nominal
HPBWis 7\farcs8.  For scan maps a typical HPBW at 850 $\mu$m is $\sim$
15\arcsec, while it is $\sim$ 8\farcs5 at 450 $\mu$m. For jiggle-maps
our model beam is composed of three symmetric Gaussians: a main beam, a
near-error lobe, and an extended, low amplitude, far-errror lobe. For
scan maps we only use a two component model, i.e., a main beam and a
near-error beam. At 850 $\mu$m the  compact near-error beam has a HPBW
of $\sim$ 40\arcsec\ -- 55\arcsec\ with an amplitude of 1 -- 2\% of the
beam peak and an extended error lobe with a HPBW $\ge$ 100\arcsec\ and
an amplitude of $\sim$ 0.1\% of the maximum.  Because the amplitude of
the extended error lobe is so small at 850 $\mu$m, we can ignore it.  At
450 $\mu$m we find the near-error beam to have a HPBW  in the range
25\arcsec\ --- 34\arcsec\ with an amplitude 3 - 7\% of the beam peak,
i.e., much larger variations than those we see at 850 $\mu$m. The
extended error lobe for jiggle-maps at 450 $\mu$m has a HPBW of $\sim$
120\arcsec\ and an amplitude of $\leq$ 1 \%.

In almost all cases, we have restored our CLEAN images back to
14\arcsec\ resolution at 850 $\mu$m and to 8\arcsec\ resolution at 450
$\mu$m. In a few cases, where the signal-to-noise is marginal, we have
restored the maps back to a resolution (15\arcsec\ - 16\arcsec\ for 850
$\mu$m; 10\arcsec\ - 12\arcsec\ for 450 $\mu$m)   which improves the
signal-to-noise without degrading the resolution too much. Then, using
the task IMFIT in MIRIAD, we fit a two-component elliptical Gaussian,
one for the  HAEBE star, and one for the surrounding cloud core. The fit
of the cloud core is mainly used to provide a good subtraction of the
extended emission and is not used to estimate the flux density of the
surrounding cloud. For blended sources like LkH$\alpha$\,198 and
LkH$\alpha$\,198\,mm we used constrained fits with multiple components,
increasing the uncertainty in the derived parameters. The results of
these Gaussian fits for the detected HAEBE stars  are given in
Table~\ref{tbl-2a}. Note that the quoted flux densities are background
corrected flux densities, i.e., the surrounding cloud emission has been
subtracted before we computed the integrated flux density. Additional 
sub-millimeter photometric measurements with UKT14 are presented
in Table~\ref{tbl-2b}. Limits for non-detections, all 3-$\sigma$ upper limits, are given in
Table~\ref{tbl-2c}.

%MWC\,1080 is not listed
%in Table~\ref{tbl-2a}, because the emission is completely dominated by nearby
%sources and/or emission from the surrounding nebulosity; therefore we
%could not separate emission from the star from that of its surroundings
%(see Section~\ref{emb}).

\section{Archive {\it Spitzer} observations}

We searched the  {\it Spitzer} archive for all MIPS 24 and 70 $\mu$m
observations of our target  stars and retrieved and analyzed all photometric (image)
data. A few stars were covered by the c2d legacy project \citep{Evans03}. For
these stars we retrieved the results  from the IRSA General Catalog Query Engine, 
Gator (http://irsa.ipac.caltech.edu/applications/Gator/). If the MIPS pbcd-data appeared to be 
of good quality, we extracted photometry with Mopex, which was developed by the {\it Spitzer} Science
center; otherwise we first reduced the images from the Archive BCD-data using Apex, another
 software package developed by the  {\it Spitzer} Science
center. The results of the {\it Spitzer} MIPS photometry are given in Table~\ref{tbl-3}.

\section{Analysis}\label{anal}

We fit the observed total fluxes with a simple graybody fit, see e.g.,
\citet{Sandell00a,Sandell01a}. For ``isolated'' HAEBE stars, i.e., stars for which we do
not see any emission from the surrounding cloud or from nearby sources,
we make use of IRAS, ISO or KAO far infrared data longward of 60 $\mu$m
in addition to our own sub-millimeter data. In a few cases, where we have no 100 $\mu$m data or where there is
a clear excess at 100 $\mu$m, we assume  a dust temperature or explore a plausible range in dust temperature.
 The fits for these stars are shown in Figure~\ref{fig-tr1} and Figure~\ref{fig-tr2}.

For confused regions, i.e., where nearby sources or strong
emission from the surrounding cloud core are present, we do not use the far-infrared
data, because we have no way to accurately partition the contribution of
each component to the far infrared flux measured in a much larger beam.
In these cases we also assume a plausible dust temperature based on all the available information we have for for these
sources.. Since we have measured the source size (Table~\ref{tbl-2a}), we
constrain the size and only fit for the dust emissivity, $\beta$, the
dust temperature, T$_d$ (if we have far infrared flux density measurements), and the
dust optical depth at 850 $\mu$m. To derive masses we assume standard
Hildebrand opacities ($\kappa_{\rm 1200GHz}$ = 0.1 cm$^2$ g$^{-1}$) and
a gas-to-dust ratio of 100 \citep{Hildebrand83}. The results of the
graybody fits are  given in Table~\ref{tbl-4}. In the same table we also
list the spectral index of the dust emission $\alpha$ (F$_\nu \propto \nu^\alpha$), after subtraction of free-free emission, which 
was derived by least-squares fitting all millimeter and sub-millimeter data for each source. For optically thin dust in the 
Rayleigh-Jeans regime, the spectral index, $\alpha$ = $\beta$ + 2, where $\beta$ is the dust emissivity index.

\section{Results}\label{res}

All maps of the sources listed in Table~ 1, except one, show emission in
the SCUBA field of view, although the observed emission is not
always associated with the HAEBE stars. Early B stars are luminous enough to heat up their surrounding
clouds and, with the exception of the very distant HAEBE star V431 Sct
(MWC 300), invariably show extended cloud emission.This is the only field where we see no emission
at all. We have only detected three HAEBE stars of spectral type early  B or
late O.  The failure to detect early type stars does not, however, mean
that massive stars do not form disks; rather, in this sample it
appears to be a selection effect.  When Herbig Be stars
are bright enough to illuminate reflection nebulae, they may already
have freed themselves from the cloud cores in which they were formed,
i.e., they are already close to the main sequence and appear to have
dispersed the disks with which they were born.

\subsection{Isolated HAEBE stars}
\label{Isolated}

Isolated HAEBE stars have the same  characteristics as ``classical''
HAEBE stars, i.e., they are often located near, but not in star forming
regions, they have H$\alpha$ in emission, IR excesses due to thermal
emission from dust, but differ from them due to the absence of
nebulosities \citep{Grinin91}.  The absence of nebulosities suggests
that these stars on the average are older than ``classical'' HAEBE
stars, because they have already freed themselves from the clouds in
which they were born. At least three (HD\,135344\,B, HD\,141569, and
HD\,169142), possibly more of our isolated HAEBE stars are transition
objects \citep{Najita07}. These are stars which show significant disk
evolution. The transition disks are characterized by an inner gap, which
is still gas rich but contains very little dust, and a cold outer disk.
Due to the inner gap they have very little near-infrared excess, but they
have strong mid- and far-infrared excess  due to the outer disk.

\citet{Meeus01} did ISO spectroscopy
of 14 isolated HAEBE stars and separated them into two groups based on
their infrared SEDs: group I  and group II. For group I the infrared to
sub-millimeter continuum could be modeled by a power-law and a
blackbody, suggesting that stars of this group have optically thick and
geometrically thin disks (power-law component) and an optically thin
flared region (black body component). Group II objects only needed a
power-law to fit their continuum, which \citet{Meeus01} suggested
indicates that they are more evolved stars with partially optically thin
inner disks protecting the outer disk from flaring. We have
sub-millimeter data on many of the stars observed by \citet{Meeus01}.
Observations of several of them made with millimeter arrays
confirm that the infrared excess is due to circumstellar disks
\citep{Mannings97a,Mannings97,Mannings00,Natta04,Hamidouche06,Isella07}.
In the millimeter and sub-millimeter regime there is very little
difference, if any, between \citeauthor{Meeus01} group I and group II
objects.

\indent {\bf MWC 480} (HD\,31648) is a bright, isolated Ae star in
the Taurus-Auriga complex with a Hipparchos parallax distance of 131 pc
\citep{Ancker98}. It has a spectral class of A3 - A5 Ve and an age of 4
- 8 Myr \citep{Simon00,Pietu06,Pietu07}. The star has never been
searched for free-free emission, but if it has free-free emission, it is
presumably weak. It has been extensively studied with mm-arrays and is
surrounded by a large circumstellar disk in Keplerian rotation
\citep{Mannings97,Simon00,Pietu06,Pietu07,Hughes08}. The disk is well
resolved both in CO and continuum. At 1.4 mm \citet{Hamidouche06} find a
disk size of  0\ptsec8 $\times$ 0\ptsec7 at a P.A. of 143\degr\ $\pm$
5\degr, while the CO disk is much larger with a diameter of $\sim$
8\arcsec\ at a P.A. of $\sim$ 150\degr\  \citep{Simon00,Pietu07}. Both
continuum and CO images give a similar inclination  angle for the disk,
{\it i} =  37\degr\ \citep{Hamidouche06,Pietu07}. Even though the disk
is quite large, it is not seen in coronagraphic near-infrared images
\citep{Augereau01}. \citet{Pietu07} derive $\sim$1.8  \Msun\  for the
mass of the central star from analysis of the Keplerian rotation of the
disk in the J = $1 \to 0$ and J = $2 \to 1$ transitions of CO and its
isotopomers  as well as from HCO$^+$ J = $1 \to 0$. The derived mass
would suggest that the star is of spectral type A5.

Our SCUBA observations (Table \ref{tbl-2a}, Figure \ref{fig-mwc480})
resolve the disk around MWC\,480 both at 850 and 450 $\mu$m.  Our
observations show an  emission feature  extending to the south from the
disk. This emission is nearly at the same P.A. as the observed jet-like
emission by \citet{Hamidouche06} and could therefore be real.  Our
isothermal graybody fit to the millimeter and 100 $\mu$m data   predict
a cold dust disk, T$_d$ $\sim$ 28~K and a dust emissivity index, $\beta
\sim$ 0.8 (Table~\ref{tbl-4}), As we can see from Figure~\ref{fig-tr1}
the dust emissivity is well constrained, but the fit underestimates the
observed flux densities at 60 $\mu$m, which suggests that one needs a
warm dust component as well in order to explain the 60 $\mu$m emission.

\indent {\bf HD 34282} is of spectral type A3 Ve \citep{Mora01,Merin04}
with an age of 6.4 Myr. It has no detectable free-free emission
\citep{Natta04}. Its disk was resolved at 1.3~mm with the IRAM Plateau
de Bure interferometer (PdB) with a size of 1\ptsec7 $\times$ 0\ptsec9
\citep{Pietu03}. They find from imaging the disk in CO J = $2\to1$ that
the disk is in Keplerian rotation and seen at an inclination of 56\degr\
$\pm$ 3\degr. Based on the dynamical mass they argue that the previously
reported distance estimate, from Hipparcos parallax observations, 160 
$\pm$ 42 pc \citep{Ancker98} must be in error and that the most likely
distance is  400~pc. The detailed modeling of the stellar parameters by
\citet{Merin04} gives 348 pc, We will therefore use the distance 350 pc.
The disk was also imaged with the OVRO interferometer by
\citet{Mannings00} in the continuum at 2.6 mm and by \citet{Natta04}
with the PbB interferometer at 3.2 and 1.3~mm.  The observed continuum
fluxes agree well with the results of  \citet{Pietu03}.

We find the star to be unresolved at 850 $\mu$m and  450 $\mu$m (Table
\ref{tbl-2a}). These SCUBA data have also been published by
\citet{Sheret04}, who get somewhat discrepant results at 450 $\mu$m. In
this analysis we use our results. The star was included in the CO  J = $
3 \to 2$ single dish JCMT survey by  \citet{Dent05} and shows a CO
profile similar to what was observed with the PdB interferometer,
confirming that this is an isolated HAEBE star. Since the star appears
compact in both single dish, coronographic imaging \citep{Doering07}
and aperture synthesis observations, we combined all millimeter and
sub-millimeter data in our isothermal modeling. The dust emissivity we
derive, $\beta$ = 1.29, is very similar to what \citet{Pietu03} derived
from their detailed modeling. However, in this case we find the dust
emission to be marginally optically thick at 850 $\mu$m, $\tau_{850}
\sim$ 0.2, and we get a slightly higher mass, 0.18 \Msun (Table
\ref{tbl-4}), than \citeauthor{Pietu03}. The mass of the disk is much
higher than we find for any other isolated Ae star.

{\bf \indent HD\,35187} is a visual binary star with a separation of
1\farcs38 in Taurus at a distance of 150 pc \citep{Dunkin98}.
\citeauthor{Dunkin98} found the stars have spectral types of A2 V
(HD\,35187\,B) and A7 V (HD\,35187\,A). Only HD\,35187\,B shows
H$\alpha$ in emission and has an infrared excess, suggesting it is
surrounded by an accretion disk. The age of the system is $\sim$ 10 Myr.

HD\,35187 was detected at 1.1 mm and 800 $\mu$m by \citet{Sylvester96}
with the single pixel bolometer UKT14 on JCMT. It is a relatively strong
radio source with a flux density of 1 mJy at 3.6 cm \citep{Natta04}.
\citet{Natta04} imaged the star in continuum with the VLA at 7 mm and
with the PdB interferometer at 3.6 and 1.3~mm. Most of the emission at 7
mm is due to free-free emission. The free-free emission is still
significant at 3.6~mm, but at 1.3 mm and shorter wavelengths the
emission is  dominated by dust emission.

Our 850 and 450 $\mu$m SCUBA maps show an unresolved point source. These
SCUBA maps were previously published by \citet{Sheret04}. We have
reduced them independently and find somewhat higher flux densities (16\%
and 27\% at 850 $\mu$m and 450 $\mu$m, respectively) than
\citeauthor{Sheret04}, but in good agreement with other published data
on this star. A graybody isothermal fit  to all (sub-)millimeter data
and the IRAS 100 $\mu$m data point (Figure~\ref{fig-tr1}) gives a dust
temperature of 44 K and a  dust emissivity index, $\beta$ =
0.87.  The mass of the disk is relatively low, 0.002 \Msun\
(Table~\ref{tbl-4}).

{\bf \indent HD\,36112} is an isolated HAEBE star located in the
outskirts of the Taurus-Auriga complex.  \citet{Chapillon08} adopted a distance of 140~pc, arguing that the star
is likely to be at the same distance as the Taurus-Auriga complex. \citet{Beskrovnaya99}
classified it as A8 Ve with an age of 5 - 10 Myrs. A three component fit
to low resolution UV spectra give a spectral class of A7 IIIe, with an
age of 4.4 Myr and a distance of 200 pc \citep{Blondel06}. The latter
agrees with the Hipparcos parallax distance \citep{Ancker98}. We
assume 200 pc. 

The star has
not been detected in free-free emission at centimeter wavelengths. It
was first observed by \citet{Mannings97} with the  OVRO array in
continuum at 2.7~mm and in the $^{13}$CO J = $1 \to 0$ line. The dust
disk was unresolved in continuum and barely detected in $^{13}$CO.
Follow-up observations with the OVRO array \citep{Mannings00} observed
the star in continuum at 1.3~mm, but did not resolve it.
\citet{Chapillon08} observed the star with the IRAM PdB interferometer
in continuum at 3.4~mm, 2.6~mm, and 1.3~mm, as well is the $^{12}$CO J =
$2 \to 1$ and $1 \to 0$ transitions. They resolve the disk in continuum
at 1.3 and 2.6~mm with a size of $\sim$ 1\arcsec\ (150 AU), and find the
disk slightly more extended in CO (r $\sim$ 250 AU). \citet{Dent05} detected the disk in the CO J = $3 \to 2$
line emission. By modeling the CO profile they found the disk to be
almost face-on with an inclination $<$ 10\degr\  and deduced an outer
radius of 170$\pm$ 30 AU. Their results agree quite well with the far
better constrained modeling by \citeauthor{Chapillon08}, who determine a
disk inclination of 16\degr\ - 18\degr\ from their CO observations. 

We
do not resolve the dust emission, which agrees well with the the PdB
results. A graybody fit to all (sub-)millimeter and FIR data to 100
$\mu$m, including ISO data from \citet{Elia05},  give a  dust emissivity
index, $\beta$ = 1.25, a dust temperature of 43 K, and a disk mass of
0.014 \Msun. Our deduced dust emissivity is slightly higher, but well
constrained by our fit (Figure~\ref{fig-tr1}),  and our disk mass is a
factor of two lower than what was found by \citet{Chapillon08}, They
derived  a dust emissivity index, $\beta$ = 1.0 $\pm$  0.15, and a disk
mass of 0.027 \Msun.

\indent {\bf HD\,135344\,B}  is an isolated  F4 Ve  star
\citep{Dunkin97,Grady09} in the Sco~OB\,2 - 3 association at a distance
of $\sim$ 140 pc \citep{Coulson95,Boekel05}. The star has been
extensively studied from X-rays to millimeter wavelengths \citep[and
references therein]{Grady09}. It has strong millimeter and far-infrared
excesses \citep{Walker88,Coulson95,Sylvester96}, and \citet{Grady09}
conclude that is definitely a PMS star with an age of $\sim$ 8 Myr. As
deduced both from modeling \citep{Dent05,Brown07} and direct imaging in
the mid-infrared and in the sub-millimeter
\citep{Pontoppidan08,Brown09}, the disk is seen approximately face-on
and has a large inner gap. The star  therefore has a transitional disk,
where the inner disk is largely free of dust. Yet the disk appears
surprisingly gas rich \citep{Dent05}.

The star is well studied in the (sub-)millimeter wavelength regime. Both
\citet{Coulson95} and \citet{Sylvester96} obtained photometry of the
star with UKT14 on JCMT, \citet{Walker95} observed it at Caltech
sub-millimeter Observatory (CSO) at 1.3 mm and 800 $\mu$m, and
\citet{Brown09} imaged it with high spatial resolution with  the
Smithsonian Submillimeter Array (SMA) at 880 $\mu$m. \citeauthor{Brown09}
resolved the dust disk with a radius of $\sim$ 0\farcs9 (125 AU) and
found an inner hole with an outer radius of 39 AU. We have deep images
with SCUBA at  850 $\mu$m and 450 $\mu$m, which show that the disk is
unresolved in an 8\arcsec\ beam (Table \ref{tbl-2a}). This is consistent
with the observed size (1\farcs8) at 880 $\mu$m \citep{Brown09},
although our observed flux density is $\sim$ 50\% higher than what
\citeauthor{Brown09} find in their high spatial resolution SMA
observations, suggesting they may filter out some of the extended
emission. Our isothermal graybody fit, including  all millimeter,
sub-millimeter as well as the IRAS 100   $\mu$m flux densities gives a
dust emissivity index, $\beta$ = 1.38, and a dust temperature of 32 K.
We did not include the ISO 200 $\mu$m flux density \citep{Walker00},
because it  appears anomalously low compared to all the sub-millimeter
and  IRAS data. We derive a much lower dust temperature than
\citet{Coulson95}, because they fitted the SED with a blackbody. As with
all evolved disks,  however, the graybody fit to the millimeter and
sub-millimeter data underestimate the flux densities shortward of 100
$\mu$m, which require a warm dust component as well. If we assume normal
gas-to-dust ratio, we get a disk mass of  0.025 \Msun\
(Table~\ref{tbl-4}).

\indent {\bf HD\,141569} is a nearby, isolated A0 Ve/B9.5 Ve star, which
is the primary member of a triple system in the outskirts of the Sco-Cen
system with an age of 3 - 5 Myrs \citep{Dunkin97,Weinberger00}.
\citet{Merin04} determined a more precise age for the primary by fitting
surface gravity, metallicity and effective temperature to high
resolution spectra and photometry. They find an age of 4.7 Myr, a
spectral type of B9.5 Ve and a distance of 108 pc, which agrees within
errors with the Hipparchos distance, 99 pc.  Coronagraphic imaging in
the near-infrared \citep{Augereau99,Weinberger99} shows a large (r $\sim$
400 AU), slightly asymmetric  dust disk with depression (gap) at 250 AU
and a 30 AU inner hole \citep{Marsh02}.  The disk has
also been resolved in the mid-infrared \citep{Fisher00,Marsh02} with a
size of 2\ptsec2 at 10 $\mu$m. HD\,141569 is often referred to
as a debris disk; however,
\citet{Dent05} detected CO J = $3 \to 2$, and all debris disks are gas
poor.  The only supposed debris disk star known to have a molecular gas disk is
49 Ceti \citep{Hughes08b}, and  it is quite possible that the disk
surrounding 49 Ceti is primordial, rather than a debris disk. In addition {\it
Spitzer} IRS spectra also show that  HD\,141569 has strong PAH emission
\citep{Sloan05,Keller08}, which is not seen in debris disks.
Furthermore, at an age of only 5 Myr, HD\,141569 is likely too 
young to be a debris disk star. Thus, its youth, the presence of a gas in the disk, and the detection
of PAH emission from the disk
all strongly suggest that HD\,141569 has a primordial transition disk,
not a debris disk.

For this star the the SCUBA data presented in Table \ref{tbl-2a} were
obtained in photometry mode and are based on the same data that were
published by \citet{Sheret04}. The flux densities that we obtain agree
quite well with those of \citeauthor{Sheret04}. The star was also
observed by SCUBA at 1.35~mm by \citet{Sylvester01} and at CSO and IRAM
at 1.3~mm by \citep{Walker95}, but the latter have poor signal-to-noise
and have not been used in our analysis. The star was observed in
photometry mode by {\it Spitzer} MIPS at 24, 70, and 160 $\mu$m. The
photometry at 24 and 70 $\mu$m is given in Table~\ref{tbl-3} and agrees
with results by Weinberger (private communication). The MIPS images only
show emission from HD\,141569 and no emission is seen from the
companions. Our graybody fit severely underestimates the observed flux
at 100 and 70 $\mu$m, suggesting that the disk must have a warm dust
component as well. The temperature of the cold outer  disk is therefore
uncertain. In Table~\ref{tbl-4} we give the results for an assumed dust
temperature of 40 K, which results in a $\beta$ index of 0.5  and a disk
mass of 9$\times$10$^{-5}$ \Msun.  Because we assumed  a temperature
rather than derived it, this results in much higher uncertainties for
the dust emissivity and disk mass. By varying the temperature from 25 --
70 K, we find that varies by less than  $\pm$ 0.2, while  the mass could
be off by a factor of two.

\indent {\bf HD\,142666} is located in the Sco OB2-2 association at a
distance of 145 pc \citep{Boekel05}. It has a spectral type of A8 Ve
\citep{Dunkin97,Mora01,Boekel05,Guimaraes06} with an estimated age $>$
10 Myr \citep{Natta04}. It was not detected in free-free emission at 3.6
cm \citep{Natta04}. It was included in the ISO survey by \citet{Meeus01}
where it was classified as a group IIa object, i.e. a star having an
optically thick disk in the FIR. The disk was detected in the CO J = $3
\to 2$ survey by \citep{Dent05}. Even though the CO emission is clearly
detected, the signal-to-noise is poor and \citeauthor{Dent05} fitted it
with a single broad line predicting that the disk is seen nearly face-on
({\it i} = 18 $\pm$ 5\degr{}) and a radius of 45 AU, which would suggest
a disk size $<$ 1\arcsec.

The star was observed with a 7-channel bolometer array on IRAM at 1.2~mm
\citep{Bockelee94} and with UKT14 on JCMT at wavelengths from 2~mm to
450 $\mu$m \citep{vanderVeen94,Sylvester96}.  These observations were
all photometric observations. HD\,142666 was also observed with the IRAM
PdB interferometer in continuum at 3.3, 3.1, and 1.2 mm. Although
\citet{Natta04} give no size information, the 1.2 mm flux density from
PbB appears somewhat low compared to single dish photometry, suggesting
that the disk could be extended. The disk is unresolved in our SCUBA
images at 850 $\mu$m, although at 450 $\mu$m we obtain a size of
2\ptsec2. Since the size is not verified at 850 $\mu$m, we consider this
an upper limit. Far infrared flux densities observed with ISO
\citep{Elia05} appear overestimated when compared to IRAS data, and we
have not included them in our analysis. If we only use the
(sub-)millimeter observations and the IRAS 100 $\mu$m flux density, our
graybody fit gives a relatively low dust emissivity, $\beta$ = 0.72 for
a dust temperature of 31 K. The isothermal fit cannot reproduce flux
density at 60 $\mu$m (Figure~\ref{fig-tr1}). To fit the millimeter and
far-infrared SED we need a warm dust component as well, which is true
for most isolated HAEBEs.

\indent {\bf HD\,144432} is a binary star with a late type
companion in the Sco OB2-2 association at a distance of $\sim$ 145 pc
\citep{Perez04}. The spectral type is somewhat uncertain.
\citet{Dunkin97} classified it as A9/F0 Ve, while \citet{Perez04} and
\citet{Blondel06} assign  it a spectral stype A8 Ve or A9 IIIe with an
age of 6.5 - 6.8 Myr. It was included in the ISO survey of isolated
HAEBE stars \citep{Meeus01}, where it was classified as group IIa
object. It was detected at 1.2 mm by \citet{Bockelee94} and at 1.3 mm
and 800 $\mu$m by \citet{Sylvester96}.

We have only SCUBA photometry of this star (Table~\ref{tbl-2a}). A
graybody fit to all (sub-)millimeter and IRAS 100 $\mu$m data give
$\beta$ =  0.59 and a disk mass of 0.002 \Msun. The fit greatly
underestimates the flux density at 60 $\mu$m (Figure~\ref{fig-tr2}),
suggesting that we need a warm dust component as well.

{\bf \indent HD\,150193} is located in Ophiuchus
at a distance of $\sim$ 150 pc. \citet{Mora01} classified it as an A2
IVe star, while  \citet{Blondel06} found it to have a spectral type A3
IIIe from analysis of UV-spectra. \citet{Elias78} classified it as A0.
It is a visual binary with a separation of 1\farcs1 \citep{Reipurth93}. 
The companion is quite faint. \citet{Carmona07} find the secondary to be
a T Tauri star with a spectral type F9Ve.  \citet{Carmona07} find a
common age for the binary system of $\sim$ 10 Myr.  The star was not
detected in the CO J = $3 \to 2$ survey by \citet{Dent05}, confirming
that it is an isolated HAEBE star, which is not associated with any
molecular  cloud. It also appears to have a gas-poor disk. HD\,150193 is
associated with faint free-free emission, 0.2 mJy at 3.6 cm 
\citep{Skinner93,Natta04}, almost certainly originating from a thermal
wind.

The star has strong infrared excess \citep{Elias78} and was first
detected in sub-millimeter continuum  at 800 and 450 $\mu$m with the
single pixel bolometer UKT14 on JCMT by \citet{Jensen96}. It was imaged
with the OVRO interferometer in the 2.6 mm dust continuum and in the CO
J = $1 \to 0$ line \citep{Mannings97}, who found the dust continuum to
be unresolved and  did not detect the disk in CO. It was also imaged by
\citet{Natta04} with the VLA at 7~mm in their study of evolved dust
disks in HAEBE stars.  We report SCUBA observations at 850 $\mu$m (Table
\ref{tbl-2a}), which show an unresolved point-source.  The star was also
imaged by {\it Spitzer} MIPS, which show an isolated point source at 70
$\mu$m (Table~\ref{tbl-3}). It was not detected by IRAS at 100 $\mu$m,
but at 60 $\mu$m the IRAS flux density is more than twice as large as
what we see with MIPS at 70 $\mu$m suggesting that IRAS picked up some
extended emission in its much larger beam.  Fitting an isothermal
graybody disk model to all available millimeter  and sub-millimeter data
and the MIPS 70 $\mu$m data point indicates that the disk is marginally
optically thick at 850 $\mu$m ($\tau \sim$ 0.2) with a temperature of
44~K (Table~\ref{tbl-4}) and a mass of 0.002~ \Msun. The dust
emissivity we derive, $\beta$ = 0.65,  is lower than what
\citet{Natta04} found ( $\beta$ = 1.0 -- 1.6) from their more limited
sampling of the SED. Figure~\ref{fig-tr2} shows that our fit is
reasonably well constrained with an uncertainty in $\beta$ of $\pm$ 0.2.

\indent {\bf KK~Oph} is in the outskirts of the Ophiuchi cloud complex.
It was classified as A8 Vev by \citet{Mora01}, while \citet{Blondel06}
classify it as a  A5 Ve, with a range A5 - A7 Ve from modeling UV
spectra and photometry. It is a binary  with a separation of 1\farcs5
\citep{Pirzkal97} The secondary is a T Tauri star with a spectral type
G6 Ve \citep{Carmona07}. The distance is not well determined, but
\citet{Blondel06} quote $<$ 250 pc from \citet{Wallenquist37}, while
\citet{Leinert04} estimate the distance to be  160 $\pm$  30 pc, which
we adopted in this paper. \citeauthor{Blondel06} estimate an age of
$\sim$ 7.9 Myr, which agrees well with \citet{Carmona07} who find an age
of 7 Myr. Even though {\it Spitzer} MIPS archive images at 24 and 70
\mum\ do not show any sign of nebulosity or cloud emission around the
star, the flux density observed by MIPS at 70 $\mu$m is  rather faint
when compared to the IRAS 60 $\mu$m flux density. The MIPS flux density
at 70 \mum, even without color correction, is only 2.88 $\pm$ 0.02 Jy,
while the ADDSCAN IRAS flux density at  60 $\mu$m is 6.7 $\pm$ 0.7 Jy
\citep{Hillenbrand92}. Since both IRAS and {\it Spitzer} produce well
calibrated data, it would appear that the much broader IRAS beam picks
up some extended emission around the star. The star has weak free-free
emission, 0.25 mJy at 3.6~cm \citep{Skinner93}.

KK~Oph was detected 1.3~mm by \citep{Hillenbrand92} with broad beam
photometry, but there are no other published (sub-)millimeter
observations. We obtained 800 $\mu$m photometry with UKT14 (Table
\ref{tbl-2b}) and mapped it with SCUBA at 850 \mum.  We find KK~Oph to
be faint and unresolved (Table \ref{tbl-2a}). For KK~Oph, we derive a
fairly typical mass of 0.002\Msun\ and a rather shallow $\beta$ index,
0.74; however, the $\beta$ index is not well constrained, since we only
have data covering a narrow wavelength regime. 

{\bf \indent HD\,163296} is an  early A star at a distance of 122 pc and
with an age of $\sim$ 4 Myr (Table~\ref{tbl-1}). The star is isolated,
i.e., there is no sign of the molecular cloud in which the star was
presumably formed \citep{Dent05}. However, \citet{Grady00} show that
HD\,163296 is still an active Ae star, suggesting that there are still
some remnants left of the cloud out of which it was formed. Single-dish
observations of HD\,163296 show strong dust emission at millimeter and
sub-millimeter wavelengths
\citep{Hillenbrand92,Mannings94,Henning94,Henning98}. The star is very
faint in free-free emission \citep{Natta04}. Therefore, the millimeter
emission is completely dominated by thermal emission from dust.
\citet{Isella07} have observed HD\,163296 in continuum with the SMA at
870 \mum, with PdB at 1.3 mm and 2.8 mm, and with the VLA at 7 mm. They
resolve the disk at all wavelengths except 7 mm, where the VLA is less
sensitive or has resolved out the cold, extended, outer part of the
disk. At 870 \mum\ they derive a disk size of 3\ptsec6 $\times$
2\ptsec6. They also imaged the disk in emission from several molecular
lines. From the  $^{12}$CO J = $3 \to 2$ and $^{12}$CO J = $2 \to 1$
lines, they deduce that the velocity pattern is well fitted with a
Keplerian disk with an inclination angle 46\degr\ $\pm$ 4\degr\ and a
stellar mass of 2.6\Msun, while \citet{Grady00} found an inclination
angle of $\sim$ 60\degr\ $\pm$ 5\degr\ from their coronagraphic imaging
of the reflected light from the disk. The radius of the disk observed in
CO, $\sim$ 540 AU, is much larger than in continuum, $\sim$ 200 AU; such
a behavior can be explained by an exponential falloff of the surface
density at large radii \citep{Hughes08}.

We find HD\,163296 to be extended and well resolved at both 850 and 450
$\mu$m (Table \ref{tbl-2a}; Figure \ref{fig-hd163296}). Although there
are many observations of HD\,163296 in the JCMT archive, very few of the
observations were done in good sky conditions with sufficient
calibration observations to enable us to obtain well-calibrated data
sets with good beam characterization. We retained only three data sets
taken in engineering time in 2004 and 2005. Even though we find the dust
emission around HD\,163296 to be slightly elliptical, the elongation is
not large enough to permit us to determine the position angle of the
disk with high accuracy at either 850 or 450 $\mu$m. For our 450 $\mu$m
data sets, where the HPBW is $\sim$ 8\arcsec, we created a model beam by
rotating the measured beam profiles to the average position angle of
each map of HD\,163296, since the beam is always somewhat elongated in
the chop direction (see Section \ref{obs}). We added the three rotated
beam profiles together to form a beam profile which best approximates
the beam of the coadded 450 $\mu$m map. After cleaning the 450 $\mu$m
map with this model beam, we find a FWHM = 4\ptsec6 $\times$ 3\ptsec2,
P.A. = $-$13\degr\ $\pm$ 27\degr, with an uncertainty of $\sim$
1\arcsec\ (Table \ref{tbl-2a}). The size of the 450 $\mu$m disk is
slightly larger than what \citet{Isella07} determined from their 870
$\mu$m SMA observations, but both results agree well within
observational errors. The total flux densities derived by
\citeauthor{Isella07} also agree very well with single dish results at
1.3~mm and 870 \mum, confirming that the aperture synthesis observation
have recovered all the continuum emission from the disk. We can
therefore use both interferometric data and single dish data to
determine the emissivity index, $\beta$, and disk mass. An isothermal
graybody fit to these data gives T = 28 K and  $\beta$ = 0.94. By
assuming a gas-to-dust mass ratio of 100, we deduce a disk mass $\sim$
0.065 \Msun. The graybody fit underestimates the IRAS flux density at 60
\mum, suggesting that there is an additional contribution from warm dust
at 60 \mum, although its contribution to the disk mass is negligible.
Our results are very consistent with the results of \citet{Isella07}.
These studies indicate that some grain growth has already occurred in
the HD\,163296 disk.

\indent {\bf HD\,169142} was classified as A5 Ve \citep{Dunkin97,Boekel05}, A7Ve
\citep{Blondel06} and as A9 III/IVe \citep{Guimaraes06,Vieira03}. Here
we adopt the spectral class A7 Ve, which is based on
fitting low resolution UV spectra with a three component model, the
stellar photosphere, an optically thick accretion disk, and  a boundary
layer between the disk and the star. The analysis of \citet{Blondel06} indicate that the
disk is seen approximately face-on, which agrees with agrees with other
findings (see below) and predicts a distance of 150~pc and an age of $\sim$ 12 Myr.
This distance agrees very well with the distance of 145~pc derived by \citet{Boekel05}, which we
have adopted here.
\citet{Dent06} did not detect any free-free emission at 3.6~cm, although
 the flux density they observed at 7~mm appears anomalously high and is
most likely due to free-free emission or a radio flare (see below). It
was included in the ISO survey of isolated HAEBEs \citep{Meeus01} and
classified as a Ib type HAEBE star. HD\,169142 shows a narrow emission
line in CO  \citep{Dent05} and in [OI] at 6300 \AA \citep{Acke05},
suggesting that the disk must be seen nearly face-on. \citet{Meeus10} modeled
the disk including new {\it Herschel} data, and found that in order to explain the
weak 10 $\mu$m excess emission, the disk must have a gap, i.e. it appears to be
a transition disk.

In the (sub-)millimeter the disk was first detected in photometry mode at
1.2 mm with the 7-channel bolometer on IRAM  \citep{Bockelee94} and at 2
mm, 1.3 mm, and 800 $\mu$m with UKT14 on JCMT \citep{Sylvester96}. Our
observed continuum flux densities (Table \ref{tbl-2a}) agree well with
the photometry results and suggest that the disk may be resolved with a
size of 2\farcs3 $\times$ 1\farcs6 (Figure~\ref{fig-hd169142}).
\citet{Raman06} observed HD\,169142 with the SMA in continuum at 1.3 mm
and in the CO J = $2 \to 1$ line. They barely resolve the continuum
emission (FWHM $\sim$ 0\ptsec9) but find that the CO emission traces a
Keplerian disk around the star. Their best fit disk model gives a radius
$\sim$ 235 AU (1\ptsec6), a disk mass of $\sim$0.02\Msun, and an
inclination of 13\degr. This result agrees well with the SED modeling by
\citet{Dent06},  who find a disk radius of 300 AU, a disk mass of 0.04
\Msun\ and an inclination angle of $\sim$ 30\degr. Our observed disk
size, r  $\sim$ 140 AU, is consistent with these previous measurements.
The star was also observed by the PACS imager on  {\it Herschel} in
continuum at 160 and 70 $\mu$m \citep{Meeus10}. Our graybody fit 
includes all (sub-)millimeter and far infrared data up to  100 $\mu$m,
except the 200 $\mu$m ISO value  \citep{Walker00}, which appears
anomalously low, and the 7 mm VLA data point, which is too high to be
due only to dust. It must therefore include a significant contribution
from free-free emission. When we include the accurate 160 $\mu$m flux
density from PACS, we get  $\beta$ = 1.30, a dust temperature of 41 K,
and a disk mass of 0.03 \Msun\ (Table~\ref{tbl-4} and
Figure~\ref{fig-tr2}). The latter is in good agreement with the results
of \citeauthor{Raman06}. The dust emissivity that we obtained, $\beta$ =
1.30, is more than twice as large as what \citet{Dent06} found, because
they did a fit to a more limited set of data and assumed that all the
flux density at 7 mm was due to dust emission. As we can see from
Figure~\ref{fig-tr2}, the fit underestimates the observed flux densities
shortward of a 100 $\mu$m, requiring a warm dust component as well.
Since we have reliable far-infrared data on HD\,169142 we also fitted
the the millimeter and far-infrared SED with a two-component graybody
model, where we allowed the dust properties, dust temperature and size of 
the emitting region to vary for each component. However, we set the dust emissivity for the warm component equal to two, although the fit is not very sensitive to the emissivity of the warm gas.  The two-component fit results in a slightly higher dust temperature for the cold dust, 52 K vs. 41 K. The dust emissivity for the cold dust component is now 1.38 compared to 1.30 for the single temperature fit, i.e., a marginal change.  The temperature of the warm dust is 155 K. The two-component graybody fit  now
matches the 70,  60, and 25 $\mu$m flux densities very well, see Figure~\ref{fig-tr2}.

\indent {\bf MWC\,349} is a remarkable binary system at a distance of
1.2 kpc with a luminosity of $\sim$ 3 $\times$ 10$^4$\Lsun\ \citep{Cohen85}. 
MWC\,349\,A was classified as a Be star by \citet{Merrill33}. Its
optical and infrared spectrum is complicated. The star has very strong
hydrogen and He I lines  \citep{Thompson77,Cohen85}.  Many lines are
double peaked, suggesting they are formed in a Keplerian disk viewed
edge-on \citep{Hamann86,Hamann88}. The secondary is at a separation of 
$\sim$2\farcs4 and has a spectral type B0 III \citep{Cohen85}.
MWC\,349\,A is the brightest known radio star at centimeter wavelengths and the
only known hydrogen recombination line maser source. Maser action is seen in
hydrogen recombination lines from  $\sim$ 100 GHz (H39$\alpha$) to 1800
GHz (H10$\alpha$), with the all masering lines being double peaked and with
the two peaks coming from the receding and approaching parts of the  disk,
due to Keplerian rotation \citep{Strelnitski96}.  Direct imaging of the
H30$\alpha$ recombination line with the SMA at 231.9 GHz confirm that
the maser spots originate in an approximately edge-on disk with a
rotation curve that is somewhat steeper than that expected for a simple Keplerian
rotation  \citep{Weintroub08}.

The radio emission  of MWC\,349A is approximately proportional to
$\nu^{0.6}$, consistent with what is  expected from an ionized stellar
wind \citep{Olnon75}. High resolution VLA imaging shows that the radio
emission is bipolar, with a dark lane separating the two lobes
\citep{White85,Cohen85}. \citet{Tafoya04} argue that the dark lane with
no free-free emission must be due to neutral gas. \citeauthor{Tafoya04}
show that the flux density increases with frequency as $\nu^{0.67 \pm 0.03}$,
and the angular size decreases with frequency as $\nu^{-0.74 \pm 0.03}$,
confirming the presence of a biconical thermal wind that expands at
constant velocity. The slow ($\sim$ 50~km~s$^{-1}$) bipolar wind is
believed to originate from  photoevaporation of the neutral disk
\citep{Hamann86,Lugo04}.  Analysis of  VLA data over a 20 year
time period \citep{Tafoya04,Rodriguez07b} suggest that the radio emission
is fading by $\sim$ 1.5\% per decade. The free-free emission still
dominates at millimeter-wavelengths and perhaps even at 100 $\mu$m
\citep{Harvey79}. 

Deep H$\alpha$ imaging of MWC\,349 shows a faint shell to the north of
the star with a radius of 75\arcsec\ \citep{Marston08}. This shell is
seen  to the east and north-east of the star in a MIPS 70 $\mu$m archive
image, suggesting that MWC\,349\,A is illuminating a faint low surface
brightness \ion{H}{2} region.

Our SCUBA observations at 850 $\mu$m and 450 $\mu$m show a
point-like source with flux densities that are consistent with free-free
emission (Table~\ref{tbl-2a}, Figure~\ref{fig-mwc349sed}). However, if we
assume that the wind is launched at $\sim$ 15 AU from the star
\citep{Tafoya04}, the wind should become optically thin at $\sim$ 450
$\mu$m, and the free-free emission will then be proportional to 
$\nu^{-0.1}$. We may therefore start to pick up some dust emission
shortward of 450 $\mu$m. We currently have no observations that
enable us to measure the mass of dust and neutral gas in the disk.

\subsection{Embedded HAEBE Stars}\label{emb}

All embedded HAEBE stars are essentially ''classical'' HAEBEs, i.e., they
illuminate reflection nebulae and are still located within the clouds in
which they were formed. On average they are younger than the
isolated HAEBEs and many drive bipolar outflows and can be deeply embedded.
Their spectral classifications, and hence the ages, are also more
uncertain, because some of them are very faint or invisible in the UV
and visible parts of the spectrum. Some of them also have much higher
accretion rates and veiling than isolated HAEBEs. Many of the embedded
stars that we have observed are early to mid-B stars illuminating large
reflection nebulae. With one or two exceptions they are
non-detections, i.e., they no longer have observable disks in the
sub-millimeter. These non-detections will be discussed in the next section (Section
\ref{non}).

\indent {\bf Lkh$\alpha$\,198} and {\bf V\,376\,Cas} are late B and A or F  stars
at a distance of $\sim$ 600 pc (Table \ref{tbl-1}). Both stars have been
well studied at optical, near-infrared, far-infrared and sub-millimeter
wavelengths. LkH$\alpha$\,198 has a deeply embedded infrared companion,
LkH$\alpha$\,198\,B, $\sim$ 6\arcsec\ to the north
\citep{Lagage93,Koresko97} with a luminosity of $\sim$ 100 \Lsun, i.e.,
roughly similar to that of LkH$\alpha$\,198. LkH$\alpha$\,198 drives a
bipolar molecular outflow \citep{Canto84,Nakano90} and excites several
Herbig--Haro objects. A deeply embedded, cold protostellar object,
LkH$\alpha$\,198-mm, $\sim$ 19\arcsec\ to the northwest, dominates the
sub-millimeter emission in this region \citep{Sandell94a,Henning98} and
may power the large bipolar molecular outflow  \citep{Canto84,Nakano90}
that originally was believed to originate from LkH$\alpha$\,198.
LkH$\alpha$\,198 and LkH$\alpha$\,198\,B lie inside an elliptical
limb-brightened nebulosity. Both stars appear to excite Herbig-Haro
objects \citep{Corcoran95,Koresko97}. \citet{Koresko97},  using adaptive
optics compensated speckle observations in the H band, found a bar-like
structure extending to 3\arcsec\ on each side of LkH$\alpha$\,198. They
found LkH$\alpha$\,198\,B to be extended at J and H as well, suggesting
that both stars may be surrounded by circumstellar disks.
\citet{Perrin04} used laser guide star adaptive optics and near-infrared
dual-channel imaging polarimetry  to uncover a strongly polarized
biconical nebula around LkH$\alpha$\,198 with a radius of 5\arcsec, as
well as a polarized, extremely blue, jet-like feature emanating from
LkH$\alpha$\,198\,B. Although these results strongly suggest that
LkH$\alpha$\,198 has a disk, they still largely trace the surrounding
infalling envelope. V\,376\,Cas has a bar feature, it illuminates a
sickle-shaped nebulosity, is very highly polarized and excites several
Herbig-Haro knots \citep{Piirola92,Corcoran95}. All of this suggests
that it is also surrounded by a circumstellar disk.  Such a disk has
been confirmed by speckle interferometry  at H and K' \citep{Smith04},
which show that V\,376\,Cas is surrounded by a compact, nearly edge-on
flared disk with a radius of 100 mas.

CO J = $2 \to 1$ mapping of the central part of the CO outflow with
20\arcsec\ spatial resolution \citep{Sandell94a} suggested that there is
more than one outflow in the vicinity of LkH$\alpha$\,198 and
V\,376\,Cas. LkH$\alpha$\,198 and/or LkH$\alpha$\,198\,B could be
driving one outflow, V\,376\,Cas another, and LkH$\alpha$\,198-mm could
be the source for the large scale CO outflow. High spatial resolution CO
J = $1 \to 0$ imaging  ($\sim$ 6\arcsec{}) with the Berkeley Illinois
Maryland Array (BIMA) supplemented by single dish mapping to fill in
missing short spacings \citep{Matthews07} confirm that there are several
outflows. V\,376\,Cas drives one outflow, LkH$\alpha$\,198-mm drives
another, while there appear to be two outflows  associated with
LkH$\alpha$\,198. LkH$\alpha$\,198 definitely powers one outflow. The
second outflow is closer to LkH$\alpha$\,198\,B,  but since it does not
show any high velocity wings, they associate the driver of the outflow
with a yet unseen source between LkH$\alpha$\,198 and
LkH$\alpha$\,198\,B. 

Our 850 $\mu$m map (Figure~\ref{fig-lkha198}) shows
extended dust emission over $\sim$ 3\arcmin. In the northeast, the
emission becomes negative due to more extended and possibly stronger
emission to the west, i.e., an artifact due to chopping. The 850 $\mu$m
dust emission is dominated by the protostellar source
LkH$\alpha$\,198-mm. There is a tail of emission to the southeast from
LkH$\alpha$\,198-mm  and a fainter ridge extending north towards
V\,376\,Cas.  The southwestern extension can be explained by an extended
source, $\sim$ 11\farcs4 $\times$ 5\farcs6,  centered on
LkH$\alpha$\,198 and its infrared companion with a position angle
similar to that of the binary system (Table~\ref{tbl-2a}). The peak of
emission is close to LkH$\alpha$\,198\,B, which agrees with the 1.3 mm
map by \citet{Henning98}, which has slightly better angular resolution
and shows a secondary peak on LkH$\alpha$\,198. Both the 1.3 mm map and
the 850 $\mu$m map suggest that  LkH$\alpha$\,198 and LkH$\alpha$\,198B
are surrounded by dust disks. We
estimate that  at most one third of the emission from the
LkH$\alpha$\,198 binary is due to LkH$\alpha$\,198. If we assume a dust
temperature of 50~K and a $\beta$ = 1.5, we find that LkH$\alpha$\,198 has a disk mass of
$\leq$ 0.15 \Msun. 

Interferometer observations
\citep{Francesco97,Matthews07} at 2.7 mm did not have enough sensitivity
to detect LkH$\alpha$\,198 or V\,376\,Cas, although
\citeauthor{Francesco97} did detect LkH$\alpha$\,198-mm. They found it
to be unresolved or marginally resolved, while our SCUBA observations
show it to be quite extended (FWHM $\sim$ 20\arcsec{}) with  an
integrated flux density of 1.78 Jy.  We do see an enhancement in
continuum level near V\,376\,Cas, but the position we deduce is off by
3\arcsec\ from the optical position and it could simply be due to
emission from the surrounding dust cloud. However, observations at 450
or 350 $\mu$m at JCMT or CSO would have enough spatial resolution to
resolve the HAEBE binary from the protostellar source and to determine
how much of the emission originates from LkH$\alpha$\,198-IR and how
much originates from LkH$\alpha$\,198.

\indent {\bf V\,892~Tau} is a heavily obscured (A$_V$
$\sim$ 10 mag)  late B star with a strong infrared excess
\citep{Elias78,Strom94}.  It was first discovered by \citet{Elias78} in
a near-infrared survey of the Taurus dark cloud complex. Inspection of H$_2$
column density maps by \citet{Goldsmith08} show it to be located in an
opaque area of L\,1495. Multiepoch trigonometric parallax observations
with the  Very Long Baseline Array (VLBA) \citep{Torres09} indicate that
L\,1495 is on the near side of the Taurus complex at a distance of
$\sim$ 130 pc. \citet{Hernandez04} derived  a spectral type of B8 Ve for
V\,892~Tau, while \citet{Strom94} classified it as B9. Spectral types
as late as A6 have been reported in the literature \citep{Berrilli92}.
near-infrared speckle observations and  mid-infrared interferometry show that it is
a close  binary with a separation of 50 mas \citep{Smith05,Monnier08}. 
\citet{Monnier08} analyzed the SED for the binary and deduced a
bolometric luminosity of 350 \Lsun\ (corrected to 130 pc), with both
stars having the same spectral type. The inferred luminosity is too low
for two B8 V stars, and the stars are more likely spectral type B9 V.
Mid-infrared segment-tilting interferometry \citep{Monnier08} shows a
circumbinary disk inclined at 60\degr\ with an inner hole diameter of 33
AU. 

V\,892~Tau  was first detected  in dust continuum at 1.3 mm by
\citet{Beckwith90}. \citet{Mannings94} carried out extensive
(sub-)millimeter photometry (1.3 and 1.1 mm, 850, 800, 750, 450, and 350
$\mu$m) of V\,892~Tau with the single pixel detector UKT14 on JCMT, and
found it to have a very flat spectral index, $\alpha$ = 2.18 $\pm$ 0.10.
It was mapped at 1.3~mm by \citet{Henning98}, who found it compact and
unresolved. \citet{Francesco97} detected it in the dust continuum with
the VLA at 7 mm and at 2.7 mm with the BIMA array, but  did not resolve
the continuum emission. Recent CARMA observations by \citet{Hamidouche10}
at 2.7 and 1.3 mm resolve the disk with a size of 0\farcs4 at 2.7 mm.

V\,892~Tau is a very bright X-ray source \citep{Strom94}. It is only
seen in hard X-rays, suggesting that it has low or no accretion
\citep{Hamidouche08}. However, it is a thermal radio source
\citep{Skinner93,Francesco97}, suggesting that there is still some
accretion going on, although it is not known to drive a molecular
outflow. \citet{Strom94} deduced an age of $\sim$ 3 Myr by comparing
its location on the H-R diagram with two different evolutionary track
models. \citet{Hernandez04} assumed it to be a B8 V star and found it
to lie on the main sequence. \citet{Strom94} found it somewhat puzzling
that the star would appear older than nearby T Tauri stars, when it is
still deeply embedded in the molecular cloud core and illuminates a
reflection nebula. \citet{Monnier08} find that it has an SED resembling a transition
disk object  and a clear inner gap in the disk, which
is one of the main signatures of an evolved disk. However, since it is a binary, the 
gap in the disk is most likely caused by the binary.

Our SCUBA images show that V\,892~Tau is unresolved at both 850 and 450
$\mu$m (Figure~\ref{fig-V892_SCUBA}), consistent with the small size found
by \citet{Hamidouche10}.  We also retrieved and analyzed archive {\it Spitzer}
MIPS images at 24 and 70 $\mu$m. At 24 $\mu$m the star is heavily
saturated and we could not deduce a flux density.  The 70 $\mu$m image
is of excellent quality and shows extended warm nebulosity  east and
north-west of the star (Figure~\ref{fig-V892_MIPS}). We did psf
photometry of this image and deduced a color corrected flux density at
70 $\mu$m of 27.0 $\pm$ 0.4 Jy.  However, examination of the residuals
from psf-fitting indicate that the star is mildly saturated. The deduced
flux is therefore strictly a lower limit, although probably not by much.
\citet{Francesco98} mapped V\,892~Tau with the Kuiper Airborne
Observatory (KAO) at 50 and 100 $\mu$m. They found much lower flux
densities than IRAS. At 50 $\mu$m they found a peak flux density  of
59.1 $\pm$ 5.5 Jy and a source size of 25\arcsec, resulting in an
integrated flux of  86 $\pm$ 45 Jy. Although the star dominates at 50
$\mu$m, the KAO observations probably include some contribution from the
extended warm nebulosity, which we see in the MIPS 70 $\mu$m image.

Isothermal graybody fits fit the millimeter and sub-millimeter data very
well, but underestimate the flux at 100 $\mu$m and shorter wavelengths.
Since the KAO beam at a 100 $\mu$m is $\sim$ 30\arcsec, the observed 100
$\mu$m flux  density is an upper limit.  The temperature of the dust
emission is therefore somewhat uncertain; 40 K is probably a reasonable
assumption. For an assumed dust temperature of  40 K we find a dust
emissivity, $\beta$ = 0.52, and a disk mass of 0.009 \Msun.  If the dust temperature is
60 K we get $\beta$ = 0.40, and a dust mass of  0.004 \Msun. In both cases the dust is somewhat optically thick at 850 $\mu$m, but less so for warmer dust.
As we can see from Figure~\ref{fig-V892_sed} we can fit the millimeter
and sub-millimeter data extremely well, but the 70 $\mu$m flux density
is underestimated. The V\,892~Tau disk therefore appears very similar to the
isolated HAEBE stars, which all require a warm dust component. The dust
emissivity, which is reasonably well constrained, is rather low,
suggesting that grain growth has already occurred in the disk.
V\,892~Tau has all the characteristics of an evolved disk.

%\citet{Hillenbrand92}, however, 
%classified it as a group II object, i.e. a young HAEBE star surrounded by gas and dust not confined to
%a disk. 
%It is a binary system with a separation of
%$\sim$ 4\arcsec, and the companion is a T Tauri star \citep{Skinner93}. 

\indent {\bf LkH$\alpha$\,101} is a heavily obscured (A$_V$ $\sim$ 10
mag), infrared bright Herbig Be star with a spectral type $\sim$ B0 Ve and
luminosity of 4 $\times$ 10$^4$ \Lsun, that is embedded in the dark cloud L\,1482 at a
distance of $\sim$ 700~pc \citep{Herbig04}. It illuminates the
reflection nebula NGC\,1579 and ionizes the \ion{H}{2} region S\,222
\citep{Herbig60,Herbig71,Herbig04}. It is the brightest member of a
deeply embedded cluster of heavily reddened near-infrared sources
\citep{Barsony91,Aspin94b} and faint cm-radio sources
\citep{Becker88,Stine98}, most of which are probably young PMS stars.
Interferometric imaging in the near- and mid-infrared
\citep{Tuthill02a,Tuthill02b} shows that LkH$\alpha$\,101 surrounded by
a  dust disk with a hot inner wall facing the central star. The disk is
seen nearly face on and shows a central hole or cavity.  At 11.15 $\mu$m
the size of the disk is $\sim$ 0\ptsec06 (10 AU). They also find that
the star is a binary system with a low luminosity companion $\sim$
0\ptsec18 to the northwest; this result has  been independently confirmed by
speckle interferometry \citep{Alvarez04}. LkH$\alpha$\,101 is a strong
radio source with a spectrum  consistent with a spherically symmetric
constant velocity  wind \citep[and references therein]{Becker88,Gibb07}.
Although the star has been well studied in the centimeter regime, there is a lot of
scatter in the free-free flux densities and size, even if we only look
at VLA A-configuration data (Figure~\ref{fig-lkha101sed}). This may be
due to contamination from the surrounding \ion{H}{2} region and because
different array configurations resolve out more or less of the extended
emission. It is also possible that the free-free emission is variable. If we use all
high-spatial resolution observations (i.e., VLA A-array and MERLIN) available in
the literature we get a spectral index $\alpha$ = 0.71 $\pm$ 0.07, which is
steeper than but consistent with what \citet{Gibb07} found, i.e., $\alpha$ = 0.61.

In the sub-millimeter (Figure~\ref{fig_lkha101_850}), LkH$\alpha$\,101 is located at the
southwestern boundary of an extended  elliptical cloud with a size of
3\arcmin\ $\times$ 1\arcmin. Although the map covers five of the deeply
embedded VLA sources seen by \citet{Stine98}, only LkH$\alpha$\,101 has
a sub-millimeter counterpart. At 850 $\mu$m the star appears somewhat
extended with a size of $\sim$ 6\farcs5 $\times$ 2\farcs6 or $\sim$ 2800 AU. This size is
probably an over-estimate, since the mapping was done in scan-map mode with
relatively poor sky transmission. The dust disk, however, is not very massive, since
most of the emission at 850 $\mu$m is due to free-free emission. In the
low signal-to-noise 450 $\mu$m map the emission is completely dominated
by the surrounding dust cloud and we cannot derive a flux density for the star. After
subtracting the free-free emission we are left with $\sim$ 180 mJy at
850 $\mu$m. If we assume  $\beta$ = 1 and a dust temperature of 50 K we
find that LkH$\alpha$\,101 could have a  disk mass of $\sim$ 0.005
\Msun, which is similar to the  disks of isolated Ae stars.

{\bf \indent AB~Aur} is a bright, optically visible A1 Ve star located on
the front side of the dark cloud L\,1517 in the Taurus-Auriga dark cloud
complex at a distance of 144 pc \citep{Ancker98,Hernandez04}. It is $\sim$
3\arcmin\ away from the classical T Tauri star SU~Aur (spectral class G2
IIIe), which is also surrounded by an accretion disk
\citep{Weintraub89,Andrews05}. AB Aur was included in the
\citet{Meeus01} sample of isolated HAEBEs. However, AB Aur is not an
isolated HAEBE star,  since it is still located within the dark cloud in which it
was formed. Its age, however, 4 $\pm$ 1 Myr \citep{Dewarf03} is similar
to some of the younger isolated HAEBEs. Near-infrared coronagraphic imaging with
adaptive optics on Subaru \citep{Fukagawa04} show extended emission as far as
$\sim$ 4\arcsec\ from the star and resolve the emission into a
spiral-like structure with at least two spiral arms.

AB Aur has been extensively studied in the (sub-)millimeter, both with
single dish telescopes and interferometers \citep{Mannings94,
Mannings97, Pietu05, Semenov05, Corder05, Lin06}. The star is detected
in free-free emission \citep{Gudel89, Skinner93, Rodriguez07a} but is
very faint, emitting only 0.2 $\pm$ 0.02 mJy at 3.6~cm. High spatial
resolution (0\ptsec6 - 2\arcsec{}) interferometric imaging at millimeter
wavelengths \citep{Pietu05,Corder05,Lin06} in various CO transitions and
their isotopomers show that disk is seen nearly face-on with an
inclination angle between 20\degr\ and  40\degr\ and appears to be
non-Keplerian. \citet{Pietu05} and \citet{Corder05} resolve the
continuum emission and find it asymmetric, possibly tracing the spiral
structure seen in near-infrared.  \citet{Lin06}, who observed AB~Aur  in
$^{12}$CO J=$3 \to 2$ with the SMA, confirm by tracing the inner spiral
arm seen in the near-infrared in several velocity channels that the
disk deviates from Keplerian rotation. Their 870 \mum\ continuum
observations show a central depression towards the star, suggesting the presence of a
large inner dust-free hole around the star. \citeauthor{Lin06} suggest
that both the central depression and the spiral structure with
non-Keplerian motions  could be caused by a giant planet forming in the
disk. This view is also supported by \citet{Oppenheimer08}, who did
adaptive optics coronagraphy and polarimetry of the AB Aur inner disk.
They find an azimuthal clearing in the dust  at  a radius of $\sim$ 102
AU along with a clearing of dust at radii inside this annulus,
suggesting the formation of at least one planetary body at $\sim$ 100
AU. \citet{Rodriguez07a} also detected faint radio emission from a
source located about 40 AU from AB Aur, suggesting that the star could
have a low mass stellar companion.

We find AB Aur unresolved both at 850 $\mu$m and 450 $\mu$m (Figure
\ref{fig-abaur}), even though  \citet{Lin06} resolved it at 870 \mum\
with a size of 3\farcs1 $\times$ 1\farcs9. The star also appears compact
in a MIPS 70 $\mu$m image, but the star  is somewhat saturated and we
could not retrieve a flux density.  Our flux density measurements with
SCUBA (Table~\ref{tbl-2a}) agree within errors with our less accurate
UKT\,14 photometry (Table \ref{tbl-2b}).  Figure \ref{fig-abaurfit}
shows a least-squares graybody fit to all reliable (sub-)millimeter and
far-infrared data of AB Aur, resulting in a dust temperature of 66 K
and a dust emissivity index, $\beta$ = 1.05. If one assumes a normal
gas-to-dust ratio, the mass of the disk is  $\sim$ 0.006 \Msun. This
disk mass is consistent with the results of \citet{Lin06} and
\citet{Corder05}, but  lower by a factor of three compared to the mass
obtained by \citet{Pietu05}.
%AB Aur binary Baines et al 2006

\indent {\bf VY~Mon} is a late B to early A star at the southern edge of
the bright reflection nebula IC\,446  at a distance of 800 to 900 pc
\citep{Casey90}. The star is heavily reddened and variable with a
bolometric luminosity of  $\sim$ 870\Lsun\  for a distance of 800~pc.
\citet{The94} assigned it a spectral class B8:e, while \citet{Mora01}
classified it as A5 Vep. The observed luminosity appears too high for an
A5 star, and we have therefore adopt the spectral classification by
\citeauthor{The94}, i.e., B8. The star is a faint radio source
\citep{Felli82}. VY Mon  drives a high-velocity outflow, which we mapped
in CO J = $2 \to 1$ on JCMT\footnote{The original data for these CO observations are missing in the
JCMT archive and in our personal archive,
   but we have preserved
   in our records and present in this paper the maps derived
previously from those data.}.
Figure~\ref{fig-vymon_spec} shows high velocity emission from about
$-$28 to $+$18 km~s$^{-1}$. The outflow is very compact and appears to
be seen pole on, since the blue and red-shifted outflow lobes are
completely overlapping (Figure~\ref{fig-vymon_map}). \citet{Wang07}
analyzed near-infrared and {\it Spitzer} IRAC observations of IC\,446 and find
that VY Mon is part of a compact group of 26 YSOs within a radius of 0.5
pc.

VY~Mon was mapped at 1.3 mm by \citet{Henning98}, who found strong,
compact 1.3~mm emission centered on the star and faint extended emission
from the dark cloud south of VY~Mon. We have no SCUBA observations of VY
Mon, and our single beam photometry with UKT14 only goes down to
800~$\mu$m (Table~\ref{tbl-2b}). At 1.3~mm our observed flux density is
higher than what  \citet{Henning98} measured from their higher spatial
resolution IRAM observations, suggesting that we pick up some extended
emission from the surrounding cloud in our larger beam. A graybody fit
to the (sub-)millimeter data underestimates the far-infrared flux
densities observed by \citet{Casey90} and gives $\beta$ = 1.9 and a disk
mass of 1.9 \Msun. This mass is almost certainly too high to be from the disk alone and
suggests that our single dish observations may  include a substantial
fraction from the surrounding infalling envelope. The high value of
$\beta$ is also consistent with unevolved (small) dust grains, which is
additional evidence that this flux is capturing emission from the
infalling envelope.

\indent {\bf AFGL\,961} is a deeply embedded (A$_V$ $\geq$ 18 mag),
young, high-luminosity, double star system with a separation of $\sim$
6\arcsec\ \citep{Castelaz85}. Both stars appear to be early B stars and
are surrounded by extended nebulosity in the near-infrared
\citep{Castelaz85,Aspin98}. AFGL\,961 is southwest of the the \ion{H}{2}
region NGC\,2244 (the Rosette nebula) and part of the Mon OB2
association, which lies at a distance of 1.4 - 1.6 kpc
\citep{Turner76,Hensberge00}. Here we adopt 1.5 kpc for the distance to
AFGL\,961. The double star system is located at the center of a large
bipolar molecular outflow \citep{Lada82,Dent09}. The main outflow has an
extent of 5\arcmin\ at a P.A. of 45\degr, but \citeauthor{Dent09} found
at least one other, more compact outflow orthogonal to the main outflow.
Early millimeter continuum observations \citep{Henning90} easily
detected AFGL\,961, but had insufficient spatial resolution to identify
which star is responsible for the dust emission. The same was true with
a more recent (sub-)millimeter imaging survey by \citet{Klein05}, who
showed that AFGL\,961 is embedded in a dense sub-millimeter core with a
size of 0.3 pc and a mass of $\sim$ 550 \Msun.

The eastern component, AFGL\,961\,A (also known as AFGL\,961\,E) is 
more deeply embedded and a faint free-free emission source
\citep{Bally83}, with stronger thermal excess in the near-infrared than
AFGL\,961\,B (also known as AFGL\,961\,W), although both stars have
Br$\alpha$ in emission \citep{Aspin98}. AFGL\,961\,B  shows the 2.3
$\mu$m CO bandheads in emission, while the eastern source shows a
featureless spectrum except for the Br$\alpha$ emission. Analysis of
2MASS and deeper near-infrared imaging \citep{Li05,Roman-Zuniga08} shows that
the majority of stars in the Rosette molecular cloud complex are
contained in clusters. In the near-infrared AFGL\,961 with its two
early B stars, appears to be a very small cluster that is barely detectable
in the deep near-infrared imaging  by \citet{Roman-Zuniga08}, who
identify it as PL06. Recent  high-spatial resolution mid-infrared and
sub-millimeter imaging with the SMA  \citep{Williams09} show that PL06
is a deeply embedded, very young cluster. They find three continuum
sources at 1.4~mm. The strongest one, SMA\,1 coincides with
AFGL\,961\,A, while two of them have no near- or mid-infrared
counterparts. SMA\,2 appears to be starless, but accreting, i.e., a
pre-stellar, collapsing dense core, while SMA\,3 drives an outflow, and
is possibly driving the large NE-SW outflow. SMA\,3 is therefore a Class
0 protostar. They also identify a third mid-infrared source,
AFGL\,961\,C, $\sim$ 30\arcsec\ west of A, with strong mid-infrared
excess, but they did not detect it in continuum at 1.4~mm. AFGL\,961\,A
was also detected with the VLA at  7~mm, with a flux density of 8.1
$\pm$ 0.1 mJy \citep{Zapata09}.

%de Wit 2009
At 850 $\mu$m we find  a strong  elliptical source centered slightly
northwest of AFGL\,961\,A and superposed on relatively strong, extended
emission from the surrounding cloud (Figure~\ref{fig-afgl961}). At 450
$\mu$m the elliptical source is resolved into two compact sources. The
eastern one is approximately centered on AFGL\,961\,A, possibly with
some contribution from SMA\,2. The second 450 $\mu$m source,
AFGL\,961\,SMM overlaps with SMA\,3, but is definitely centered east of
SMA\,3, suggesting that there are even more sub-millimeter sources in the
AFGL\,961 cloud core. To learn more about this complicated young
cluster, we therefore also retrieved and reduced {\it Spitzer} archive
MIPS 24 and 70 $\mu$m images of AFGL\,961. The 70 $\mu$m image is
plotted in gray scale in Figure~\ref{fig-afgl961} and overlaid with the
850 $\mu$m SCUBA image in contours. In Figure~\ref{fig-afgl961} we also
separately plot both the 70 and the 24 $\mu$m images over the same area
mapped by SCUBA. AFGL\,961\, A and B are heavily saturated both at 24
and 70 $\mu$m, but SMA\,3 coincides with a clear 70 $\mu$m peak. There
is, however, a clear extension to the southeast towards A \& B confirming
that the  western 450 $\mu$m  source has a 70 $\mu$m counterpart. What
we see at 450 and 70 $\mu$m are probably two partially blended sources,
one of which coincides with SMA\,3. At 24 $\mu$m  AFGL\,961\,C is also
seen as a double source, but with the southeastern component stronger
than C. The second source (which we call 961\,D) is also seen by
\citet{Williams09} at 20 $\mu$m as a faint extension to the southwest.
At 70 $\mu$m there is no clear peak at C and D although both sources lie
within the relatively bright 70 $\mu$m core. A much stronger ridge of
emission is seen protruding to the south-southeast of A \& B. This ridge
of emission is also seen in the SCUBA images, and suggests that there
may be yet  more deeply embedded sources hidden in the AFGL\,961 cloud
core.

Analysis of the embedded cluster population is outside the scope of this
paper, but AFGL\,961\,A has all the characteristics of a very young
Herbig Be star, which most likely is surrounded by an accretion disk and
an infalling envelope. Although \citet{Williams09} did not find an
outflow from the star, the spectral index of the free-free emission
between 5 and 15 GHz is consistent with the free-free emission
originating in an ionized wind or jet \citep{Snell86}. A graybody fit to
all (sub-)millimeter data gives a dust temperature of $\sim$ 37 K, a
dust emissivity, $\beta$ $\sim$ 1.51, resulting in a total mass of the
disk/envelope of 12.5 \Msun. This fit overestimates slightly the flux density of
SMA\,1 at 1.4~mm, but is rather close to the sum of SMA\,1 and 2,
suggesting that we pick up emission from both sources in our much larger
SCUBA beam. On the other hand the fit severely underestimates the 7~mm flux
density, which only comes from A (SMA\,1), suggesting that part of the
emission may be free-free emission, rather than thermal dust emission.

To get a better idea of the true luminosity and mass of AFGL\,961\,A we
have  additionally explored the archive of two-dimensional axisymmetric
radiative transfer models of protostars calculated for a large range of
protostellar masses, accretion rates, disk masses, and disk orientations
created by \citet{Robitaille06}. This archive also provides an on-line
regression tool for selecting all the best fit SEDs. We use  all near-,
mid-infrared and millimeter photometry that resolve the star as well as
a {\it Spitzer} IRS Short-High (SH, i.e., 9.9 - 19.6 $\mu$m) spectrum
from the {\it Spitzer} archive. Our results are similar to what
\citet{Williams09} found using the same tool. The best fit models
predict a stellar mass of 10 - 12 \Msun, obscured by 12 - 18 mag of
visual extinction. These models result in a disk mass of 0.02 - 0.4
\Msun, an envelope mass of 20 - 30 \Msun, an envelope accretion rate of
10$^{-6}$ \Msun~yr$^{-1} $, a stellar age of $\sim$ 10$^4$ yr and a
bolometric luminosity of $\sim$ 3,000 \Lsun. \citet{Williams09} find a
slightly higher luminosity, L$_{bol}$ $\sim$ 4000 \Lsun, but otherwise 
our results are very similar. \citet{deWit09} modeled the observed SED
of AFGL\,961\,A with a spherical model with a radial density profile
described by a simple power law using partially the same data as we do.
Their best fit model predicts a power law index of 0.5 and gives a total
mass of 25 \Msun\ and a luminosity of $\sim$ 6000 \Lsun. However, some
of the data, like MIPS 70 $\mu$m and our SCUBA data included in their
SED, includes contributions from the nearby sources SMA\,2 and
AFGL961\,B and therefore the derived luminosity is  over-estimated.

AFGL\,961\,A appears to be a very young star driving an ionized outflow
(and possibly a molecular outflow) that is surrounded by a large accretion
disk. The modeling results by us and \citeauthor{Williams09} predict a protostellar
mass of 10 - 12 \Msun\ and a luminosity of 3000 - 4000 \Lsun, which
correspond to a B2 star. Since the star is still heavily accreting, it
may end up as perhaps a B1 star by the time it reaches the main
sequence.

\indent {\bf R~Mon}  is a deeply embedded HAEBE star at the apex of the
NGC\,2261 nebula, also known as Hubble's Variable Nebula. In the
catalogue by \citet{The94} the spectral type is given as B0e. This is
clearly incorrect, since the bolometric luminosity of the star is only
$\sim$ 740 \Lsun\ at a distance of 800 pc  \citep{Cohen85}.
\citet{Mora01} classified it as B8 IIIev, which is consistent with the
stellar luminosity of $\sim$ 450 \Lsun, determined by \citet{Natta93}.
It is a binary system with a low-mass companion located $\sim$  0\farcs7
to the  northwest \citep{Close97}. The star drives a prominent molecular
outflow and excites the HH object HH\,39, located in a small anonymous
dark cloud $\sim$ 7\arcmin\ north of R~Mon \citep{Canto81,
Jones82,Brugel84}. R~Mon was the first HAEBE star to be detected with a
millimeter-interferometer \citep{Sargent87}. They did not have enough
sensitivity to detect the star  with the OVRO three-antenna array in
continuum at 3 mm, but detected $^{13}$CO  J = $1 \to 0$ in the envelope
surrounding the star. The star is a faint VLA source, 0.29 $\pm$ 0.03
mJy at 6 cm \citep{Skinner93}, with a spectral index in the radio regime
of $\alpha$ = 0.9 $\pm$ 0.4, consistent with wind excited free-free
emission. The first (sub-)millimeter (1300, 1100, 800, 450, and 350
$\mu$m) detection of R~Mon was reported by \citet{Mannings94}, who also
determined that R~Mon had a disk/envelope with a substantial mass of
0.24 \Msun.  Recently \citet{Fuente03} detected the continuum emission
from R~Mon at 2.7 mm using the PdB interferometer. They found a total
flux of  6.4 mJy and claimed that the emission was extended on the
3\arcsec\ - 4\arcsec\ level. They also observed R~Mon with the VLA at
1.3 and 0.7 cm, but do not report integrated flux densities. In a
follow-up paper \citep{Fuente06}, they observed R~Mon with the PdB
interferometer in the continuum at 2.7 and 1.3 mm and in the rotational
lines of  $^{12}$CO  J = $1 \to 0$ and J = $2 \to 1$. Although they find
a Keplerian signature in their CO images and derive a mass for the
central star of 8 \Msun, based on the bolometric luminosity R~Mon is a
late B (see above) and cannot be as massive as 8 \Msun. The derived CO
rotation curve is most likely heavily contaminated by the strong outflow
from the star.

We have observed R~Mon at various wavelengths between 350 and 1300 $\mu$m
in the dust continuum with the single pixel bolometer UKT14.
The star was detected at all wavelengths (Table~\ref{tbl-2b}). Peak-up
observations of the star suggest that the emission is compact, which
agrees with the results by \citet{Fuente06}, who barely resolved it at
1.3~mm with a size of 0\farcs3. Fitting an isothermal graybody model
provides a dust opacity index $\beta$ = 1.3 and a disk mass
$\sim$0.1\Msun\ (Table~\ref{tbl-4}). The disk mass we derive is much
larger than found by \citet{Fuente06}, because we find a much higher
dust emissivity, 1.3 vs  0.3 - 0.5 deduced by \citeauthor {Fuente06} Our
observations cover a much larger frequency range than
\citeauthor{Fuente06}, and, since the free-free emission is negligible in
the sub-millimeter regime, our deduced dust emissivity is rather well
constrained (see Figure~\ref{fig-rmon_sedfit}). It is possible, however, that
the envelope may contribute to the observed dust emissivity, especially since 
most of our data come from single pixel observations.
  
\indent {\bf HR\,5999} (HD\,144568) illuminates a bright reflection
nebula in  the central part of the Lupus 3 dark cloud complex at a
distance of 208 pc \citep{Preibisch06}. The star varies strongly and
irregularly on time scales of weeks \citep{Perez92}. HR\,5999 was
classified as A6 III  with an error of about one subclass
\citep{Blondel06} and forms a $\sim$ 45\arcsec\ wide proper-motion
binary system with the A1.5 star HR\,6000  \citep{Preibisch06}.
\citet{Boekel05} found a similar spectral type as
\citeauthor{Blondel06}, A5-7 III/IVe,  and deduced a stellar mass of 3.2
$\pm$ 0.5 \Msun\ and an age of 0.5 Myr. The star has a strong far-infrared
excess \citep{Hillenbrand92} and shows broad silicate emission in the
mid-infrared \citep{Siebenmorgen00}, but it is a rather faint millimeter
source and it was not detected at 1.3~mm  \citep{Henning94}.
\citet{Preibisch06} observed the star in the mid-infrared with MIDI at
the ESO Very Large Telescope Interferometer (VLTI) and was able to
resolve the circumstellar disk  around the star. They found that the
disk is highly inclined, at 58$^\circ$, and truncated at 2.6 AU, likely
due to a close binary. We barely detect the disk as an unresolved point
source at 850 $\mu$m (Table \ref{tbl-2a}). A graybody fit to
the 1.3~mm upper limit by \citet{Henning94},  our 850 $\mu$m flux
density, the 100 $\mu$m IRAS upper limit, and the IRAS 60 $\mu$m flux
density,  yields a dust temperature of 41 K, a $\beta$ = 1.75,  and a
dust mass of $\sim$ 0.009 \Msun, which is comparable to the mass
estimated by \citet{Siebenmorgen00}.

\indent {\bf MWC\,297} is a highly reddened (A$_V$ $\sim$ 8 mag) early
type star of spectral class B1.5 Ve, with a luminosity of 3 $\times$
10$^3$~\Lsun\ and a distance of 250~pc \citep{Drew97}. It has very
strong and variable Balmer line emission \citep{Finkenzeller84} and 
is a relatively strong radio source \citep{Skinner93,Drew97}. 
\citet{Francesco98} found the far-infrared emission from MWC\,297 to be
extended  with a size of about one arcminute at 100 $\mu$m. However, at
1.3~mm the map of MWC\,297 \citep{Henning98} shows that the millimeter
emission from the star appears to be compact, and that there is strong
emission from the surrounding dust cloud to the northeast of the cloud.
The 1.3~mm  map also shows compact emission from a very red  2MASS
source (J18273709-034938), with a K magnitude of 8, located about
40\arcsec\ west of MWC\,297.  \citet{Manoj08}, using the SMA, 
found the 1.3~mm continuum emission to be
unresolved with a size of $\le$ 80 AU. They did not detect molecular gas
emission in the J=$2\to 1$ rotational transitions of
$^{12}$CO,$^{13}$CO, and C$^{18}$O. They deduced a very low dust
emissivity index, $\beta$ = 0.1 -- 0.3, and suggested that the low
$\beta$ index could be due to grain growth if the dust is optically
thin. \citet{Alonso-Albi09} observed the star with the PdB
interferometer at 1.3 and 2.6~mm and with the VLA at  1.3~cm and 7~mm.
They also found the millimeter emission  to be compact and modeled the
SED with a disk with a radius of $\sim$ 28 AU seen approximately
face-on, consistent with near-infrared interferometric observations
\citep{Acke08}, who found that the disk is seen with an inclination of
$\le$ 40\degr.

Our sub-millimeter maps (Figure~\ref{fig_mwc297scuba}) agree with the
1.3~mm map. MWC\,297 dominates the emission at 850 $\mu$m, while the
star is quite faint at 450 $\mu$m. At 450 $\mu$m the emission from the
surrounding cloud  is much stronger than from MWC\,297, suggesting that the
sub-millimeter SED is exceptionally flat. Follow-up observations by
Sandell et al. (in prep) with BIMA and the VLA, show that the observed
shallow SED at millimeter wavelengths is due to free-free emission. They
observed MWC\,297 with the VLA in the AnB configuration at 6 and 2.0 cm
and also re-analyzed VLA data obtained by \citet{Skinner93}.  The
\citeauthor{Skinner93} CnD data show that MWC\,297 powers a large
bipolar ionized jet.  A fit to the VLA data shows that the radio
emission has a spectral index, $\alpha \sim$ 0.97, while the size
decreases as $\nu^{-0.8}$, as is typically seen in collimated thermal
jets. If we include data at wavelengths out  to 3 mm, we get a spectral
index, $\alpha \sim$ 1.03, i.e.,  essentially the same as we got from
the more uncertain fit from VLA data alone
(Figure~\ref{fig-mwc297sed}).     The thermal emission from the jet can
therefore explain all the emission at 3~mm and there is no evidence for
any excess due to cold dust emission. There is a hint of a dust excess
at 850 $\mu$m and almost certainly some excess at 450 $\mu$m, suggesting
that the wind becomes optically thin somewhere around 300 - 600 GHz.
MWC\,297 looks therefore very similar  to MWC\,349, in that the emission
in the millimeter and sub-millimeter  regime is dominated by free-free
emission. The bipolar morphology seen in the VLA maps suggests that the
disk is seen nearly edge-on, as is the case for MWC\,349. Both stars
appear to be surrounded by disks, which are largely ionized and only
contain a small fraction of neutral gas and dust.

\indent {\bf The BD+40$^\circ$4124 region} forms a small
pre-main-sequence stellar cluster with three bright emission line stars.
 \citet{Herbig60} identified the first two, BD+40$^\circ$4124  and
V\,1686 Cyg, as HAEBE stars. The third star,  V\,1318~Cyg, is a deeply
embedded binary system with a separation of $\sim$ 5\arcsec\ and a total
luminosity of $\sim$ 1600 \Lsun \citep{Aspin94}. The southern member of
this binary, V\,1318~Cyg\,S, is more deeply embedded  with a visual
extinction of 10 mag or more. \citet{Aspin94} found that the  position of V\,1318~Cyg\,S  coincides with
the position of a compact sub-millimeter source that has a luminosity similar to that of a HAEBE
star. None of the stars in the BD+40$^\circ$4124 cluster show any
free-free emission at centimeter wavelengths \citep{Skinner93,Palla95},
except perhaps BD+40$^\circ$4124, near which \citeauthor{Skinner93}
detected faint emission at  3.6~cm and 6~cm.

\citet{Palla95} discovered a compact molecular outflow and  an H$_2$O
maser, which appear to be excited by V\,1318~Cyg\,S. \citet{Looney06},
however, who observed the region in the 3.1~mm continuum with BIMA,
found that the dust emission is offset by 1\farcs1 to the north-east
from V\,1318~Cyg\,S, and argue that the dust emission originates from a
deeply embedded intermediate mass protostar V\,1318~Cyg\,S-mm, rather than from
either component of V\,1318~Cyg\,S identified by \citet{Aspin94}. These results are supported by the less
accurate results by \citet{Francesco97}, who detected the star in
continuum at 2.7 mm with the PdB interferometer, and found a similar
offset, 0\farcs9,  from the southern component. More importantly,  VLBA
observations of the H$_2$O masers toward 
BD$+$40$^\circ$4124\citep{Marvel05} indicate that it is associated with
the sub-millimeter source rather than with the optically visible star.
It is therefore the invisible millimeter source that drives the outflow,
not V\,1318~Cyg\,S.

Figure~\ref{fig-BD40} shows the 850 and 450 $\mu$m SCUBA images of the
BD+40$^\circ$4124 region. It is clear from these images that neither
BD$+$40$^\circ$4124 and V\,1686\,Cyg are associated with any significant
sub-millimeter continuum emission. Because of the extended emission from
the molecular cloud in which these stars were born, we cannot place a
very stringent limit on the dust emission; at 850 $\mu$m our upper limit
is $\sim$ 35 mJy~beam$^{-1}$.  Examination of MIPS 24 and 70 $\mu$m
images show that all HAEBE stars in this region are saturated at 24 $\mu$m
(Table~\ref{tbl-3}). At 70 $\mu$m there is some extended emission near
BD$+$40$^\circ$4124, most likely heated dust from the surrounding
reflection nebulosity. There is strong, compact, 70 $\mu$m emission from the vicinity of
V\,1318~Cyg\,S (Table~\ref{tbl-3}), almost certainly coming from the
deeply embedded protostar, V\,1318~Cyg\,S-mm. This is where we find a strong, extended
(8\arcsec\ $\times$ 3\arcsec{}) sub-millimeter source
(Table~\ref{tbl-2a}). Although it is possible that some of the emission
could be due to V\,1318\,Cyg~S, we assume that all the emission comes
from the deeply embedded millimeter source. A graybody fit to the SCUBA
and MIPS data give a dust emissivity of 0.85 and a total mass of 1.9
\Msun\ (Table~\ref{tbl-4}), suggesting that we see the disk and envelope
of a young high-mass protostar.

\indent {\bf PV~Cep} is an irregularly variable star at the head of the
cometary nebula GM\,29 that illuminates this fan-shaped nebula; it is
located at a distance of 500 pc \citep{Cohen81}. Its spectral
classification is uncertain, because the optical spectrum is strongly
veiled \citep{Hernandez04}. \citet{Cohen81} suggest a spectral type 
between B5 and F2,  perhaps A5. The latter would be consistent with the
observed bolometric luminosity of $\sim$ 80 - 175 \Lsun\ 
\citep{Evans86,Abraham00}. It drives a giant parsec-scale Herbig-Haro
outflow \citep{Reipurth97} and molecular outflow
\citep{Levreault84,Arce02a,Arce02b}. The outflow appears to be
precessing \citep{Arce02b}, suggesting that PV~Cep could be a binary
system, although no companion has yet been detected. \citet{Torrelles86}
discovered an H$_2$O maser 5\arcsec\ to the south of PV Cep, which could
mark the location of another young star, because H$_2$O masers excited
by low-luminosity stars are usually close to the star. More recent
observations  show H$_2$O maser activity close to PV Cep
\citep{Marvel05}. PV Cep coincides with a faint VLA source at 3.5 cm
\citep{Anglada92}.

%free-free Beltran et al 2001, Anglada et al. 1992
PV~Cep was mapped at 1.3~mm by \citep{Fuente98a}, who found a compact
source centered on the star. It was recently image on CARMA by
\citet{Hamidouche10}, who resolved it at 2.7 and 1.3 mm with a size of 
$\sim$ 0\farcs5 (250 AU). Our maps at 850 $\mu$m and 450 $\mu$m
(Figure~\ref{fig-pvcep}) go deeper than the single dish map at 1.3~mm 
by \citet{Fuente98a}. We resolve PV~Cep in the sub-millimeter with a
size of 1200 AU $\times$ 700 AU   at PA 74\degr\ (Table~\ref{tbl-2a}),
which is roughly perpendicular to to the outflow.  We also see faint,
extended emission with a diameter of $\sim$ 1\arcmin\ from the dense
surrounding cloud core at both 850 and 450 $\mu$m. PV~Cep was found to
be unresolved at 50 and 100 $\mu$m \citep{Natta93}, which is in good
agreement with the compact size that we find in the sub-millimeter. Our
observed source size, however, is much larger than what
\citet{Hamidouche10} found with CARMA and the 1.3 mm flux density found
by \citet{Fuente98a} is about twice as large as what is seen by CARMA, 
suggesting that the array filters out some of the extended emission. A
graybody fit to the (sub-)millimeter single dish data, ISO
\citep{Abraham00}, and KAO far-infrared data
(Figure~\ref{fig-pvcep_sedfit}) gives a dust temperature of 29 K, a dust
emissivity index, $\beta$ = 1.59, resulting in a mass of $\sim$ 1.0
\Msun. This mass estimate is much larger than what one would expect to
find even for a young circumstellar disk. It therefore seems that the
single dish data maybe dominated by the compact infalling envelope,
although \citet{Hamidouche10} finds 0.8 \Msun, suggesting that even
array observations are dominated by the envelope.
 
\indent {\bf V\,645~Cyg} powers a low velocity molecular outflow in the
north-south direction \citep{Schulz89,Verdes91}. The star coincides with
a faint, somewhat extended VLA source \citep{Girart02} roughly aligned
with the outflow. \citet{Girart02} also found a faint extended, $\sim$
6\arcmin\ $\times$ 2\arcmin, \ion{H}{2} region close to the star, which
could be excited by V\,645~Cyg. Since V\,645~Cyg is known to power a
1665 MHz OH maser \citep{Morris82}, and a Class II CH$_3$OH maser
\citep{Slysh99},  it must be a high-mass star, since these masers are
only excited by high mass stars.  \citet{Clarke06}  classified it as a
transition object between a massive YSO and a young Oe type star in a
weak \ion{H}{2} region.

\citet{Francesco97} marginally detected V\,645 Cyg at 2.7 mm with the
IRAM array. We find V\,645\,Cyg to coincide with a bright sub-millimeter
source embedded in narrow east-west cloud ridge which turns mostly
northward east of the star. The emission is clearly extended
(Table~\ref{tbl-2a}), but at a distance of 4.2 kpc it is clear that the
sub-millimeter emission is dominated by the dense  envelope, rather than
a circumstellar disk. An isothermal fit to the data in
Table~\ref{tbl-2a} gives a mass of 85 \Msun\ for the elliptical
sub-millimeter source, which is far more than what one would expect even
for a very young circumstellar disk.

\indent {\bf MWC\,1080\,A} must be an early B star, because it has
strong Balmer and \ion{Fe}{2} lines showing P Cygni  type emission
profiles \citep{Finkenzeller84}. It drives a prominent bipolar outflow
seen pole on \citep{Canto84}. Distance estimates vary. Here we have
adopted 2.2 kpc (Table~\ref{tbl-1}) from \citet{Canto84}. MWC\,1080 is a
hierarchical multiple system. The primary, MWC\,1080\,A, is an eclipsing
binary. The period of 2.9 days \citep{Shevchenko94}. MWC\,1080\,B is
also a PMS star and more heavily reddened than the primary
\citep{Leinert97,Polomski02}. A third companion is found $\sim$ 5\ptsec2
east of the primary \citep{Polomski02,Pirzkal97}. MWC\,1080 is a strong
far-infrared source with a source size  of $\sim$ 50\arcsec\ --
60\arcsec\ \citep{Evans86,Francesco98,Abraham00}. VLA observations
\citep{Girart02} show three free-free emission sources in the vicinity
of MWC\,1080 with VLA\,4 coinciding with MWC\,1080\,A. It has a flux
density of 0.2 mJy at 6~cm and was also detected by \citet{Skinner93} at
3.6 cm with a flux density of 0.18 mJy. \citet{Fuente98a} and
\citet{Henning98} both observed MWC\,1080 with the IRAM 30 m telescope
and found strong extended continuum emission towards MWC\,1080\,A and B.

Our short integration 850 $\mu$m map (Figure~\ref{fig-mwc1080}) is
dominated by strong extended emission from the reflection nebula
surrounding MWC\,1080 and perhaps nearby embedded sources. 
MWC\,1080\,A, i.e.,  VLA\,4,  coincides with a faint emission peak in an
extended east-west ridge. The other two VLA sources lack sub-millimeter
counterparts. We consider MWC\,1080\,A a marginal detection with a flux
density of $\sim$ 250 $\pm$ 50 mJy. This appears plausible since deep
IRAM interferometer observations at  2.7 mm and 1.4 mm \citep{Fuente03}
show extended emission towards MWC\,1080\,A. At 1.4 mm
\citeauthor{Fuente03} find that MWC\,1080\,A coincides with a faint
unresolved source. The observed flux density is much less than what
\citet{Fuente98a} fouund from their single dish observations, suggesting
that interferometer has resolved out most of the emission. If we assume
that the  observed  850 $\mu$m flux density comes from a circumstellar
disk surrounding  MWC\,1080\,A we can make some simple assumptions to
estimate the mass of the disk. For a dust temperature of 50 K and a
typical dust emissivity, $\beta$ = 1.5, we get a disk mass of 0.009
\Msun.

\section{Non-detections}\label{non}

Most of our non-detections are early to mid-B stars. These
stars illuminate large reflection nebulae and are strong 60
and 100 $\mu$m sources in the IRAS point source catalogue. Only a few,
however, appear to be surrounded by dusty disks that have enough cold dust 
to be detected with a single dish telescope in the sub-millimeter. Our sample also includes two stars
which are not HAEBE stars but instead appear to be supergiants.
Neither of these two, HD\,316285 and MWC\,300, are associated with any dust emission. 
These are discussed at the end of this section.

\indent {\bf MWC\,137} is a B0ep star that excites the  \ion{H}{2} region S\,266
\citep{Fich93}. Both the radio and the optical appearance are very
similar: a ring-like nebulosity with a radius of $\sim$ 30\arcsec\ and 
two spiral-like arcs connecting to the central star (Figure~\ref{fig-mwc137}). Although
MWC\,137 has long been known to be a radio source, see
e.g., \citet{Altenhoff76}, early single dish observations picked up emission
from the extended \ion{H}{2} region, not the central star. Radio
emission from MWC\,137 was first detected by \citet{Skinner93} at
3.6~cm. They found the emission to be compact and unresolved.

The star appeared to have been detected at 1.3 mm in photometric
observations  at the CSO by \citep{Hillenbrand92}. It was mapped with the IRAM 30 m
telescope at 1.3 mm by \citet{Fuente98a} and \citet{Henning98}. Both
maps show an extended narrow dust ridge extending over the nebulosity
with a stronger emission peak $\sim$ 30\arcsec\  northeast of the star
and a fainter one to the southwest, both peaks approximately coinciding
with the boundary of the \ion{H}{2} region. There is  only faint
emission towards the star itself. Our 850 $\mu$m map looks very similar,
although we see a second arc bounding the nebula to the south. We see no
enhancement in dust emission toward MWC\,137;  if anything the emission
is close to a minimum at the position of the star.  \citet{Francesco97} 
obtained a marginal detection of the star at 2.7 mm with the IRAM
interferometer, while the deep observations by \citet{Fuente03} yield
clear detections at both 2.7 mm and 1.4 mm. At 1.4 mm
\citeauthor{Fuente03} derive a source size of 1\ptsec8 $\times$ 0\ptsec8
and find a total flux density of 10 mJy. \citet{Fuente03} also detected
the star with the VLA at 1.3 cm and 7 mm, where they find it to be
compact and unresolved. The VLA data suggest that MWC\,137 has a thermal
wind with a spectral index $\alpha$ = 0.8 - 1.0, i.e. similar to
what we find for MWC\,297 and LkH$\alpha$\,101. It therefore appears
that the free-free emission completely dominates at
millimeter wavelengths. MWC\,137 appears similar to MWC\,349, MWC\,297, and
LkH$\alpha$\,101, but the free-free emission is much fainter, suggesting
that it is more evolved and has already photo-ionized most of its disk.

\indent {\bf LkH$\alpha$\,215}  illuminates the reflection nebula
NGC\,2245. It is a late B star. \citet{Valenti03} classify it as a B8
star based on low-resolution IUE spectra, while \citet{Hernandez04} give
it a spectral type B6 with an uncertainty of 2.5 subclasses. It  is part
of the young Mon R1 association, which has a distance of 800 pc
\citep{Herbst82}. Although  \citet{Mannings94} reported a detection of LkH$\alpha$\,215
with sub-millimeter photometry on JCMT, \citet{Fuente98a}, who mapped
LkH$\alpha$\,215 and the surrounding reflection nebula in $^{13}$CO J =
$1 \to 0$, CS  J = $3 \to 2$ and 1.3~mm continuum with the IRAM 30 m
telescope, did not detect  it. They found no CS emission in the vicinity
of LkH$\alpha$\,215 and even though they detected 1.3~mm dust emission,
it did not peak on the star.

Figure~\ref{fig-NGC2245} shows our 850 $\mu$m image overlaid on a POSS
image of the region. One can see a faint dust ridge near
LkH$\alpha$\,215. This emission, which is stronger east of the star,  is
probably not associated with LkH$\alpha$\,215, although we cannot
exclude the possibility that LkH$\alpha$\,215 is still surrounded by a
small accretion disk. However, we find a much stronger unresolved source, LkH$\alpha$\,215-SMM,
northwest of the reflection nebula (\mbox{$\alpha_{2000.0}$ = 06$^h$
32$^m$ 35\psec265}, \mbox{$\delta_{2000.0}$ = $+$10\degr{} 10\arcmin{}
18\ptsec3}), which has no optical or near-infrared counterpart. The flux density of this source
 is 152 $\pm$ 22 mJy at 850 $\mu$m,  220 $\pm$  80 mJy at 750 $\mu$m,
and 875 $\pm$ 250 mJy at 450 $\mu$m.

\indent {\bf  HD\,259431} is a B6 star  that
illuminates the reflection nebula NGC\,2247. This star appears similar
to LkH$\alpha$\,215. They are both part of the Mon R1 association and
probably of similar age, 0.05 - 0.1 Myr \citep{Testi98}.

\citet{Mannings94} reported a marginal 450 $\mu$m detection, but as is evident
from the 850 $\mu$m image (Figure~\ref{fig-NGC2247}) there is
extended dust emission towards NGC\,2247, but no emission associated with the position of
the star. There are two faint 850 $\mu$m continuum sources
outside the nebula; NGC\,2247-SMM1: $\alpha_{2000.0}$ = 06$^h$
33$^m$ 07\psec99, $\delta_{2000.0}$ = $+$10\degr{} 18\arcmin{}
19\arcsec, and  NGC\,2247-SMM2: $\alpha_{2000.0}$ = 06$^h$
27$^m$ 07\psec76, $\delta_{2000.0}$ = $+$10\degr{} 20\arcmin{}
22\arcsec, with integrated flux densities of  0.38 mJy and 0.16 mJy, respectively.  Neither
source has an optical or near-infrared
counterpart.

\indent {\bf LkH$\alpha$\,25}, spectral type B7 \citep{Hernandez04},  does not illuminate a
reflection nebulosity \citep{Herbig60}, nor was it detected by IRAS. {\it
Spitzer} IRAC archive images show that the star is inside a large,
roughly spherical nebula with a diameter of 6\arcmin\ - 8\arcmin.
Although the star is not centered in the nebula, it is  $\sim$ 2\arcmin\
from the  southern part of the nebula (which is also the brightest part
of the nebula at 8 $\mu$m), suggesting that it may be the illuminating
source of the nebula.  LkH$\alpha$\,25 has been observed by two groups
at 1.3~mm \citep{Henning98,Fuente98a}, but neither of them detected the
star in millimeter continuum emission.  We do not detect it either. Our sub-millimeter map (not shown)
shows that the star lies in a region free of dust emission, although
there is a small dust emission region less than one arcmin to the
southeast of the star and a large dust cloud ($\sim$ 10\arcmin\ $\times$
3\arcmin{}) about two arcminutes  west of the star extending mostly to the
north. This cloud complex is almost certainly associated with NGC\,2264.

\indent {\bf CD$-$42$^\circ$ 1172} is a well-studied, bright
B0ep star illuminating a clumpy,  arc-shaped nebula with a size of $\sim$
40\arcsec, showing both reflection and emission characteristics
\citep{Hutsemekers90}. Although several studies have questioned whether
CD$-$42$^\circ$ 1172 is a PMS star (see e.g., BorgesFernandes et al. 2007),
there should not be any doubt that this is a young star. It is embedded
in an opaque dark cloud and is the brightest member of a small cluster of
PMS stars \citep{Habart03,Boersma09}. Distance estimates vary. Here we
have adopted 1.15 kpc \citep{BorgesFernandes07}, which is based on the
interstellar absorption lines of \ion{Na}{1} and \ion{Ca}{2} K lines.
The star is associated with strong far-infrared  emission
\citep{Weintraub90,Natta93} and sub-millimeter emission
\citep{Mannings94}, but it is questionable whether the emission comes
from a circumstellar disk. \citet{Boersma09} present an IRAC color image
of the CD$-$42$^\circ$ 1172 region. The star is saturated in all IRAC
images and even a secondary peak north of the star is an artifact due to
the heavily saturated CD$-$42$^\circ$ 1172. The IRAC image shows that
the star is surrounded by a one arcminute nebulosity glowing in PAH
emission and it also shows a smaller nebulous region about 45\arcsec\ to
the northwest.

\citet{Henning98} mapped CD$-$42$^\circ$ 1172 in 1.3 mm continuum with
the IRAM 30 m telescope. They found the star to be associated with an
extended dust cloud, but there is no enhancement in the dust emission at
the position of the star. Our 850 and 450 $\mu$m images
(Figure~\ref{fig-CoD})  look very similar to the 1.3 mm map and show no
enhancement of the dust emission towards the star, only emission from
the cloud ridge in which it is embedded. We see two sub-millimeter
sources to the northwest of CoD$-$42$^\circ$ 1172, neither of which have
any near- or mid-infrared counterparts \citep{Habart03}. SMM\,1 is a point--like source offset
$-$6\ptsec8,$+$7\ptsec3 from the star  in RA and Dec, i.e., \mbox{$\alpha_{2000.0}$ = 16$^h$
59$^m$ 06\psec286}, \mbox{$\delta_{2000.0}$ = $-$42\degr{} 42\arcmin{}
00\ptsec7}, with a flux density of 670 mJy at 850 $\mu$m.  SMM\,2 is a strong (2.4 Jy),
extended source about 50\arcsec\ northwest of the star:  \mbox{$\alpha_{2000.0}$ = 16$^h$
59$^m$ 03\psec911},  $\delta_{2000.0}$ = $-$42\degr{} 41\arcmin{}
32\ptsec6.
The latter is clearly seen in the 1.3 mm map by \citet{Henning98} and
appears to coincide with the northwestern nebulosity seen in the IRAC
image shown by \citet{Boersma09}.

We therefore also downloaded and analyzed MIPS images of CD$-$42$^\circ$
1172, which has been observed as part of the MIPSGAL legacy project. The
24 and 70 $\mu$m MIPS images are shown in Figure~\ref{fig-CoD}, on which
we we overlaid the 850 $\mu$m SCUBA image with contours.  In the 24
$\mu$m image the nebulosity surrounding CD$-$42$^\circ$ 1172 is
completely saturated. One plausible explanation to why the nebulosity
would be saturated is that it is an \ion{H}{2} region ionized by the
star or by a thermal wind from the star. The star, however,  was not
detected in the VLA survey by \citet{Skinner93}, nor was it seen by 
\citet{Thompson04}, who used the Australia Compact Array.
\citeauthor{Thompson04} detected two sources at 13 cm. One is probably
extragalactic; the other one, SFO\,85a, is close to, but not coincident
with the extended SMM\,2 northwest of CD$-$42$^\circ$ 1172. The emission
is also extended in the MIPS 70 $\mu$m image, suggesting that this
region may  contain several embedded stars.

\indent {\bf TY CrA and HD\,176386} illuminate the reflection nebulae
NGC\,6727 and NGC\,6726 in the  Corona Australis molecular cloud. TY CrA
is a triple system with a close eclipsing binary \citep{Casey95}. The
primary is a B8-B9V star  with  a luminosity of 67 \Lsun, and an age of
$\sim$ 3 Myr \citep{Casey98}. HD\,176386 is a binary star with the
primary being an A0V star and the secondary a low-mass K7 star
\citep{Meyer09}.

Figure~\ref{fig-tycra} shows an 850 $\mu$m sub-image of a large SCUBA
scan map, which has also been published by \citet{Nutter05}. It is clear
from this 850 $\mu$m map that neither star has any sub-millimeter excess.
\citet{Hillenbrand92} reported a  1.3~mm detection of TY CrA, but as we
can see from Figure~\ref{fig-tycra}, the emission observed by
\citeauthor{Hillenbrand92} is from the dust cloud in which the star is
embedded. Neither TY CrA nor HD\,176386 show any near-infrared excess
\citep{Bibo92}, suggesting that they have completely dispersed their
accretion disks. The nearby Herbig Ae/Be stars, R CrA and T CrA, about 
5\arcmin\ to the southwest, appear to be much younger, and even though
neither of them have been detected in the sub-millimeter  (see Figure 2
in \citet{Groppi07}), they do show clear excess emission in the near- and
mid-infrared, suggesting that they are still surrounded with accretion disks
\citep{Bibo92}.

\indent {\bf HD\,200775} is a B3 star illuminating the reflection nebula
NGC\,7023. We do not see any dust emission towards HD\,200775 in
NGC\,7023 to a limit of $\sim$ 25 mJy~beam$^{-1}$ (Table~\ref{tbl-2c}).
NGC\,7023 was observed both in scan map and jiggle map modes.  There is
strong dust emission from the PDR region just north and northwest of the
star and fainter emission to the east and southwest at the boundary
between the reflection nebula and the surrounding dust cloud
(Figure~\ref{fig-NGC7023}).  \citet{Fuente98b} show that the cavity
around HD\,200775 has very likely been excavated by an energetic,
bipolar outflow in an earlier evolutionary stage of the star, but that
the star no longer shows any evidence of a bipolar outflow.  They
detected faint continuum emission at 3.4 mm with the IRAM
interferometer, but the observed flux density, 3 mJy, is about what one
would expect from a thermal wind.  Recently, however, \citet{Okamoto09}
detected HD\,200775 with the SMA at 350 GHz with a flux density of 35
mJy, i.e., consistent with our 3-$\sigma$ upper limit, see
Figure~\ref{fig-NGC7023}. They also found unresolved excess emission in
the mid-infrared using the  8.2 m Subaru telescope, confirming that the
star is still surrounded by a disk.

\indent {\bf HD\,316285} is listed as an extreme emission line object
in \citet{The94}; however,  HD\,316285 is a P Cygni type supergiant, possibly a
luminous blue variable, and  definitely not a young Herbig Be star
\citep{Hillier98}.  It is extremely luminous and lies at a distance of $\sim$
1.85 kpc (Table~\ref{tbl-1}). It has very little or no hot dust, but
shows strong far-infrared emission from cold dust \citep{McGregor88}.

\citet{vanderVeen94} did sub-millimeter photometry with the single pixel
bolometer UKT14 on JCMT and reported flux densities at 1.1 mm, 800 and
450 $\mu$m although the flux density at 800 $\mu$m was suspiciously low.
In Figure~\ref{fig-hd316285} we present large 850 and 450 $\mu$m images
in the direction of HD\,316285. Although there is strong sub-millimeter
emission in the field, HD\,316285, is in an area with relatively weak
dust emission. There is a marginal peak in the dust emission at 850
$\mu$m, but not at 450 $\mu$m (Figure~\ref{fig-hd316285}). We count it as
a non-detection. The upper limit we derive from these observations is
rather high, 70 mJy at 850 $\mu$m for an unresolved source
(Table~\ref{tbl-2c}).

\indent {\bf MWC\,300}  was listed as a HAEBE star by \citet{Herbig60}
and included in the catalogue by \citet{The94}. However, it has also
been classified as a B[e] star or supergiant
\citep{Allen76,Appenzeller77}. \citet{Miroshnichenko04}, who analyzed
more than ten years of optical and near-infrared spectroscopy and
mid-infrared imaging, were able to determine the distance (1.8 $\pm$ 0.2
kpc) and hence the luminosity (L = 1.3 $\times$ 10$^5$ \Lsun{}), which
together with all the other observed characteristics and with detailed
modeling identify this star as a B[e] supergiant and most likely a
binary.

\citet{Henning94} reported an approximately three-sigma detection of the
star in continuum emission at 1.3~mm.  We find no emission at all
towards MWC\,300 with a rather stringent upper limit, 7 mJy~beam$^{-1}$
(Table~\ref{tbl-2c}), suggesting that the results by
\citeauthor{Henning94} were spurious.

\section{Discussion}
\label{dis}

In this paper we present results from an extensive sub-millimeter survey
of HAEBE stars, many of which have been detected  both at 850 and 450
$\mu$m. First we note that our study includes a total of 39 objects that
were potential HAEBE stars. However, HD\,316285 and MWC\,300, which were
classified as potential HAEBE stars when we surveyed them, are now
classified as supergiants. We therefore have observed a  total of 37
objects now thought to be HAEBE stars.  Of these we have detected 27
stars (73\%)  in emission at 850 $\mu$m or 800 $\mu$m, either in SCUBA
mapping or UKT14 photometry, or both\footnote{We note, however, that
this is not an unbiased survey, most targets were chosen, because they
had a high likelihood to be surrounded by accretion disks.}. We have
also detected 17 of these stars at 450$\mu$m. Although many of these
stars have already been observed at millimeter and/or sub-millimeter
wavelengths
\citep{Hillenbrand92,Mannings94,Henning98,Fuente98a,Natta04,Alonso-
Albi09}, our study provides  the largest sub-millimeter sample of HAEBE
stars  analyzed in a consistent way. We have also  included {\it
Spitzer} MIPS photometry for those sources for which such data were
publicly available. In our analysis we fit a simple isothermal graybody
model to the observed SEDs from far-infrared to millimeter wavelengths.
This allows us to determine the temperature of the cold dust, the dust
emissivity, and the disk mass.

%One star in our sample, HD\,316285, is a P Cygni type 
%supergiant, possibly a luminous blue variable \citep{Hillier98}  and not a HAEBE star. 
%Table~\ref{tbl-2a}, \ref{tbl-2b})

We find that disks are ubiquitous in Herbig Ae and F stars, but are very
rare in early B stars. This does not mean that early B stars do not form
with disks, but instead suggests that any such disks evolve rapidly and
get destroyed quickly by photoevaporation. There are currently about
five or six early B-stars known  to have disks. The three stars  we
detected in this survey --- MWC\,297, MWC\,349, and LkH$\alpha$\,101---
look rather different from the accretion disks surrounding Herbig Ae and
T Tauri stars. The gas in these disks  is largely ionized, and the
emission at millimeter and sub-millimeter wavelengths is dominated by a
thermal wind powered by photoevaporation. Only LkH$\alpha$\,101 appears
to have a  dust excess at millimeter wavelengths, suggesting that it is
perhaps younger than the other two and has not yet had time to
photoevaporate the disk.  All three stars may have some neutral gas in
the midplanes of their disks, although there are no detections of
molecular gas toward any of these three stars at radio wavelengths.
MWC\,137 and HD\,200775, which we did not detect in our SCUBA survey,
are even fainter and possibly more evolved. Our survey also includes two
deeply embedded objects (AFGL\,961\,A and V\,1318~Cyg-mm), which have
luminosities comparable to early B-stars. The disks/envelopes around
these stars look very similar to  Class 0 or Class I protostars,
although they appear more massive than disks around young low mass
stars. Our results therefore suggest that at least some B stars form
with accretion disks in an analogous way to how low mass stars form with
disks.

Disks around HAEBE stars appear both larger and more massive than disks
around T Tauri stars. We are not aware of any accretion disk around a T
Tauri star that has been resolved with a single dish telescope in
sub-millimeter continuum emission.  Yet at least three HAEBE disks,
those around MWC\,480, HD\,163296 and PV~Cep, have been resolved with
SCUBA. The sizes that we find  agree well with  sizes derived from
millimeter-aperture synthesis observations. Overall, however, there is
no clear correlation between disk size and spectral type in our sample, 
or even any correlation between disk mass and spectral type.

What we do find, however, is a clear correlation between dust emissivity
and disk mass in the sense that a lower dust emissivity corresponds to a
lower disk mass. This is shown in Figure~\ref{fig-mass_beta}, where we
have taken disk masses and emissivities from the isothermal graybody
fits given in Table~\ref{tbl-4}.  The only data we omitted from this
plot were the two protostellar sources (AFGL\,961\,A and V\,1318~Cyg-mm)
and the very distant star V\,645~Cyg, because here the masses are
dominated by the envelopes, rather than the accretion disks. As we can
see from Figure~\ref{fig-mass_beta}  the correlation is rather good (r =
0.8). What this suggest is that when $\beta$ decreases, the mass of the
cold dust that we can detect decreases. Note that our modeling assumes
that the dust opacity is constant at 250 $\mu$m, i.e.,  the ``standard''
Hildebrand opacity, while newer studies show that large grains,
corresponding to low $\beta$-values,  have lower opacity at 250 $\mu$m,
see e.g. \citet{Draine06}.  Our disk mass estimates may therefore be
underestimated for low $\beta$-values. Comparison with other studies,
including detailed modeling of some of the disks in our sample, indicate
that the agreement in disk mass is surprisingly good, and usually within
a factor of two. Since a low $\beta$-value suggests grain growth
\citep{Natta04,Rodmann06}, the disk mass might not actually be dropping;
rather, the dust mass could be getting buried into large grains,
pebbles, and planetesimals.  Once buried, we cannot detect it anymore. 
Thus, low $\beta$, which is correlated with low ``detectable'' disk
mass, might be a good indicator of planetesimal growth rather than of
decreasing disk mass of solids. All stars that show low $\beta$-values
are isolated HAEBE stars, i.e.,  they are relatively old, 3 - 15 Myr,
which seems to be the time required for significant grain growth.

A few of these stars have SEDs that suggest  they are transition stars,
i.e., they show weak near-infrared excesses but strong
mid-to-far-infrared excesses, suggesting that they have developed gaps
largely free of dust \citep{Najita07}.  Another characteristic feature
of transition sources, which has become apparent in this study, is that
the millimeter and far-infrared regions of the SEDs for these HAEBEs
show that the outer portions of the disks around these stars have both
cold and warm dust components. All transition stars have cold outer
disks, which dominate the emission in the millimeter and sub-millimeter
regime, while most of the emission at 60 $\mu$m or shorter wavelengths
comes from warmer  dust (T$_d$ $\geq$ 100~K) in the disk, i.e.,
presumably emission from the hot surface layers of the disk or the inner
rim or ``wall''  \citep{Calvet02, Isella09}. The same trend is seen for
most of the isolated HAEBE stars, i.e., there is a clear excess  at 60
$\mu$m requiring a warmer dust component in addition to the cold outer
disk, which dominates the emission at wavelengths longer than 100
$\mu$m.  This is not seen in younger HAEBE disks, like R Mon , PV Cep or
AB~Aur, or for ex. young FU Ori disks \citep{Sandell01}, which are all
well fit with a single temperature isothermal fit down to 60 $\mu$m. A
warm dust component, producing a clear excess at 60 $\mu$m is  therefore
a clear sign of a disk approaching the transition disk stage.

The strong correlation between disk mass and dust emissivity and lack of
correlation between spectral type and disk mass suggests that  the
evolutionary age  of the disk affects  the measurable disk mass much
more strongly than does the mass of the central star. HAEBE  stars do
have somewhat larger and more massive disks than T Tauri stars.  For
example, in the large SCUBA survey  of Taurus-Auriga by
\citet{Andrews05}, the most massive T  Tauri disk is  0.2 \Msun, while
there are at least three HAEBE stars in our sample (Table~\ref{tbl-4})
that have an equally large or more massive disk.

\section{Summary}\label{sum}

We have studied a large sample of HAEBE stars in the sub-millimeter
using SCUBA on JCMT. We find that disks around Ae and F stars
ubiquitous, while disks around early B stars are rare. We have, however, 
identified several examples of disks in early B stars.  For very young
embedded stars with the luminosity of an early B star, they look
similar to other young disks, albeit more massive. In early B-stars,
which already illuminate \ion{H}{2} regions, such disks are largely
ionized and have very little cold dust. The emission in the
sub-millimeter in all of these cases is dominated by a thermal wind, presumably
driven by photoionization of the disk.

We find several examples of HAEBE stars (LkH$\alpha$\,215, HD\,259431,
CD$-$42$^\circ$ 1172, MWC\,297, and TY CrA) for which our imaging survey
clearly shows that the emission previously identified as dust emission
from circumstellar disks around these stars instead  originates from
ionized gas (as in the case of MWC\,297), or from nearby younger
objects, or from the surrounding molecular clouds in which these stars
reside. These stars are, in fact, free of detectable dust emission at
long wavelengths; we conclude that previous SED models of these stars
that derive properties of their circumstellar disks are incorrect.

Our survey includes several examples of isolated HAEBEs, which have ages
of 3 - 15 Myrs and still have gas-rich disks. A few of these are
transitional stars, which have already formed large inner gaps (perhaps
due to planet formation) and show evidence for grain growth. Most of the
isolated HAEBEs show signs of disk evolution. They have cold outer
disks, which dominate the emission in the far-infrared and
sub-millimeter, but show strong excesses at 60 $\mu$m due to warm gas in
their disks, i.e., they cannot be fitted with a single temperature
isothermal dust model.

\acknowledgements This work made extensive use of the SIMBAD
Astronomical Database at the Centre de Donnees astronomiques de
Strasbourg, France, and NASA's Astrophysics Data System Abstract
Service. This research used the facilities of the Canadian Astronomy
Data Centre operated by the National Research Council of Canada with the
support of the Canadian Space Agency. This work is based in part on
observations made with the {\it Spitzer} Space Telescope, which is
operated by the Jet Propulsion Laboratory, California Institute of
Technology under a contract with NASA. We thank Meredith Hughes for
giving us flux densities she observed with the SMA and we also
acknowledge useful discussions and help from William Vacca and Bhaswati
Mookerjea. Comments and suggestions by the referee were also
appreciated.

{}

\newpage
\begin{deluxetable}{llllccccccl}
\rotate
\tabletypesize{\scriptsize} \tablecolumns{11} \tablenum{1} \footnotesize  
\tablewidth{0pt} 
\tablecaption{Stellar parameters of our target stars} 
\label{tbl-1}
\tablehead{
\colhead{Star~~~~~~~~~~} &\colhead{Alt.~~~~~~~~~~}& \colhead{$\alpha$(2000.0)} & \colhead{$\delta$(2000.0)} & \colhead{Spectral} & ref. & \colhead{L$_{tot}$} & \colhead{ref.} & \colhead{Distance} & \colhead{ref.} & \colhead{Project ID}\\
%\colhead{} & \colhead{Name~~~~~~~~~~} & \colhead{} & \colhead{} & \colhead{Class} & \colhead{} & \colhead{luminosity} & \colhead{} & \colhead{}\\
\colhead{} & \colhead{Name~~~~~~~~~~} & \colhead{[h m s]} & \colhead{[$^\circ$ \arcmin\ \arcsec ]} & \colhead{Type} & \colhead{} & \colhead{[L$_\odot$]} & \colhead{} & \colhead{[pc]} & \colhead{} & \colhead{}
}
\startdata
LkH$\alpha$\,198 & V633\,Cas& 00 11 26.06 & $+$58 49 29.1 & B9 & 1 & 150 & 27,28 & 600   & 9        & m96bc63,m97bn21,\\
V\,376\,Cas      & HBC\,325 & 00 11 26.71 & $+$58 50 04.1 & A3-F2 & 1 & 436 & 13 & 600 & 9          & m01bi09\\
V892\,Tau     & Elias\,3$-$1& 04 18 40.62  & $+$28 19 15.5 & B8-9 & 1,29  & 400 & 11 & 130 & 12        & m99an21\\
LkH$\alpha$\,101 &HBC\,40   & 04 30 14.39  & $+$35 16 24.3 & B0 Ve & 14  & 4$\times$10$^4$ & 14 & 700 & 31 & m97bu34\\
AB\,Aur          &HD\,31293 & 04 55 45.84 & $+$30 33 04.3 & A1 Ve& 1 & 93 & 1,35 & 144 & 5  & m97bn21,nlserv\\
MWC\,480        &HD\,31648   & 04 58 46.27 & $+$29 50 37.0 & A5 V & 4 & 39 & 5,33 & 131  &  5   & m01bi09 \\
%also known as SAO 76866
HD\,34282        &V1366\,Ori& 05 16 00.48 & $-$09 48 35.3 & A3 Ve & 4,34 & 22.3 & 34  & 350    &  34     & m96bh17,m00au06\\% distance from Pietu et al03, spectral class from Gray & Corbally98  
HD\,35187       & BD$+$24$^\circ$826 & 05 24 01.17 & $+$24 57 37.6 & A2e+A7 & 58  & 23.4 & 13 & 150  & 58 & m97bc36 \\
HD\,36112        & MWC\,758 & 05 30 27.53 & $+$25 19 57.1 & A7 III & 8 & 31.3 & 13 & 200  &  5,8     & m97bc36 \\
MWC\,137 & V1308 Ori & 06 18 45.46 & $+$15 16 53.4 & B0ep & 3 & 1.5$\times$10$^3$ & 36 & 1300 &  2&m99an21,m01bi09\\
LkH$\alpha$\,215 & V699\,Mon& 06 32 41.80 & $+$10 09 33.6  & B6-8 & 1, 37 & 2.95$\times$10$^3$ & 13 & 800 & 38       & m99bu03 \\
VY\,Mon          &HBC\,202  & 06 31 06.94 & $+$10 26 05.0  & B8:e & 3  &  870 & 39 & 800 & 38         & UKT14 data \\
HD\,259431       & MWC\,147 & 06 33 05.19 & $+$10 19 20.0  & B6 & 1 & 1.07$\times$10$^4$ & 13 & 800 &   38     &m99bu03\\
AFGL\,961\,A     & IRAS\,06319$+$0415 & 06 34 37.6 & $+$04 12 43.9  & B2-B3/B5 & 6  & 3$\times$10$^3$ & 16   & 1500 & 17   & m98ai11 \\
R\,Mon           & MWC\,151 & 06 39 09.95 & $+$08 44 10.8 & B8 IIIev & 4   & 740 & 13 & 800 & 52,53 & UKT14 data \\
LkH$\alpha$\,25  & V590\,Mon& 06 40 44.64 & $+$09 48 02.1 & B7 & 1  & 63 & 13   & 800   &   13,24     &m00bc32\\
HD\,135344B    &  SAO\,206462 & 15 15 48.44& $-$37 09 15.9  & F4Ve & 7 & 8.5 & 13  & 140   &   7, 47   & m96bu90 \\
HD\,141569 & SAO\,140789 &15 49 57.75 & $-$03 55 16.4 & A0 Ve & 4 & 23 & 13 & 99 & 13,5 & m00ua06 \\
HD\,142666       & V1026\,Sco& 15 56 40.02 & $-$22 01 40.0  & A8 Ve & 25 & 17 & 13    & 145 &  13,25         & m97bc36\\
HD144432 &  SAO\,184124 &16 06 57.96 & $-$27 43 09.8 & A9 IIIe & 8 & 15 & 13 & $\sim$145 & 13,25 & m97bc36 \\
HR\,5999         & HD\,144668& 16 08 34.29 & $-$39 06 18.3  & A6 IIIe & 8  & 115 & 13  & 208 & 13 & m01bi09 \\%rms 11 mJy/beam
HD\,150193       & MWC\,863 & 16 40 17.92 & $-$23 53 45.2  & A3 IIIe & 8 & 49 & 13  & 150 &   13  & m01bi09 \\% also SAO 184536
CD$-$42$^\circ$ 11721 & V921\,Sco & 16 59 06.77 & $-$42 42 08.4 & B0ep & 40,41 & 2.34$\times$10$^4$ & 16 & 1150 & 42 & m99au36 \\
KK\,Oph          & HBC\,273 & 17 10 08.07 & $-$27 15 18.2  & A5 Ve  & 8 & 26 & 13  & 160 & 13,26 & m01bi09 \\%rms 11 mJy/beam
HD\,316285       & MWC\,272 & 17 48 14.04 & $-$28 00 53.2  & Be & 43 & 2.8$\times$10$^5$ & 43 & 1850 & 43 & m98au64 \\
HD\,163296       & MWC\,275 & 17 56 21.29 & $-$21 57 21.9  & A3Ve  & 25,50  & 35 & 49 & 122    &  5     & m97bn21,m0Xa/bec05\\
HD\,169142       & MWC 925  & 18 24 29.78 & $-$29 46 49.4 & A7 Ve & 8   & 22 & 8,49    & 145    &  25,47  & m97bc36\\
% aka SAO 186777
MWC\,297         & NZ Ser & 18 27 39.53 & $-$03 49 52.2 & B1.5 Ve & 4 & 1.07$\times$10$^4$ & 19 & 250 &  59 & m97bi09 \\
% aka SAO 185776
MWC\,300       & V431 Sct   & 18 29 25.69  & $-$06 04 37.3 & B[e] & 3 & 1.3$\times$10$^5$ & 20 & 1800 & 20        & m99au36 \\
HD\,176386        & CD$-$37$^\circ$ 13023 & 19 01 38.93 & $-$36 53 26.6 & A0V & 48 & 36 & 48 & 130 & 47 & m00au58 \\
TY\,Cr\,A        & CD$-$37$^\circ$ 13024 & 19 01 40.83 & $-$36 52 33.9 & B8V & 45,46 & 67 & 46 & 130 & 47 & m00au58 \\
BD$+$40$^\circ$4124 & V1685 Cyg & 20 20 28.37 & $+$41 21 51.4 & B3 & 1 & 6.0$\times$10$^4$ & 13 & 980 & 5 & m97bn21 \\
V\,1686\,Cyg       & LkH$\alpha$\,224 & 20 20 29.26 & $+$41 21 28.6 & F9 & 1 & 257 & 13 & 980 & 5  & m97bn21 \\
V\,1318\,Cyg\,S       & LkH$\alpha$\,225\,S & 20 20 30.59 & $+$41 21 26.0 & A5-Fe & 2  & 1600 & 22  & 980 & 5 &  m97bn21 \\
MWC\,349       &  V\,1478~Cyg   &  20 32 45.53 & $+$40 39 36.6 & Be & 44  &  3$\times$10$^4$ & 44   &1200 & 44 & m01bi09 \\
PV\,Cep          & HBC\,696 & 20 45 53.96 & $+$67 57 38.9 & A5 & 1  & 80 &  54   & 500 & 56    & m97bn21, \& UKT14 data \\
HD\,200775       & MWC\,361 & 21 01 36.92 & $+$68 09 47.8 & B3 & 1 & 1.9 $\times$ 10$^3$  & 36 & 430 & 5  & m99bu03,m01bi09 \\
V\,645\,Cyg      & RAFGL\,2789& 21 39 58.24 & $+$50 14 21.2 & B1-2 & 57 & 2 -- 6$\times$10$^4$ & 57 & 4200 & 57 & m01bi09 \\% from M99AU36 or M98au64
MWC\,1080\,A       & V628\,Cas & 23 17 25.59 & $+$60 50 43.6 & B0eq & 30 &  4.6$\times$10$^4$ & 54  & 2200 & 55 & m01bi09\\ 
\enddata
\bigskip
%\tablenotetext{a}{X = 3, 4, or 5}

\noindent \tiny{(1) Hernandez et al. (2004), (2) Hillenbrand et al.(1992), (3) Th\'e et al. (1994), (4) Mora et al. (2001), (5) van den Ancker et al. (1998), (6) Castelaz et al. (1985), (7) van Boekel et al. (1995), (8) Blondel et al. (2006), (9) Chavarria-K  (1985), (10) Natta et al. (1992), (11) Monnier et al. (2008), (12) Torres et al. (2009), (13) Manoj et al. (2006), (14) Tuthill et al. (2002), (15) Bertout et al. (1999), (16) this paper (17) Williams et al. (2009), (18) Habart et al. (2003), (19) Alonso-Albi et al. (2009), (20) Miroshnichenko et al. (2004), (21) Siebenmorgen et al. (2000), (22) Marvel (2005), (23) Reipurth et al. (1997), (24) Finkenzeller \& Mundt (1984), (25) van Boekel et al. (2005), (26) Leinert et al. (2004)}, (27) Lagage et al. (1993), (28) Natta et al. (1992), (29) Strom \& Strom (1994), (30) Cohen \& Kuhi (1979), (31) Herbig et al. (2004), (32) Meeus et al. (2001), (33) Dent et al. (2005), (34) Merin et al. (2004), (35) Acke et al. (2004), (36) Berilli et al. (1992), (37) Valenti et al. (2003), (38) Herbst et al. (1982), (39) Casey \& Harper (1990), (40) de Winter \& Th\'e (1990), (41) Shore et al. (1990), (42) Borges Fernandes et al. (2007), (43) Hillier et al. (1998), (44) Cohen et al. (1985), (45) Casey et al. (1995), (46) Casey et al. (1998), (47) de Zeeuw et al. (1999), (48) Meyer \& Wilking (2009), (49) Acke et al. (2004), (50) Gray \& Corbally (1998), (51) Cohen, Harvey \& Schwartz (1985), (52) Walker (1956), (53) Close et al. (1997), (54) Evan et al. (1986) (55) Cant\'o et al. (1994), (56) Cohen et al. (1981), (57) Miroshnichenko et al. (2009), (58) Duncin \& Crawford (1998), (59) Drew et al. (1997)

\end{deluxetable}

% here starts Table2a

\begin{deluxetable}{lllcrrr}
\tabletypesize{\scriptsize}
\tablecolumns{7}
\tablenum{2a}
\tablewidth{0pt} 
\tablecaption{Positions, deconvolved sizes and integrated flux densities of target stars}
\label{tbl-2a}
\tablehead{
\colhead{HAEBE} & \colhead{$\alpha$(2000.0)} & \colhead{$\delta$(2000.0)} & \colhead{$\theta_a$ $\times$ $\theta_b$} & \colhead{P.A.} & \colhead{S(850 $\mu$m)} & \colhead{S(450 $\mu$m)}\\ 
\colhead{star} & \colhead{[h m s]}& \colhead{[$^\circ$ \arcmin\ \arcsec ]} &  \colhead{} & \colhead{[~$^\circ$~]} & \colhead{[mJy]} & \colhead{[mJy]}
}
\startdata
LkH$\alpha$\,198\tablenotemark{a} & 00 11 26.23 & $+$58 49 32.2 & 11\ptsec4 $\times$ 5\ptsec6 & $+$29 $\pm$ 10 & 220 $\pm$ 30  & \nodata \\
V\,376\,Cas\tablenotemark{b} & 00 11 26.36 & $+$58 50 01.4 & $\leq$ 6\arcsec $\times$ 6\arcsec & \nodata  & 140 $\pm$ 20  & \nodata \\
V\,892~Tau & 04 18  40.64  & $+$28  19  16.1 & point source & \nodata & 630 $\pm$ 20 & 2250 $\pm$ 100 \\
% rms 450 100 mJy, 850 micron 20 mJy
AB~Aur & 04  55  45.89  & $+$30  33 05.1 & point source & \nodata & 350 $\pm$ 20 & 2200 $\pm$ 100 \\
%MIPS r12662784 - saturated
MWC\,480 & 04  58 46.45 &  $+$29  50  39.0 & 6\ptsec4 $\times$ 4\ptsec7& $+$138 $\pm$ \phantom{0}4 & 780 $\pm$ 20 & 3300 $\pm$ 290 \\
% FWHM 9.1 $\times$ 2.6 PA +133
% FWHM at 850 6.4 x 4.7 pa -42.4
HD\,34282  & 05 16 00.48 & $-$9 48 35.3 &  point source & \nodata & 360 $\pm$ \phantom{0}9 & 2700 $\pm$ 400\\
% 450 um, flux uncertain, FWHM lower than HPBW
HD\,36112  & 05  30  27.56 &  $+$25  19 56.9 & point source & \nodata & 197 $\pm$ 10 & 1540 $\pm$  \phantom{0}50 \\
% 850 rms 5.5 mJy/beam  450 mu, 53 mJy/beam ;m97bc36
HD\,35187         & 05 24 01.17 & $+$24 57 37.6 & point source & \nodata     & 70.6 $\pm$  \phantom{0}6  & 340 $\pm$  \phantom{0}85 \\
LkH$\alpha$\,101\tablenotemark{c} & 04 30 14.38 & $+$35 16 23.6 &  6\ptsec5 $\times$ 2\ptsec6 & $+$26 $\pm$ 20 & 818 $\pm$ 50 & \nodata \\
AFGL\,961\,A & 06 34 37.54 & $+$04 12 44.6 & 12\ptsec2 $\times$ 7\ptsec0 &  $-$41 $\pm$ \phantom{0}2 & 1950 $\pm$        60 & 14900 $\pm$ 300\\
AFGL\,961\,SMM & 06 34 36.71 & $+$04 12 50.9 & 8\ptsec8 $\times$ 5\ptsec6 &  $-$~5 $\pm$ 10 & 1330 $\pm$ 50 & 12300 $\pm$ 300 \\
HD\,135344\,B & 15 15 48.56 & $-$37 09 15.8 & point source & \nodata & 490 $\pm$ 10 & 3180 $\pm$ 230 \\
% 850 rms ~ 11 mJy/beam 
% 450 micron rms 220 mJy/beam; may need a footnote for 450 micron; poor quality; observations
% done during late morning (450 micron error  beam may be underestimated)
HD\,141569 & 15 49 57.75 & $-$03 55 16.4 & \nodata & \nodata & 10.9 $\pm$ 1.3 & 64.9  $\pm$ 13.3 \\
% photometry from m00ua06
HD\,142666 & 15  56 40.01 & $-$22  01 39.0 & 
$\leq$ 2\ptsec2 & \nodata & 313 $\pm$  \phantom{0}5  & 1140 $\pm$  \phantom{0}35 \\
% m97bc36  3.3 mJy/beam @850 29 mJy/beam @ 450 micron 
HD144432 & 16 06 57.90 & $-$27 43 09.8 & \nodata & \nodata & 129 $\pm$ \phantom{0}6 & \nodata \\
% m97bc36 one photometry set only - exceptionally dry night; but too short integration time
% and too high airmass to get a 450 micron detection
HR\,5999  & 16  08 34.42 &  $-$39 06  18.3 & point source & \nodata & 64 $\pm$ 20 & \nodata \\
% one sigma rms determined from cleaned image 20 mJy/beam
% From c2d 70 µm  3.78 +/-0.36 Jy ;  24 µm  2.89 +/- 0.72 Jy  8 µm: 3.49 +/- 0.59 Jy  6.5 µm 12.10 +/- 1.27 Jy  rest sat
% gator pos  16h08m34.37s -39d 6m 18.3s
HD\,150193 & 16  40 17.85 &  $-$23  53 44.8 & point source & \nodata & 101 $\pm$ 10 & \nodata \\
% one sigma rms determined from cleaned image 10 mJy/beam; NEAL Evans r5749248
% 17.92s ; 45.32"   J = 2.65 (0.049) H 3.35 (0.062) K 4.60 (0.068) ;3.6 0 4.5 0; 6 um 5.76 (0.62) ; 8um - ; 24mu  -, 70µm 3.05 +/- 0.29 Jy
KK\,Oph  & 17 10  08.12 &  $-$27 15 18.8 & point source & \nodata & 91 $\pm$ 15 & \nodata \\
%rms 11 mJy/beam  MIPS R14094848 ; 2.875 +/-   0.012 Jy  (ap3 2.849 +/-  0.009 ) MIPS Spectral Energy exists though (Acke)
HD\,163296  & 17 56  21.42 &  $-$21  57  21.6 & 4\ptsec6 $\times$ 3\ptsec2 & $-$13 $\pm$ 27 & 1900 $\pm$ 200 & 8700 $\pm$ 370\\
%rms 11 mJy/beam
HD\,169142
%\tablenotemark{d} 
 & 18 24 29.87 & $-$29 46 48.9 & $\sim$ 2\ptsec3 $\times$ 1\ptsec6 & \nodata & 565 $\pm$ 10 & 3340 $\pm$ 115\\
%aka SAO 186777, MWC925 rms 450 127 mJy but beam 7.4". Size from 450 um, at 850  3."02    1."5
MWC\,297  & 18 27  39.53 &  $-$03  49 50.4 & $<$4\arcsec & \nodata & 570 $\pm$  30 &  1450 $\pm$ 200 \\
V\,1318~Cyg\,S-mm  & 20  20  30.67 &  $+$41  21  29.9 & 8\arcsec\ $\times$ 3\arcsec\ & $+$62 $\pm$ ~1 & 2050 $\pm$ 50 &  10200 $\pm$ 200 \\
MWC\,349 & 20 32 45.53 & $+$40 39 36.6 & point source & \nodata & 2600 $\pm$ 70 & \phantom{0}5000 $\pm$ 1100 \\
PV\,Cep  & 20  45  54.31 &  $+$67 57 40.7 & 4\ptsec9 $\times$ 2\ptsec8 & $+$74 $\pm$ 10 & 1000 $\pm$ 20 & 6550 $\pm$ 150\\
% rms 450 micron map 120 mJy/beam; no MIPS 70 um Hector missed it
V\,645\,Cyg & 21 39 58.26 & $+$50 14 22.3 & 14\ptsec2 $\times$ 8\ptsec5 & $-$65 $\pm$ ~3 & 2400 $\pm$ 60 &    20800 $\pm$ 600 \\
%No Spitzer data
MWC\,1080\,A & 23 17 24.35 & $+$60 50 40.4 & point source & \nodata & 250 $\pm$ 50 & \nodata \\
\enddata
\tablenotetext{a}{~ LkH$\alpha$\,198  and LkH$\alpha$\,198\,B are blended with the nearby Class~0 source, LkH$\alpha$\,198-mm, resulting in uncertain size and position.  LkH$\alpha$\,198\,B appears to dominate the sub-millimeter emission.}
\tablenotetext{b}{ ~Peak offset by 3\arcsec\ from the optical position of V\,376\,Cas, possibly  emission from the surrounding cloud.}
\tablenotetext{c}{ ~Flux density at 750 $\mu$m, S(750$\mu$m) = 940 $\pm$  100 mJy}
%\tablenotetext{d}{ Observed during day time. Beam profile characterized from Neptune maps taken the same time.}
\end{deluxetable}

% here starts Table2b

\begin{deluxetable}{llcccc}
\tabletypesize{\scriptsize}
%\footnotesize
\tablecolumns{5}
\tablenum{2b}
\tablewidth{0pt} 
\tablecaption{Submillimeter photometry of compact HAEBE stars}
\label{tbl-2b}
\tablehead{
\colhead{HAEBE} & \colhead{$\lambda$} & \colhead{HPBW} & \colhead{Flux density} & \colhead{Date} \\ 
\colhead{star} & \colhead{[$\mu$m]} & \colhead{[ $''$ ]}  & \colhead{[mJy]} &  \colhead{mm/dd/yr}
}
\startdata
AB Aur & 1300\tablenotemark{a} & 32.0 & 136 $\pm$ 15 & 12/10/89 \\
             & 1100 & 18.4 &192 $\pm$ 19 & 08/20/88 \\
             & 800   & 15.8 & 534 $\pm$ 69 & 08/20/88 \\
             & 800   & 16.8 & 475 $\pm$ 56 & 08/16/90 \\
             & 450  &  17.5 & 1760 $\pm$ 360 & 08/16/90\\
R Mon & 1300\tablenotemark{a}   &32.0    & 96 $\pm$ 15  &12/09/89\\
              &1300   &19.0     &36 $\pm$18 &10/09/91\\
              &1100   &18.4	&80 $\pm$ 20   &08/19/88\\
              &850    &17.8	&171 $\pm$ 24	 &08/16/90\\
              &800    &15.8	&260 $\pm$ 30  &08/19/88\\
              &800    &16.8	&197 $\pm$ 30	 &08/16/90\\
              &450    &17.5	&1200 $\pm$ 160   &08/19/88\\
              &350    &19.0	&2600 $\pm$ 500    &08/19/90\\
VY Mon & 1300 & 19.0 & 189 $\pm$ 23 & 10/09/91 \\
             & 1100 & 18.5 & 247 $\pm$ 27 & 10/09/91 \\
             & 800   & 16.5 & 552 $\pm$ 55 & 10/09/91 \\
 KK Oph & 1100   &18.4   &36 $\pm$ 11 &08/11/91 \\
             &800    &16.8   &109 $\pm$ 20	&08/18/90\\
             &800    &16.4   &119 $\pm$ 53 &10/06/91\\
PV Cep& 1300   &32.0     &0.375 $\pm$ 0.044	 &12/10/89\\
	    & 1300   &19.5	&0.36 $\pm$ 0.023  &08/18/90\\		
	    & 1100   &18.4	&0.50 $\pm$ 0.04   &08/20/88\\		
	    & 1100   &18.4	&0.47 $\pm$ 0.03   &10/27/88\\		
  	    & 850    &17.8 	&1.01 $\pm$ 0.07   &08/18/90\\
	    & 800    &15.8	&1.3  $\pm$ 0.1    &08/20/88	\\
	    & 800    &15.8	&1.11 $\pm$ 0.09   &10/27/88\\
	    & 800    &16.8	&1.16 $\pm$ 0.06   &08/18/90 \\
	    & 800    &13.5	&1.2  $\pm$ 0.1    &08/20/88	\\
	    & 750    &17.5        &1.41 $\pm$ 0.09   &08/11/91 \\
	    & 450    &17.5	&7.91 $\pm$ 0.62   &08/18/90  \\
	    & 450    &17.5	&3.48 $\pm$ 1.20   &08/11/91 \\
	    & 450    &13.5	&10.6 $\pm$ 0.8     &08/20/88\\		  
	    & 450    &  \phantom{0}7.9	&10.5 $\pm$ 0.5    &08/20/88\\
	    & 350    &17.7	&17.6 $\pm$ 1.0    &08/20/88\\  	    	   
	    & 350    &10	         &16.2 $\pm$ 0.7    &08/20/88\\      
\enddata
\tablenotetext{a}{CSO observations}
\end{deluxetable}

% here starts Table2c

\begin{deluxetable}{lcl}
\tablecolumns{4}
\tablenum{2c}
\footnotesize  
\tablewidth{0pt} 
\tablecaption{Non-detections} 
\label{tbl-2c}
\tablehead{
\colhead{Star~~~~~~~~~~} & 850 $\mu$m upper limit & \colhead{Comments}\\ 
\colhead{}  & \colhead{[mJy/beam]} & \colhead{}
}
\startdata
MWC\,137 & 15  & cloud emission, nearby sources \\
LkH$\alpha$\,215 &22 & emission from the surrounding nebula \\
HD\,259431 & 20 & emission from the surrounding nebula \\
LkH$\alpha$\,25 & 25 & cloud emission \\
CoD$-$42$^\circ$ 11721 & 50  & nearby sources \\
HD\,316285  & 70 & cloud emission \\
MWC\,300  & 7 & empty field \\
HD\,176386 &  50 & cloud emission \\
TY~Cr\,A & 50 & cloud emission \\
%R~Cr\,A  & 5 & cloud emission, nearby sources \\
%T~Cr\,A  & 5  &cloud emission, nearby sources \\
BD$+$40$^\circ$4124 & 35 & cloud emission \\
V1686\,Cyg &  35 & cloud emission, nearby source \\
HD\,200775 & 20 & emission from the surrounding nebula \\
\enddata
\end{deluxetable}

% here starts Table3

\begin{deluxetable}{llll}
\tablecolumns{4}
\tablenum{3}
\tablewidth{0pt} 
\tablecaption{{\it Spitzer} MIPS flux densities of program stars} 
\label{tbl-3}
\tablehead{
\colhead{Star} &\colhead{24~$\mu$m}  & \colhead{70~$\mu$m} & \colhead{AOR ID}\\ 
\colhead{}  & \colhead{[Jy]} & \colhead{[Jy]} &\colhead{} \\
}
\startdata
AB Aur &  S\tablenotemark{a}  & S\tablenotemark{a}  & r12662784 \\
V\,892~Tau &  S\tablenotemark{a} &  25.41 $\pm$ 0.10\tablenotemark{b} & r11119696, r11229952  \\
AFGL\,961 & S\tablenotemark{a}  & S\tablenotemark{a} & r12234249, r12234752\\
HR\,5999    & 2.89 $\pm$ 0.72  &  3.78 $\pm$ 0.36 & c2d\\
HD\,141569 & 1.47 $\pm$ 0.01 & 4.70 $\pm$ 0.02 & r11181312 \\
HD\,150193 & S\tablenotemark{a} & 3.05 $\pm$ 0.29 & c2d \\
CD$-$42$^\circ$ 11721 & S\tablenotemark{a}  & S\tablenotemark{a} & MIPSGAL \\
KK Oph   & S\tablenotemark{a}  & S\tablenotemark{a} & r14094848\\
BD$+$40$^\circ$4124 & S\tablenotemark{a}  & \nodata & r22512128, r22512384\\
V1686\,Cyg & S\tablenotemark{a}  & \nodata & r22512128, r22512384\\
V1318\,Cyg\,S-mm  & S\tablenotemark{a}  & 85.68 $\pm$  0.66 & r22512128, r22512384\\  
MWC\,349 & S\tablenotemark{a} & 6.72 $\pm$  0.30 & r22510336 \\
\enddata
\tablenotetext{a}{Saturated}
\tablenotetext{b}{Mildly saturated}
\end{deluxetable}

% here starts Table4
\begin{deluxetable}{lllllr}
\tablecolumns{6}
\tablenum{4}  
\tablewidth{0pt} 
\tablecaption{Isothermal graybody fits to HAEBE ``disks''} 
\label{tbl-4}
\tablehead{
\colhead{Source~~~~~~~~~} & \colhead{T$_d$} &\colhead{$\alpha$} & \colhead{$\beta$}& \colhead{M$_{tot}$} & \colhead{L$_{dust}$}\\ 
\colhead{} & \colhead{[K]} &\colhead{} &\colhead{} & \colhead{[M$_{\odot}$]} & \colhead{[\Lsun]}
}
\startdata
V\,892 Tau &  40  & 2.19 $\pm$ 0.09 &  0.52  &  0.009    &   0.27    \\
AB Aur        & 66 & 2.89 $\pm$ 0.04 &  1.05 & 0.006 & 5.0 \\
MWC\,480 &28  & 2.65 $\pm$ 0.06 & 0.81 & 0.02 & 0.23\\
HD\,34282%\tablenotemark{b} 
& 29 & 3.03 $\pm$ 0.12 &1.29 & 0.18 & 1.85 \\
HD\,35187 &  44 & 2.64 $\pm$ 0.34 & 0.87 & 0.002 & 0.15 \\
HD\,36112 & 43 & 3.03 $\pm$ 0.10 & 1.25 & 0.014 & 1.0 \\
AFGL\,961\,A & 37  & 2.73 $\pm$ 0.02 & 1.51 & 12.5 & 930\\
R Mon  & 62 & 3.08 $\pm$ 0.15 & 1.3 & 0.12 & 110 \\
VY~Mon \tablenotemark{a}& 38 & 2.31 $\pm$ 0.93 & 1.9 & 1.9 & 102 \\
HD\,135344\,B & 32 & 2.94 $\pm$ 0.30 & 1.38 & 0.025 & 0.20\\
HD\,141569%\tablenotemark{c} 
& 40 & 2.21 $\pm$ 0.06 & 0.5 & 0.00009 & 0.03 \\
HD\,142666%\tablenotemark{c}
& 31 & 2.42 $\pm$ 0.07 & 0.72 & 0.005  & 0.08 \\
HD\,144432 & 40 & 2.70 $\pm$ 0.39 & 0.59 & 0.002 & 0.10\\
HD\,150193 & 44 & 3.03 $\pm$ 0.15 & 0.65 & 0.002 & 0.12 \\
KK~Oph       & 40 & 1.70 $\pm$ 2.12 & 0.74 & 0.002 & 0.13 \\
HD\,163296 & 28 & 2.66 $\pm$ 0.05 & 0.94 & 0.065 & 0.52 \\
HD\,169142 & 41 &  2.84 $\pm$ 0.06 & 1.30 & 0.03 & 0.50\\
HR\,5999 & 41 & 3.03 $\pm$ 5.18 & 1.75 & 0.009 & 1.1 \\
V\,645~Cyg  & 38 & 2.84 $\pm$ 0.06 & 1.55 & 124 & 11750 \\
V\,1318~Cyg-mm & 41 & 2.55 $\pm$ 0.06 & 0.85 & 1.9 & 118 \\
%HD\,163296 & 30 & 1.06 & 0.064 &  0.77\\
PV~Cep       & 29  & 3.09 $\pm$ 0.06 & 1.59 & 1.0 & 13.7\\
\enddata
\tablenotetext{a}{Fit to UKT14 photometry data, likely to include some contribution from the surrounding cloud.}
%\tablenotetext{b}{Fit includes data from Pietu03, Mannings97 and  Sylvester96}
%\tablenotetext{c}{Fit includes mm and submm photometry from Sylvester96}

\end{deluxetable}

\clearpage

\includegraphics[angle=0,scale=0.9]{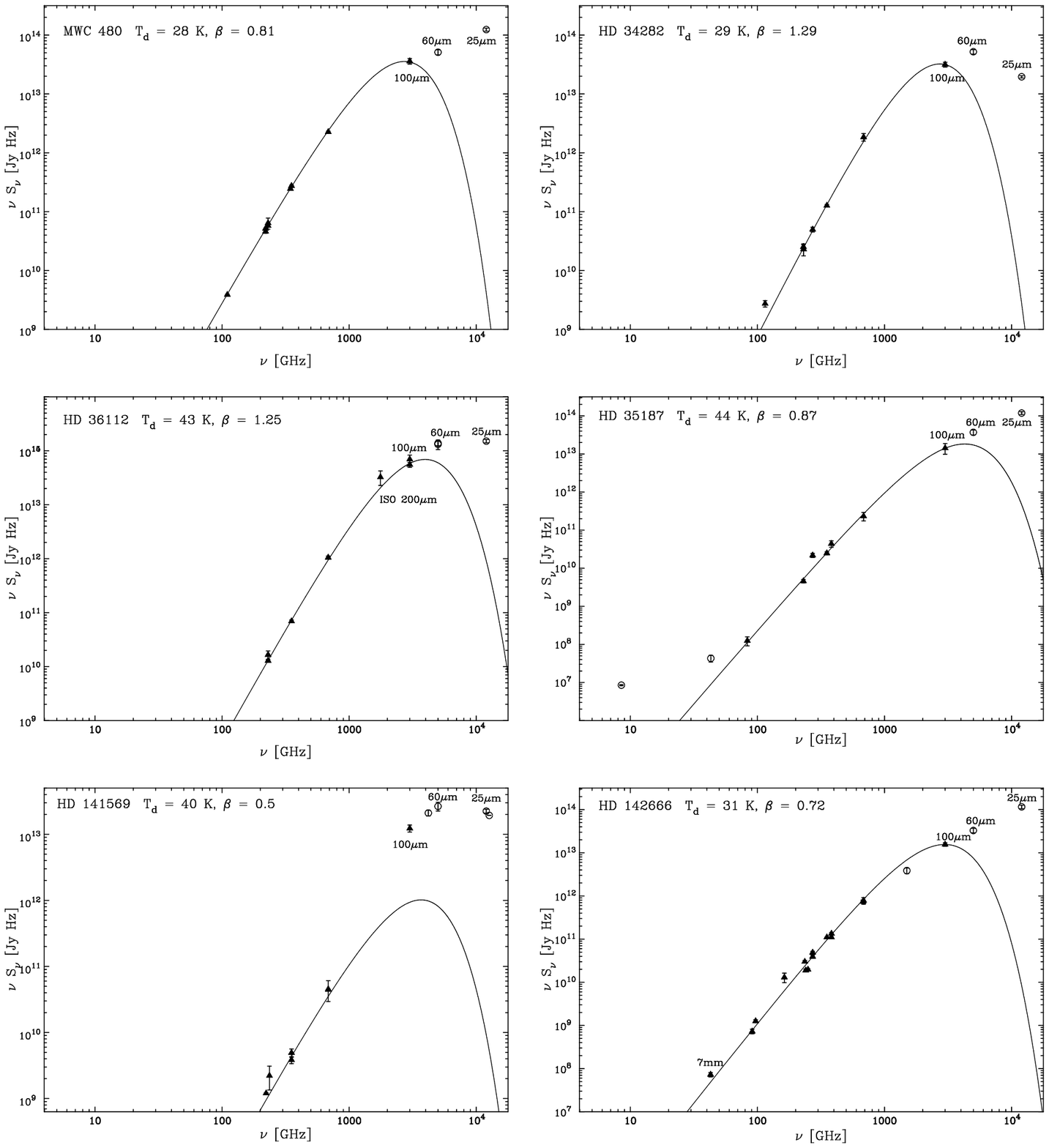}
\figcaption[]{
\label{fig-tr1}
Isothermal graybody fits to SCUBA and literature data for isolated HAEBE
stars. Flux densities used in the fits are marked by filled triangles,
while data points marked with open circles were excluded from the fits.
In the top left corner of each plot we give the name of the star and
essential fit results (T$_d$, $\beta$).}

\includegraphics[angle=0,scale=0.9]{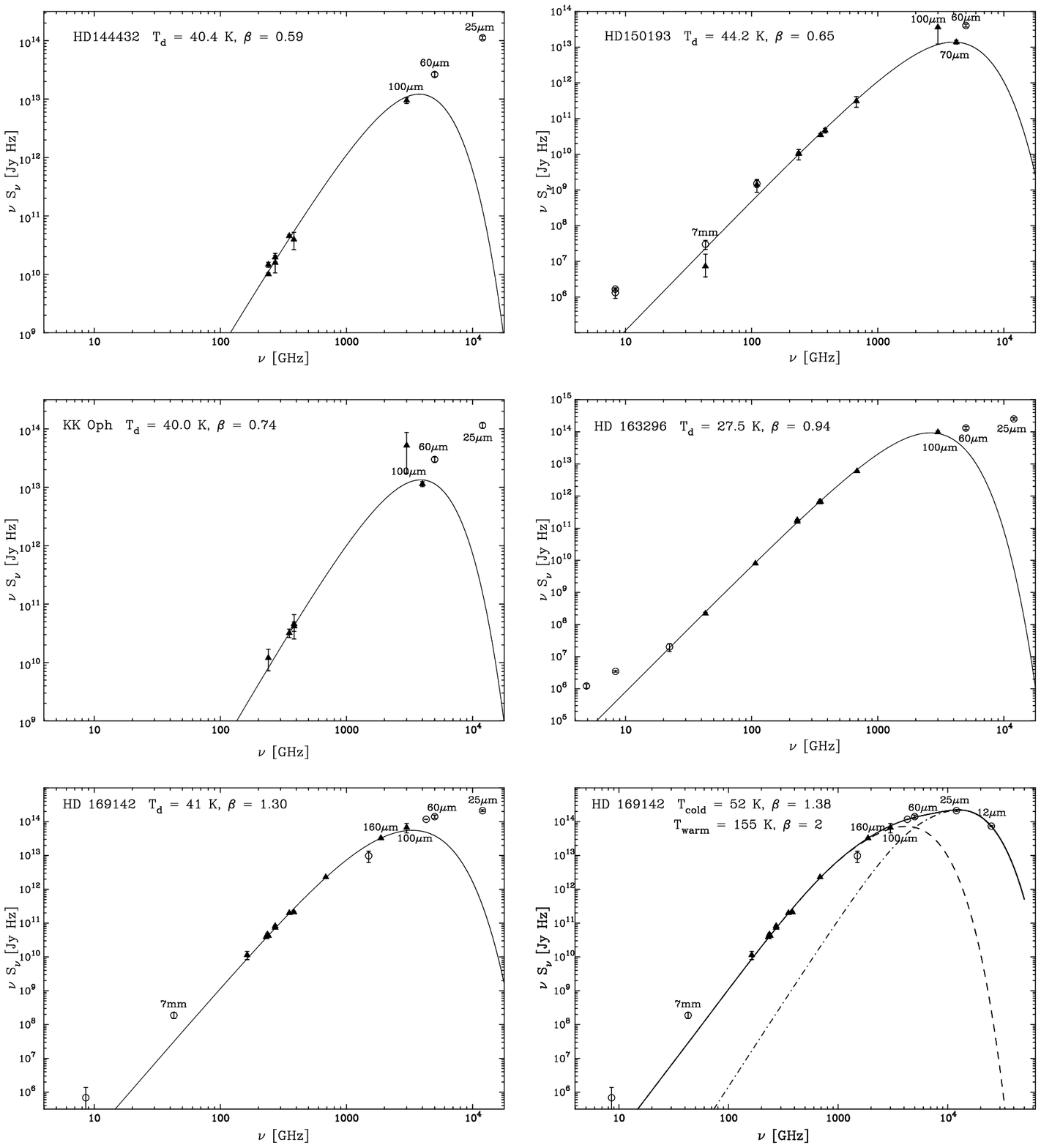}
\figcaption[b]{
\label{fig-tr2}
Isothermal graybody fits to isolated HAEBE  stars continued. Labeling as
in previous figure. In the right panel for HD\,169142, we show a  isothermal fit with two dust components at different temperatures. The cold
dust, which dominates at the long wavelengths is shown as a  dashed line, the warmer dust as a dash-dotted line, and the combined
two-temperature fit as a solid line.}

\includegraphics[angle=-90,scale=0.67]{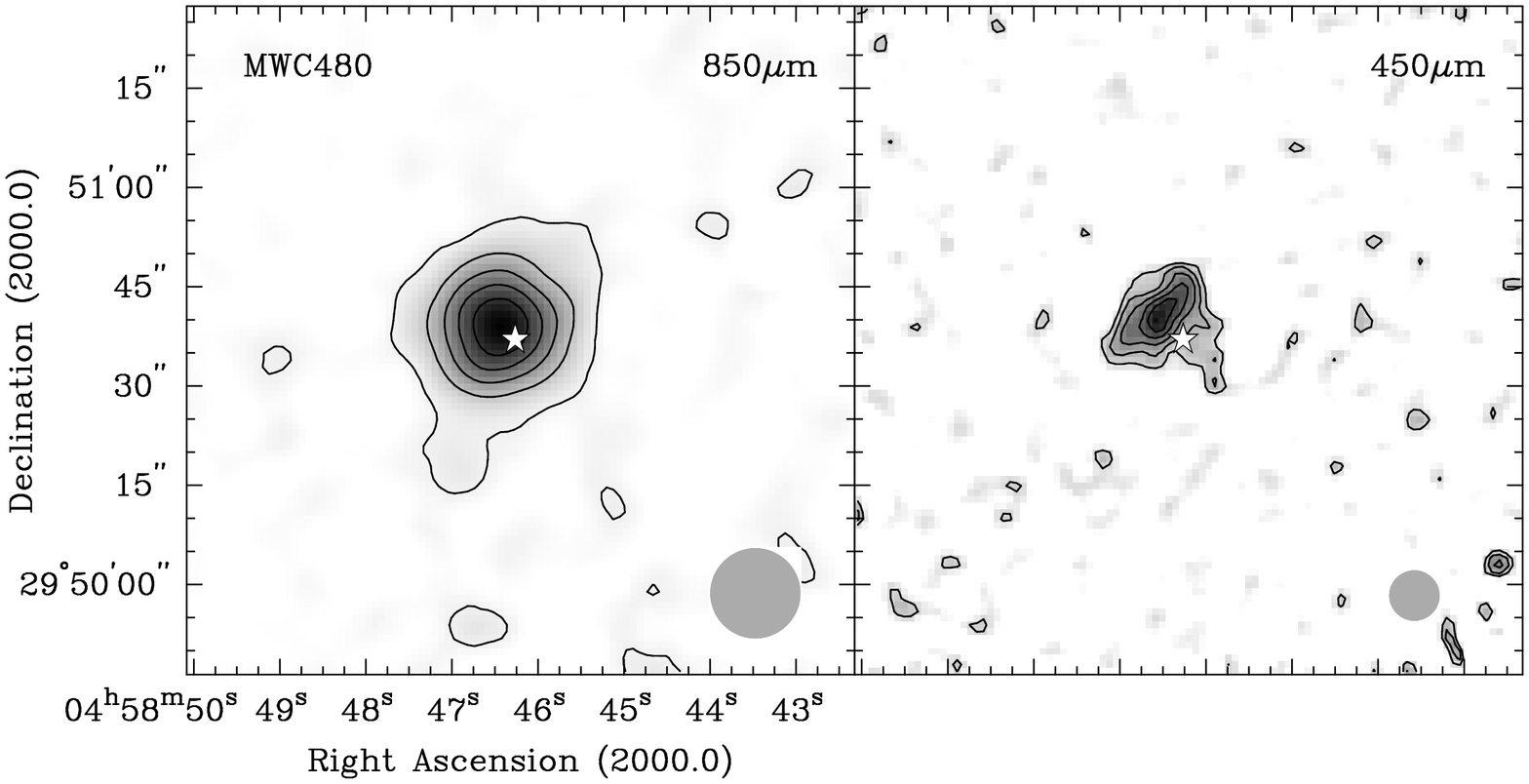}
\figcaption[]{
\label{fig-mwc480}
Deconvolved sub-millimeter images of MWC\,480 at 850 and 450 \mum\
plotted in gray-scale and overlaid with contours. The contour levels are
 linear with six contours between the lowest contour level (3-$\sigma$)
and the peak flux density. The 3-$\sigma$ level is at  54
mJy~beam$^{-1}$ and   at 840 mJy~beam$^{-1}$ for 850 and 450 $\mu$m,
respectively. The disk is resolved by SCUBA both at 850 and 450 $\mu$m.
The position of MWC\,480 is shown with a star symbol. The HPBW is
plotted in the bottom left corner of each image. }
%\end{figure}

 \newpage
% 

%\begin{figure} 
\includegraphics[angle=-90,scale=0.67]{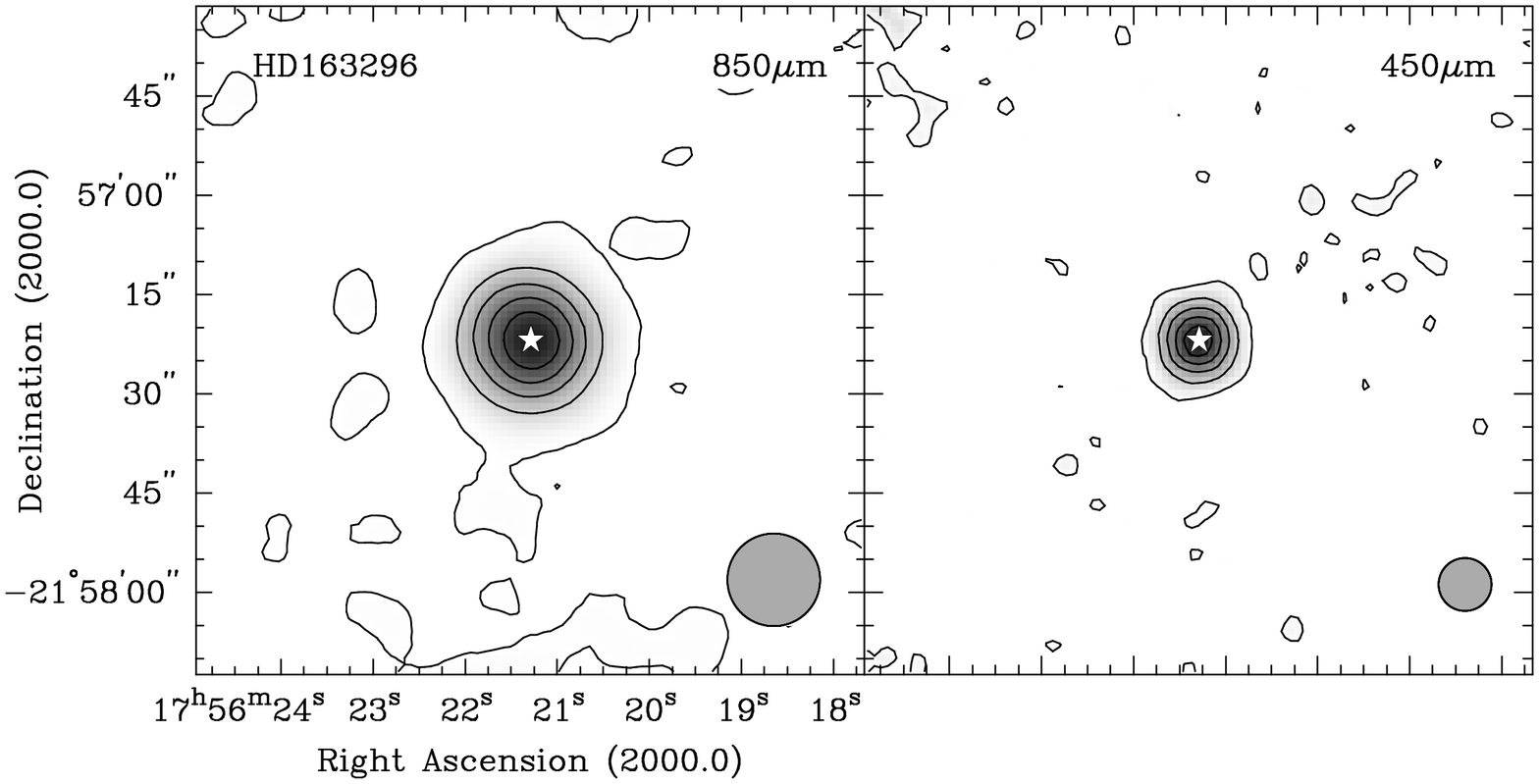}
\figcaption[]{
\label{fig-hd163296}
Deonvolved sub-millimeter images HD\,163296 at 850 and 450 $\mu$m. The
contours are plotted with  linear steps. At 850 $\mu$m we plot six
contours between the lowest contour level, 32 mJy~beam$^{-1}$
(2-$\sigma$) and the peak flux density, 1.8 Jy~beam$^{-1}$. At 450
$\mu$m we plot six contours starting from 0.45 Jy~beam$^{-1}$ to 7.24
Jy~beam$^{-1}$.  The disk is resolved by SCUBA both at 850 and 450
$\mu$m. The position of HD\,163296 is marked by a star symbol. The HPBW
is plotted in the bottom left corner of each image.}

 \newpage

\includegraphics[angle=-90,scale=0.67]{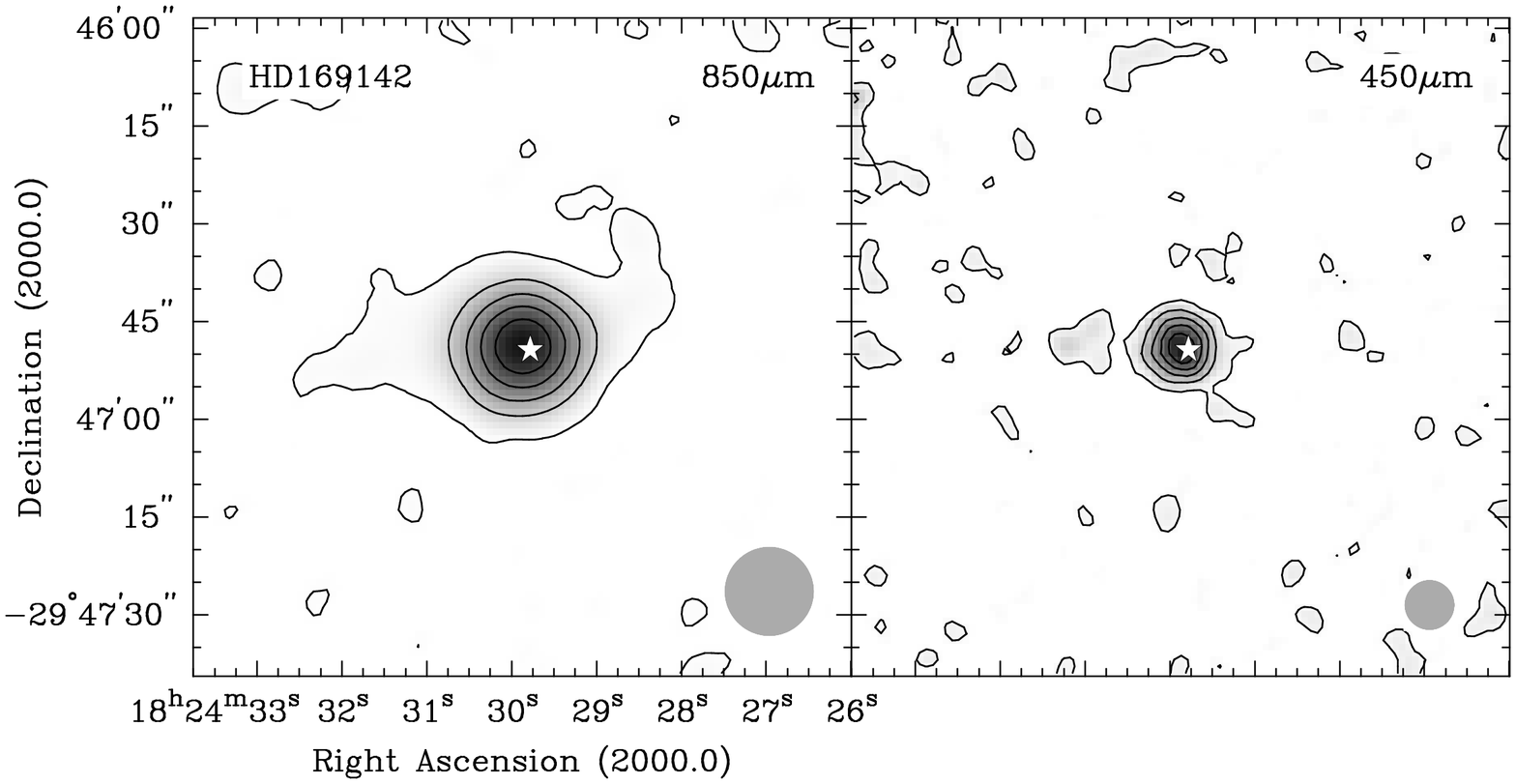}
\figcaption[]{ 
\label{fig-hd169142}
Deonvolved SCUBA images of HD\,169142 in grayscale overlaid  with
contours. The contour levels are linear for both 850 and 450 $\mu$m with
six contours starting at 2-$\sigma$. At 850 $\mu$m contours go from  15
mJy~beam$^{-1}$ to 537 mJy~beam$^{-1}$, at 450 $\mu$m from 0.28
Jy~beam$^{-1}$ to  3.17 Jy~beam$^{-1}$. The disk is marginally resolved
(see text).}

\includegraphics[angle=-90,scale=0.64]{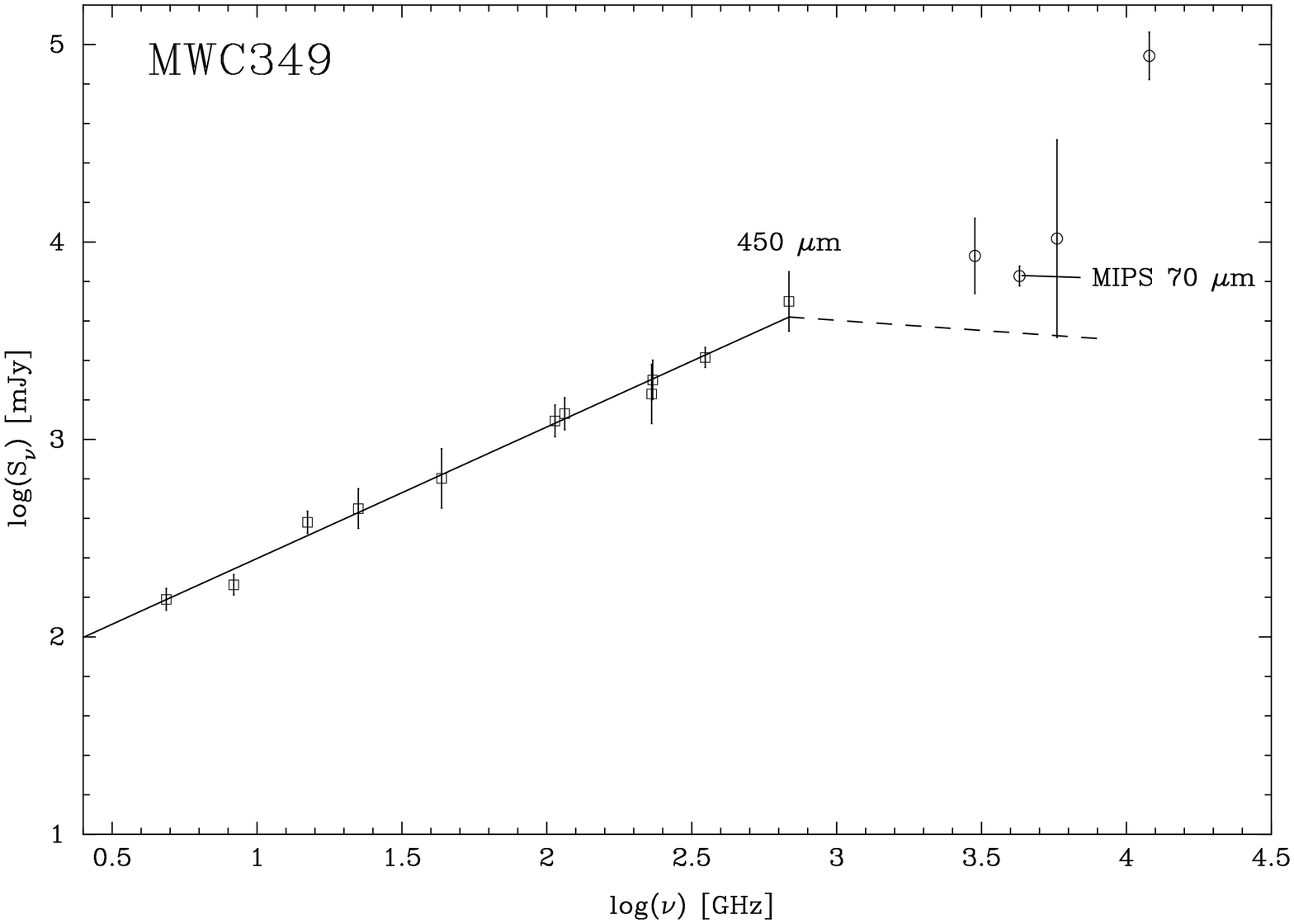}
\figcaption[]{
\label{fig-mwc349sed}
Least squares fit to radio and millimeter data for MWC\, 349 showing
that the star has a thermal wind spectrum which dominates the emission to about
450 $\mu$m with the flux density proportional to the frequency  $\nu^{0.67
\pm 0.03}$. If we assume that the wind is launched at 15 AU from the
star, the wind should become optically thin at 450 $\mu$m (see text).
}

\includegraphics[angle=-90,scale=1.0]{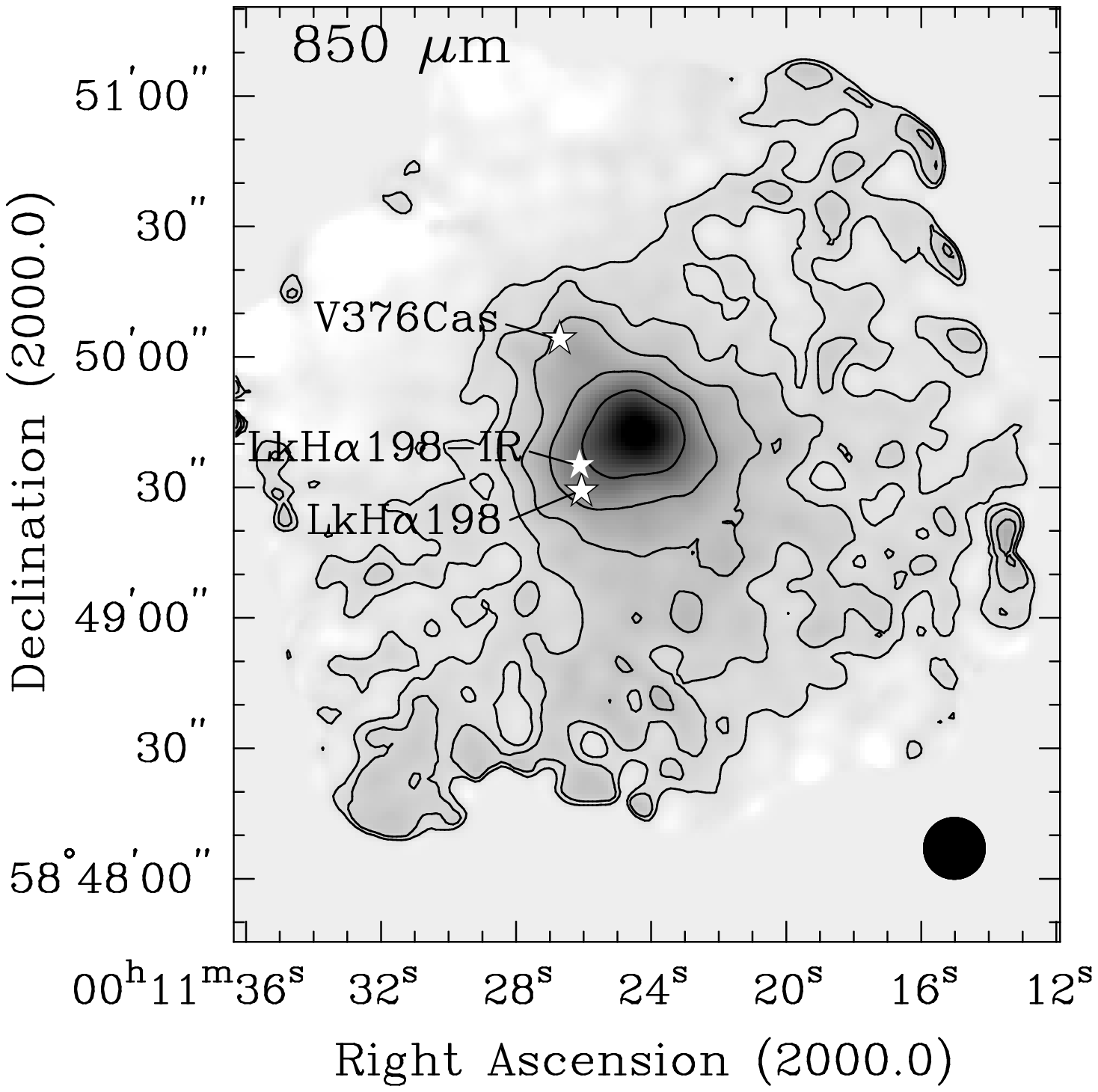}
\figcaption[]{
\label{fig-lkha198}
SCUBA 850 $\mu$m image in grayscale  overlaid with contours of  the
dense  cloud in which LkH$\alpha$\,198 and V376\,Cas are embedded. The
cloud core is quite  extended and probably more extended to the
northeast, where the image goes negative because we chopped on extended
emission. The emission is dominated by a  Class 0 source,
LkH$\alpha$\,198-mm, which does not have an optical or near-infrared
counterpart. This image suggests that both LkH$\alpha$\,198, its
infrared companion and V376\,Cas may be associated with dust emission,
but we do not have high enough spatial resolution to separate the stars
from the surrounding strong cloud emission. The contours are plotted
with six logarithmic intervals from  40 mJy~beam$^{-1}$ to 750
mJy~beam$^{-1}$. The HPBW is shown in the bottom right corner.}

\includegraphics[angle=-90,scale=0.71]{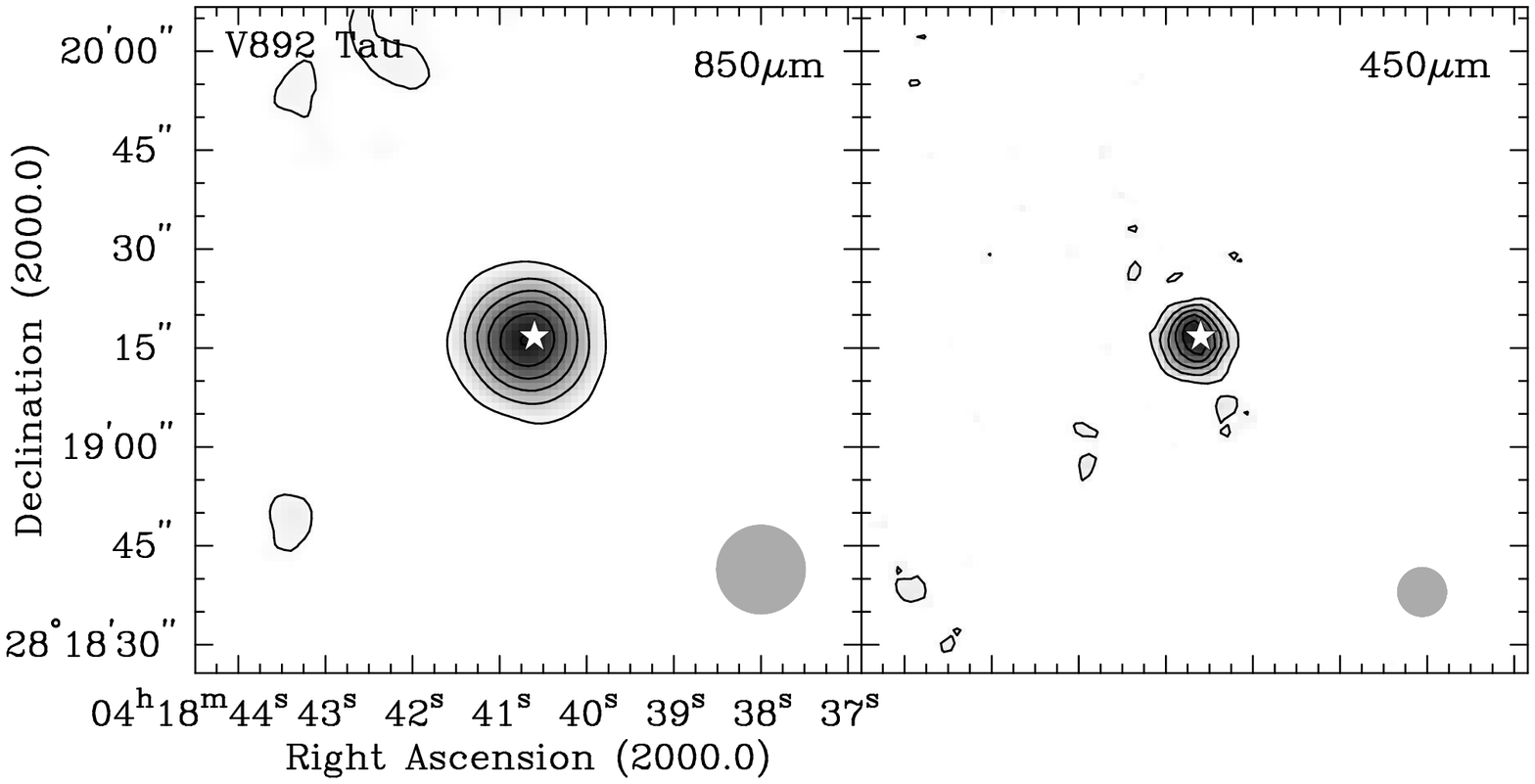}
\figcaption[]{
\label{fig-V892_SCUBA}
Deconvolved SCUBA images of   V\,892~Tau in grayscale overlaid with
contours. The contour levels are evenly spaced with six contours between the
lowest contour level, $\sim$ 2-$\sigma$ and the peak flux density. At
850 $\mu$m contours go from 36 mJy~beam$^{-1}$ to 630 mJy~beam$^{-1}$,
at 450 $\mu$m from 0.2 Jy~beam$^{-1}$ to  2.25 Jy~beam$^{-1}$.  The
position of  V\,892~Tau is marked by a star symbol. The HPBW is shown in
the bottom right corner.}

\clearpage

\includegraphics[angle=-90,scale=0.65]{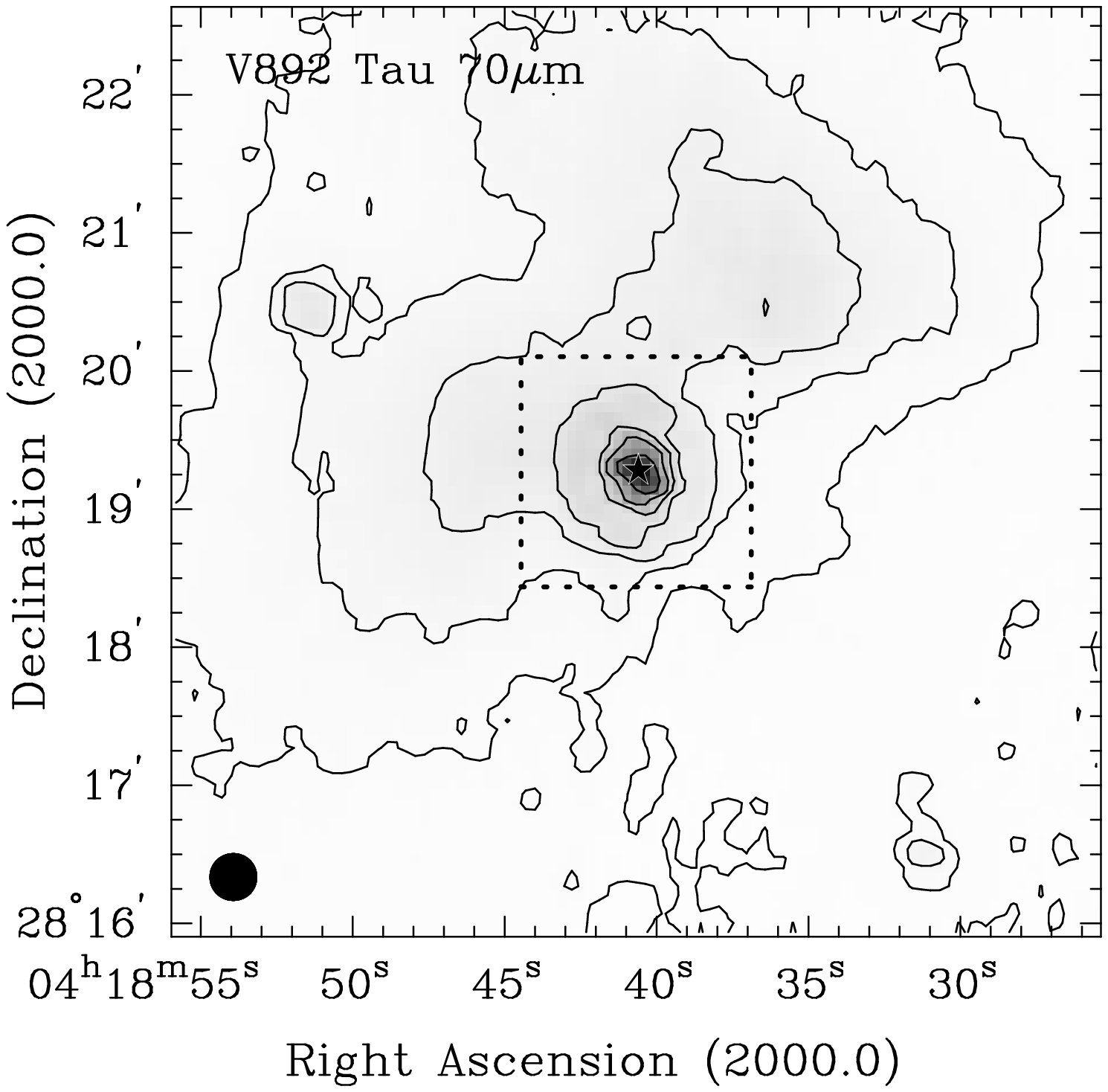}
\figcaption[]{
\label{fig-V892_MIPS}
MIPS 70 $\mu$m  image of V\,892~Tau in grayscale overlaid with logarithmic
contours and covering a much larger field than the SCUBA images. The
extent of the SCUBA images is drawn as a dotted rectangle. At 70 $\mu$m
there is extended nebulosity east and northwest of the star, but no sign
of embedded sources in the nebulosity. Two faint 70 $\mu$m sources are
visible within the displayed area. The 70 $\mu$m HPBW is shown in the
bottom left corner.}

\includegraphics[angle=-90,scale=0.63]{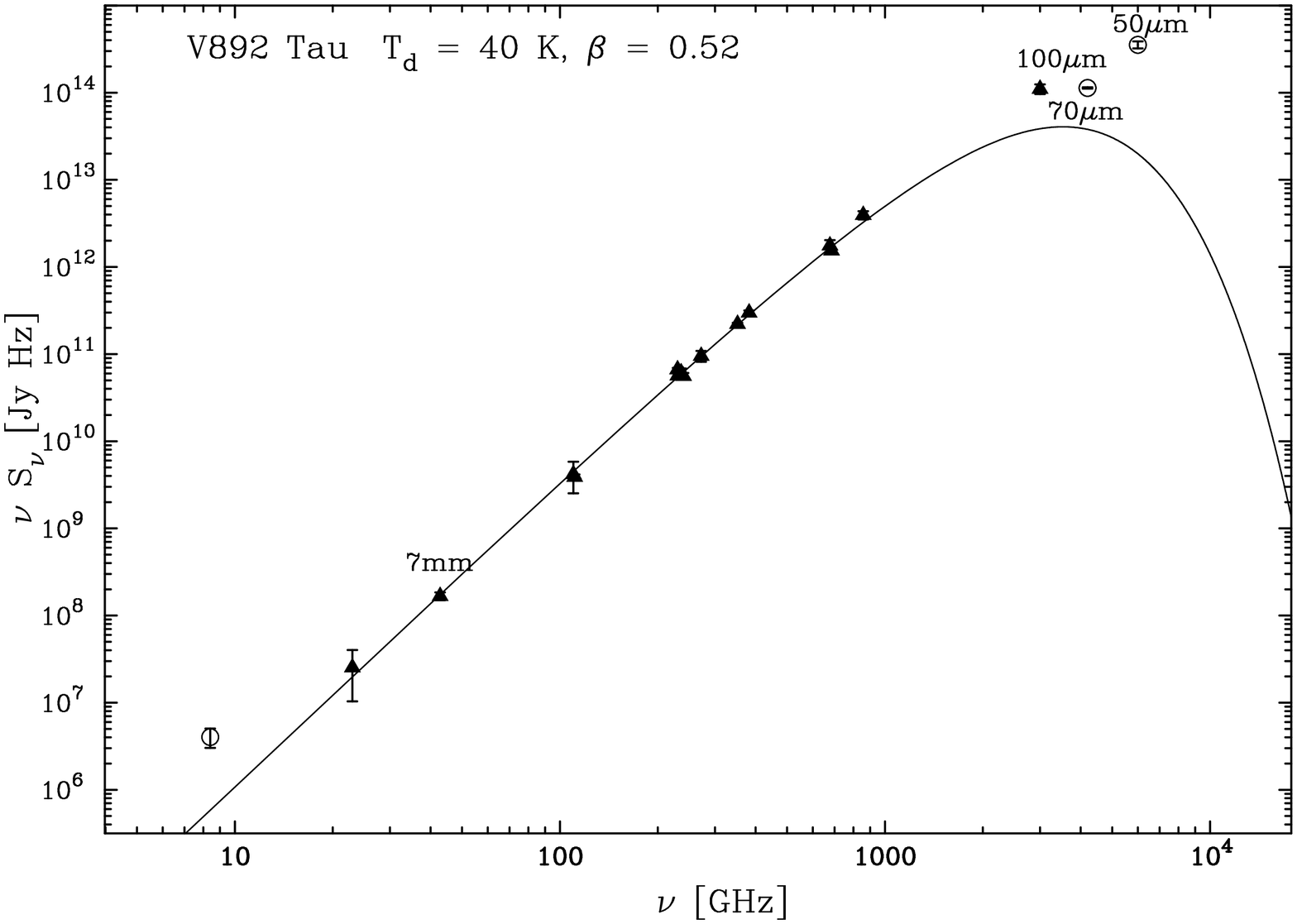}
\figcaption[]{
\label{fig-V892_sed}
Isothermal graybody fit of V\,892~Tau assuming a dust temperature of 40 K.}

\clearpage

\includegraphics[angle=-90,scale=0.65]{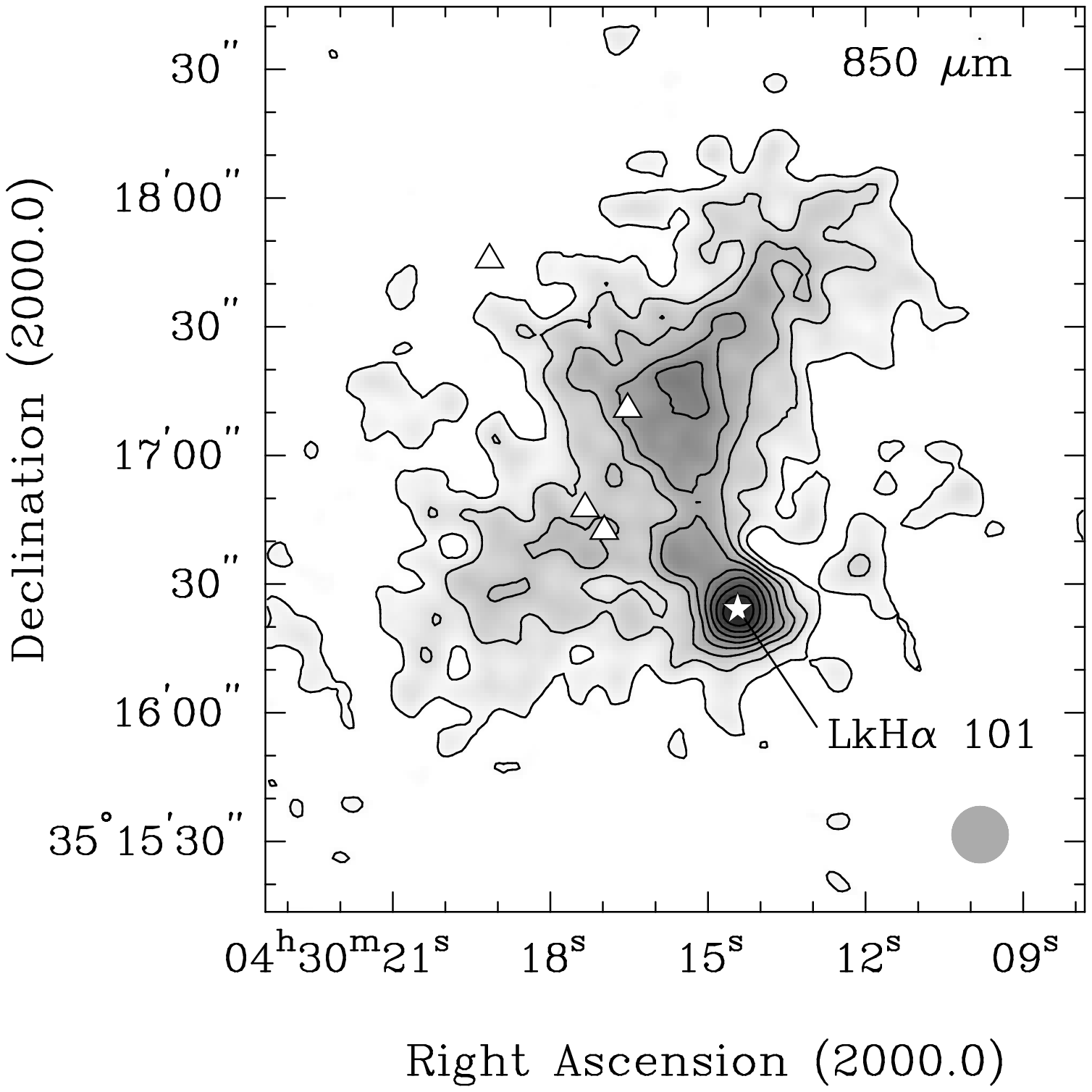}
\figcaption[]{
\label{fig_lkha101_850}
SCUBA 850 $\mu$m grayscale image of LkH$\alpha$ 101. The  contours are linear with the
lowest contour at 100 mJy~beam$^{-1}$ and a step of 100 mJy~beam$^{-1}$. The peak flux is 0.9 Jy~beam$^{-1}$. The
position of LkH$\alpha$ 101 is marked on the figure. The HPBW is shown in the bottom right corner.}

\includegraphics[angle=-90,scale=0.65]{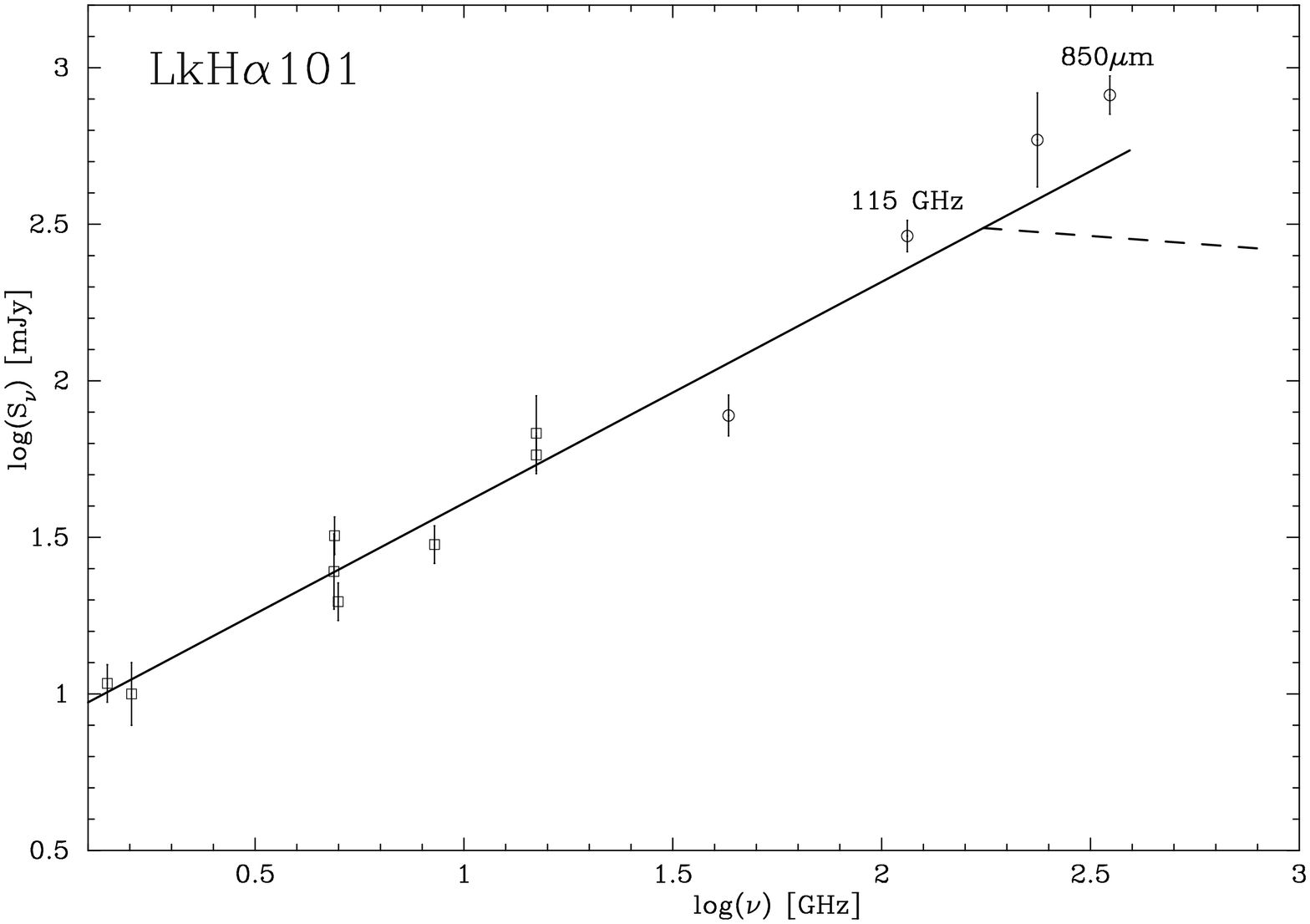}
\figcaption[]{ 
\label{fig-lkha101sed}
Power-law fit to VLA A-array data for LkH$\alpha$\,101 predicting a
spectral index, $\alpha$ = 0.71 $\pm$ 0.07 for the thermal wind. 
 If we assume that the
wind is launched at 10 AU from the star, the wind will become optically
thin at $\sim$  150 GHz.}

\includegraphics[angle=-90,scale=0.65]{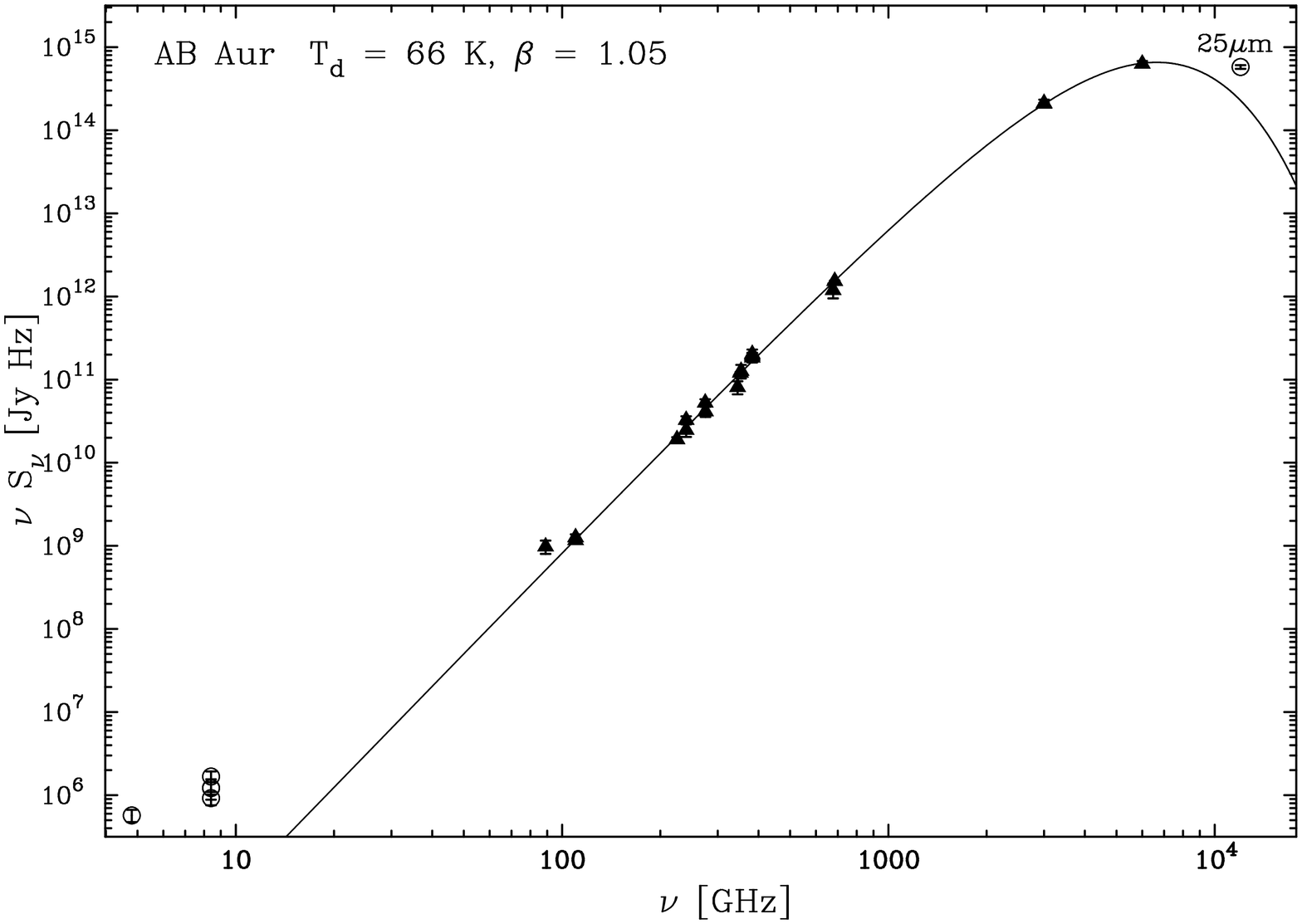}
\figcaption[]{
\label{fig-abaur}
Deconvolved sub-millimeter images of AB Aur at 850 and 450 \mum\ plotted
in gray-scale and overlaid with contours. The disk is unresolved both at
850 and 450 $\mu$m. The peak flux density is 0.349 and 2.2
Jy~beam$^{-1}$ for  HPBWs of  14\arcsec\ and  8\arcsec\  for the 850 and
450 \mum\ images, respectively. The contour levels are logarithmic with
six contours between the 2$\sigma$ noise level and peak flux density.
The  2$\sigma$ noise level is 18 mJy~beam$^{-1}$ at 850 $\mu$m and 110
mJy~beam$^{-1}$ at 450 $\mu$m. The position of AB Aur is marked by a
star symbol. The faint extension to the southwest in the 450 \mum\ image
appears to be spurious, because it does not agee in position with the
faint extension seen at  850 \mum.}
\includegraphics[angle=-90,scale=0.65]{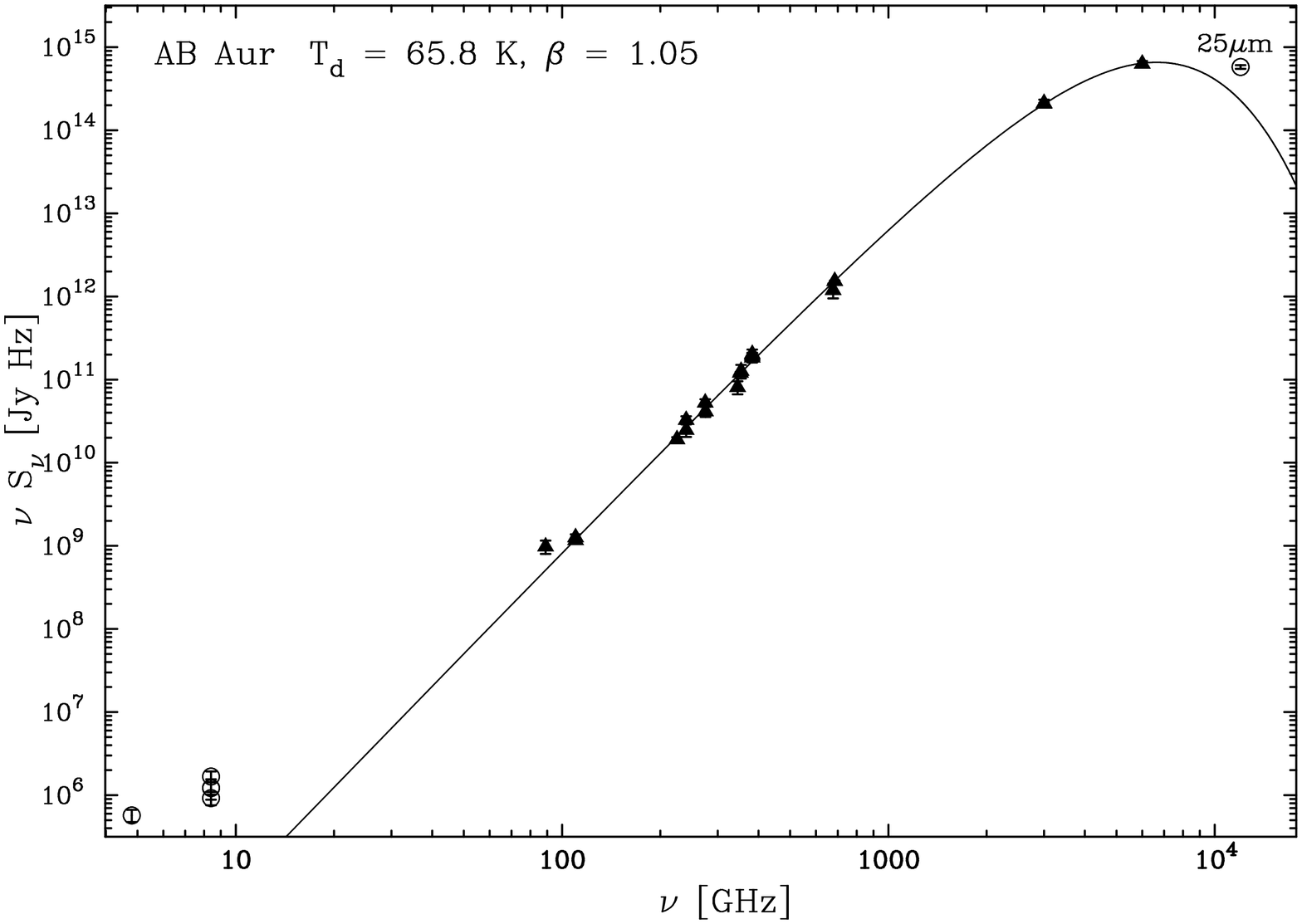}
\figcaption[]{
\label{fig-abaurfit}
An isothermal graybody fit to the SED for AB Aur.}

\includegraphics[angle=0,scale=0.29]{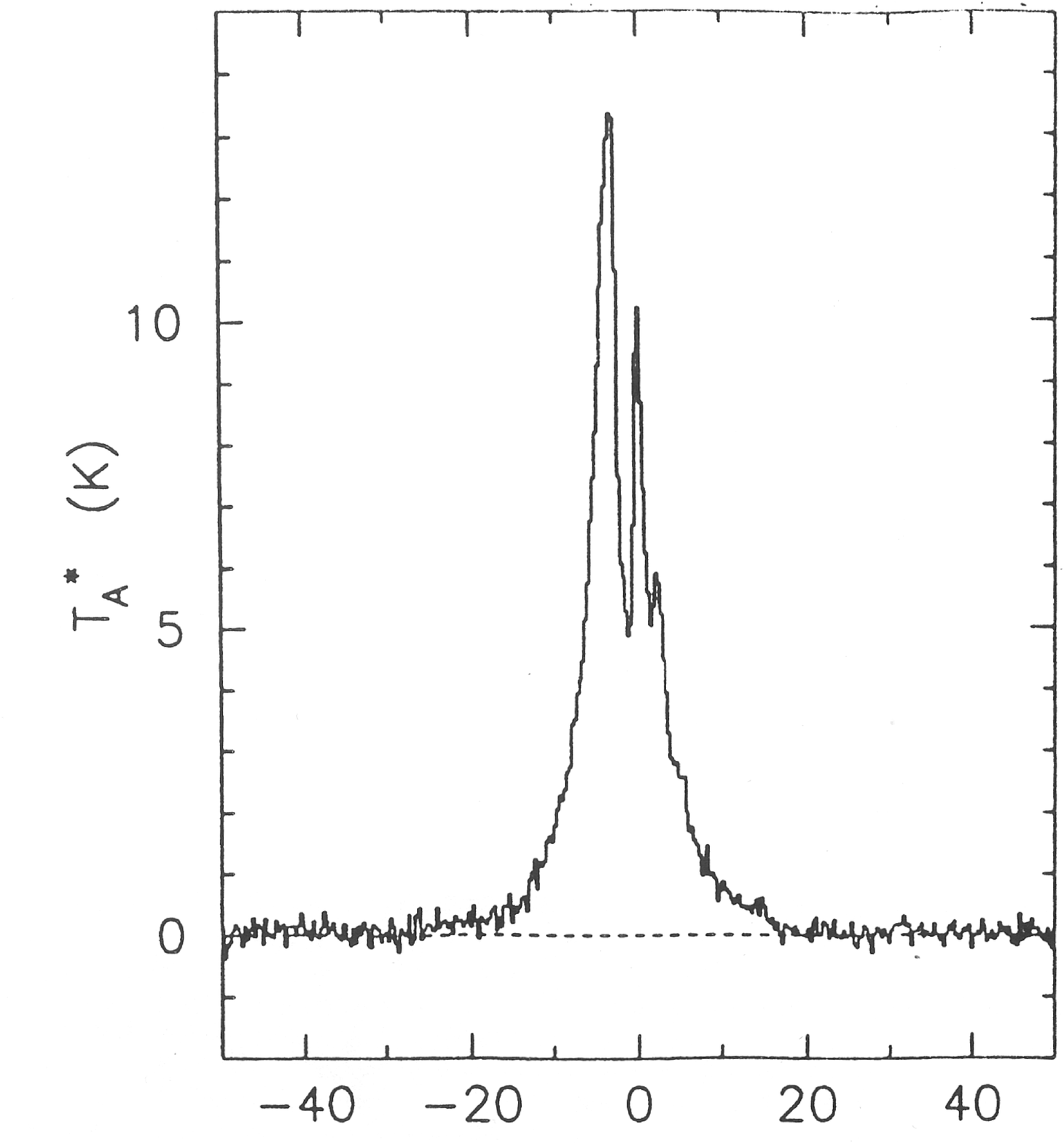}
\figcaption[]{
\label{fig-vymon_spec}
CO J = $2 \to 1$ spectrum of VY Mon showing high velocity emission from
about $-$28 to $+$18 km~s$^{-1}$.}

\includegraphics[angle=0,scale=0.29]{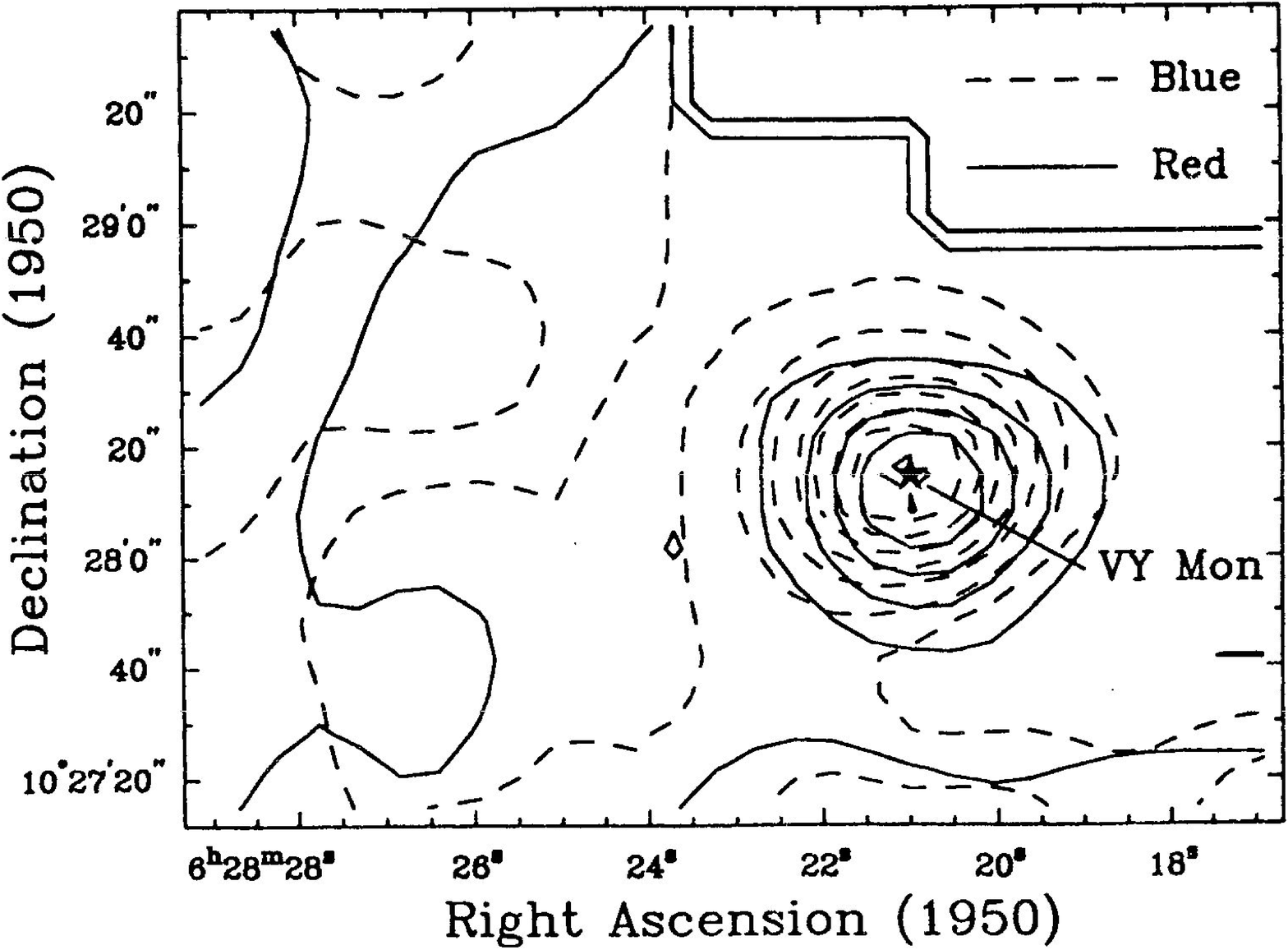}

\figcaption[]{
\label{fig-vymon_map}
Map of CO J = $2 \to 1$ high velocity emission around VY Mon. The solid
contours show the red-shifted emission integrated over the velocity
range $-$14 to $-$3 km~s$^{-1}$ with the lowest contour at 2
K~km~s$^{-1}$  and steps of 6 K~km~s$^{-1}$.  The dashed contours show
the blue-shifted emission integrated over the velocity range $+$1 to
$+$12 km~s$^{-1}$ with the lowest contour at 3 K~km~s$^{-1}$ and steps
of  5 K~km~s$^{-1}$. The bipolar outflow appears unresolved and centered
on VY Mon, which is marked by a star symbol.}

\includegraphics[angle=-90,scale=0.9]{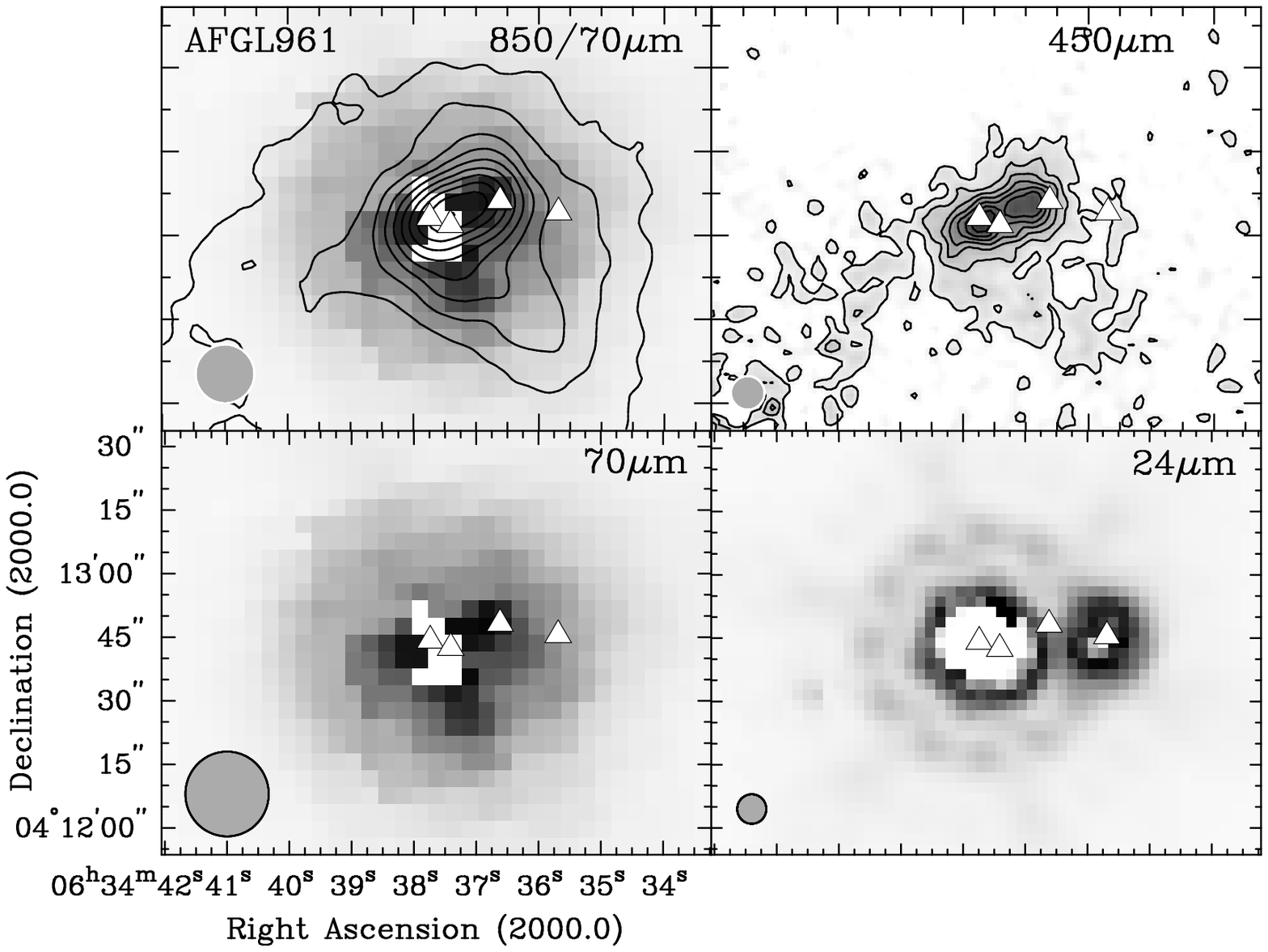}
\figcaption[]{
\label{fig-afgl961}
{\bf Upper panels:} Contour plots of 850 and 450 $\mu$m SCUBA images.
The 850 $\mu$m image (left) is overlaid on a MIPS 70 $\mu$m image in
grayscale, also shown below. The  triangles in the saturated region are
AFGL\,961\,A (furthest to the left) and AFGL\,961\,B (second from left).
The other two triangles are SMA\,3 (second from right) and source C
(furthest to the right) from \citet{Williams09}. At 850 $\mu$m we plot
ten evenly spaced  contours from  0.15 Jy~beam$^{-1}$ to 2.3 Jy~beam$^{-1}$. 
At 450 $\mu$m we plot eight evenly spaced contours between 1. 0 Jy~beam$^{-1}$ and 10
Jy~beam$^{-1}$. {\bf Lower panels:} MIPS 70 and 24 $\mu$m images in
grayscale. Both images are severely saturated (pure white in grayscale).
At 24 $\mu$m source AFGL\,961\,C breaks up into a double source. The
source immediately to the southeast of C is saturated in the 24 $\mu$m
MIPS image. The HPBWs are shown in the bottom left corner of each
image.}

\includegraphics[angle=-90,scale=0.65]{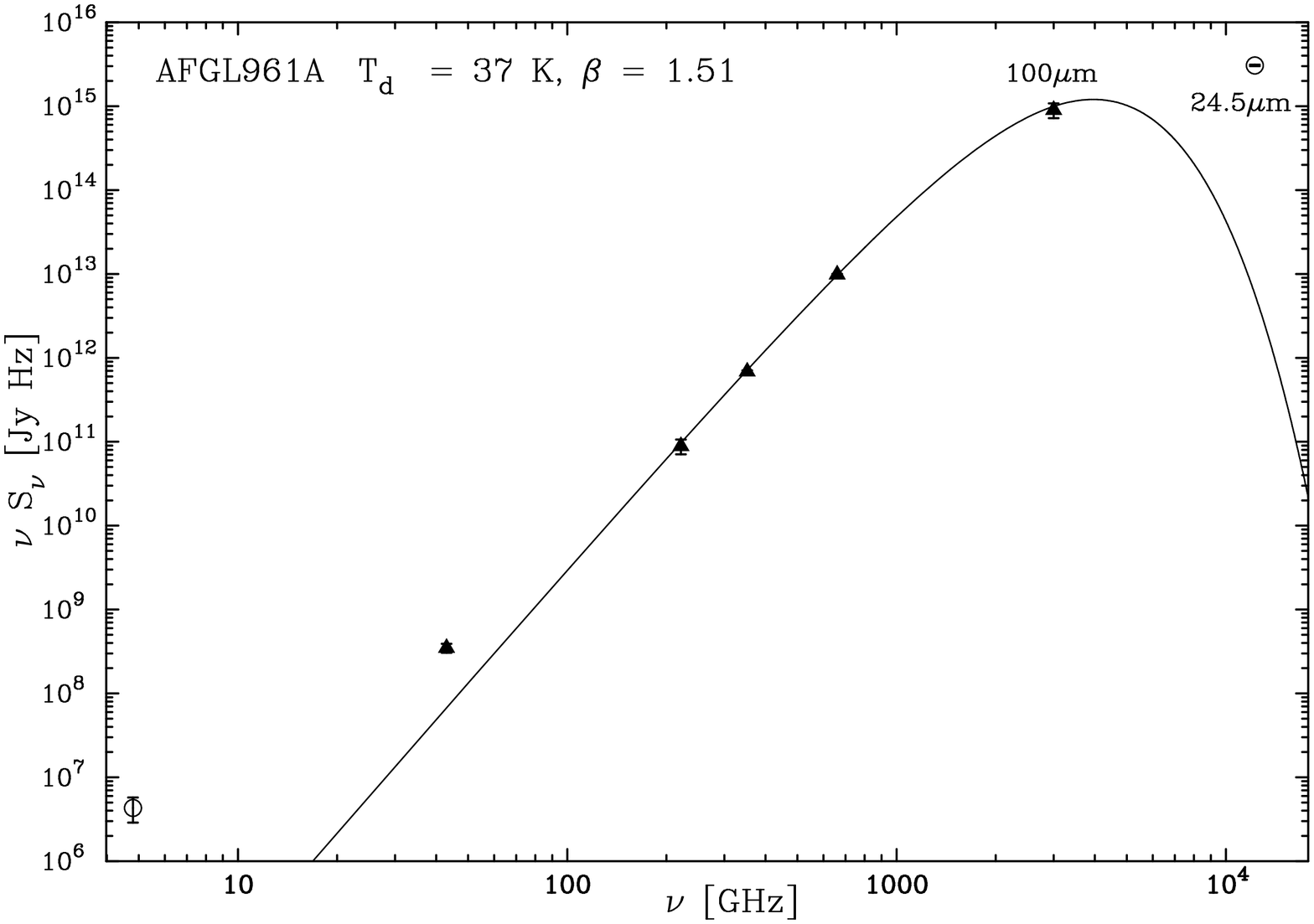}
\figcaption[]{
\label{fig-V892_sed}
Isothermal graybody fit of AFGL\,961\,A.}

\includegraphics[angle=-90,scale=0.65]{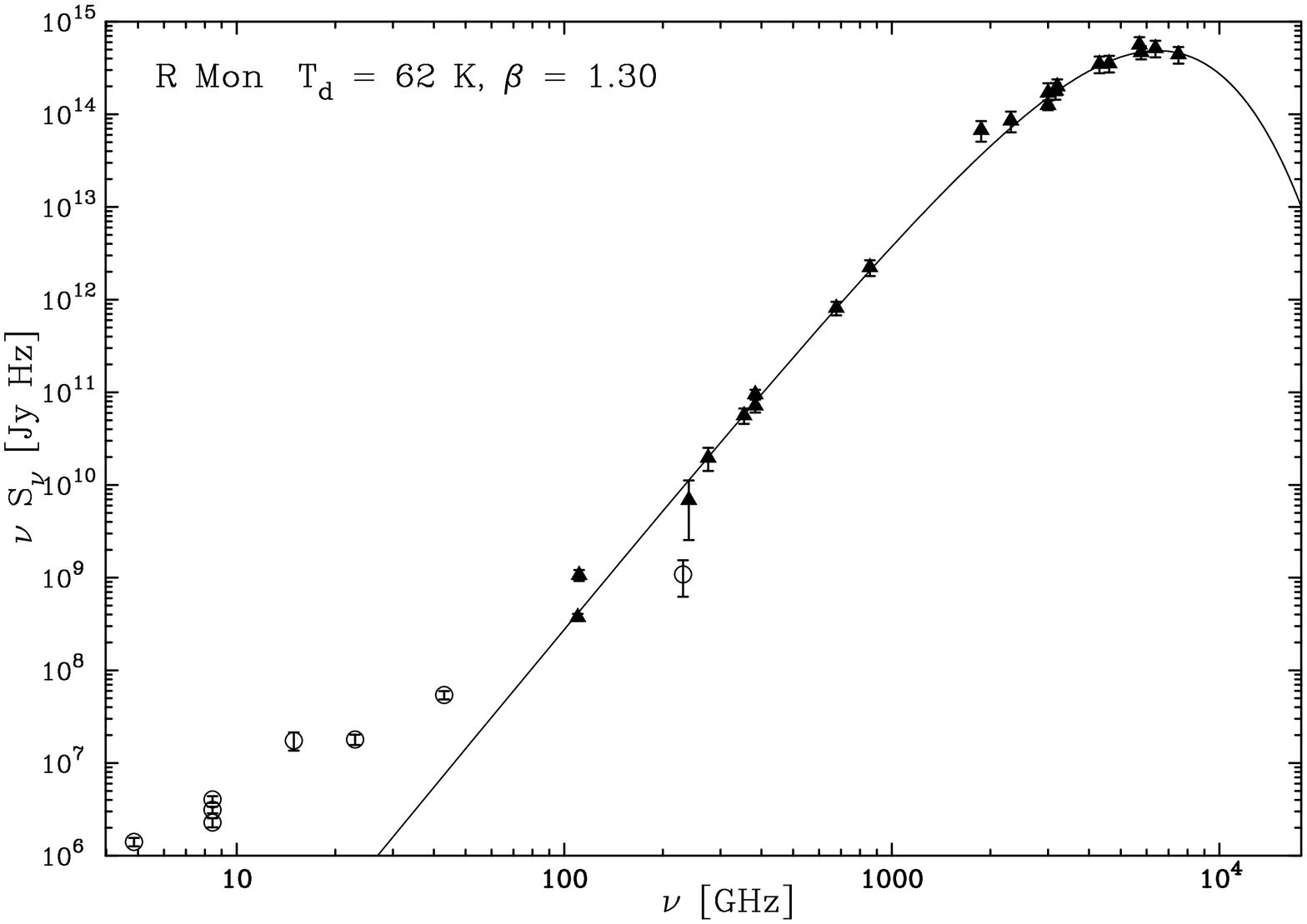}
\figcaption[]{
\label{fig-rmon_sedfit}
Isothermal graybody fit of R Mon to free-free corrected (sub-)millimeter
data and KAO far infrared data. The 1.3~mm flux density observed by
\citep{Fuente06} appears anomalously low and was not used in the fit.
}

\includegraphics[angle=-90,scale=0.65]{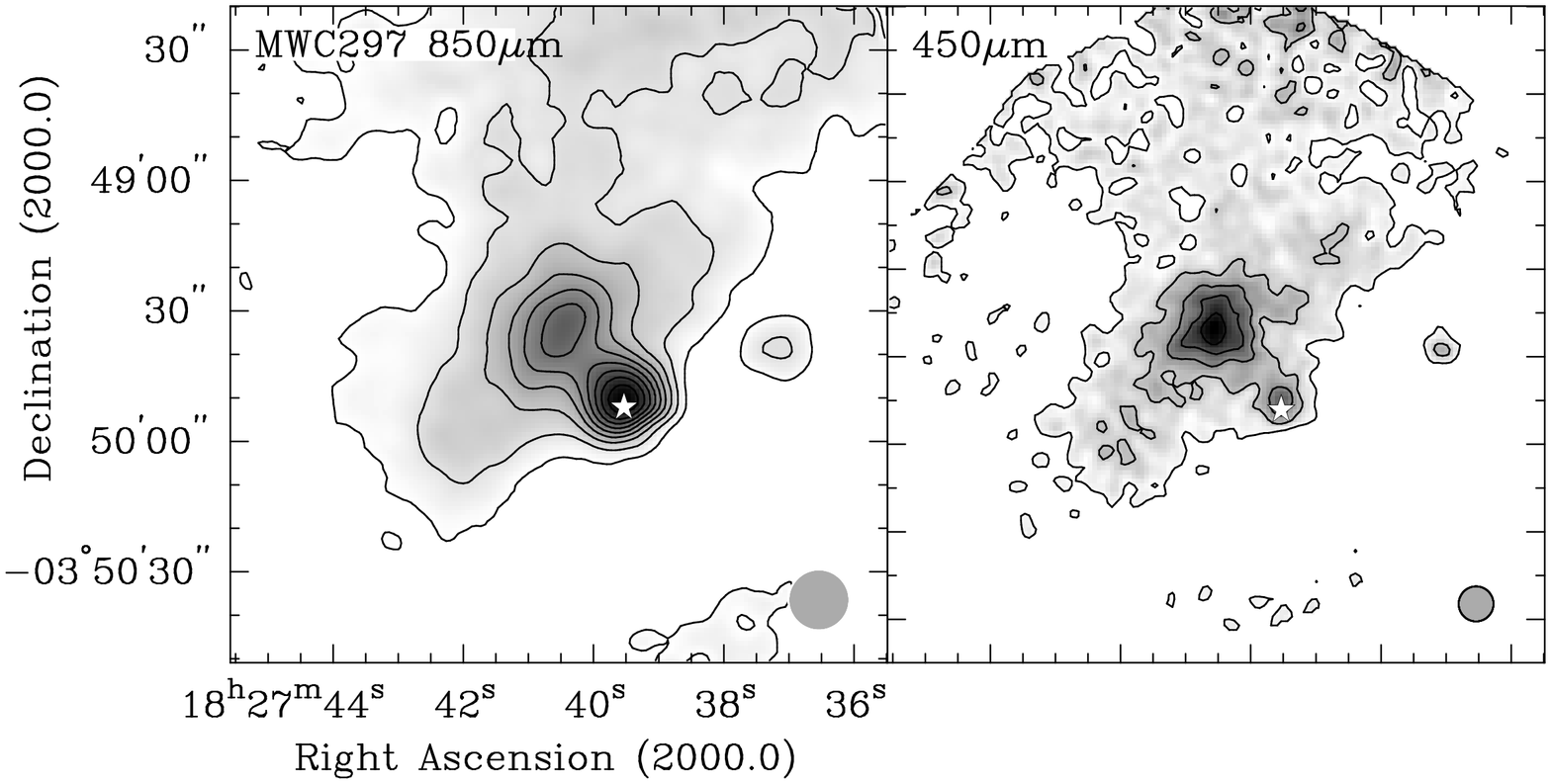}
\figcaption[]{
\label{fig_mwc297scuba}
SCUBA  850 and 450 $\mu$m images in grayscale overlaid with contours of 
the MWC\,297 region. At 850 $\mu$m we plot ten evenly spaced contours between
30 mJy~beam$^{-1}$ and 611 mJy~beam$^{-1}$, at 450 $\mu$m  six evenly spaced
contours between 0.2 Jy~beam$^{-1}$ and 2.0 Jy~beam$^{-1}$. MWC\,297 is
marked by a star symbol. The HPBW is shown in the bottom right corner of
each image. }

\includegraphics[angle=-90,scale=0.65]{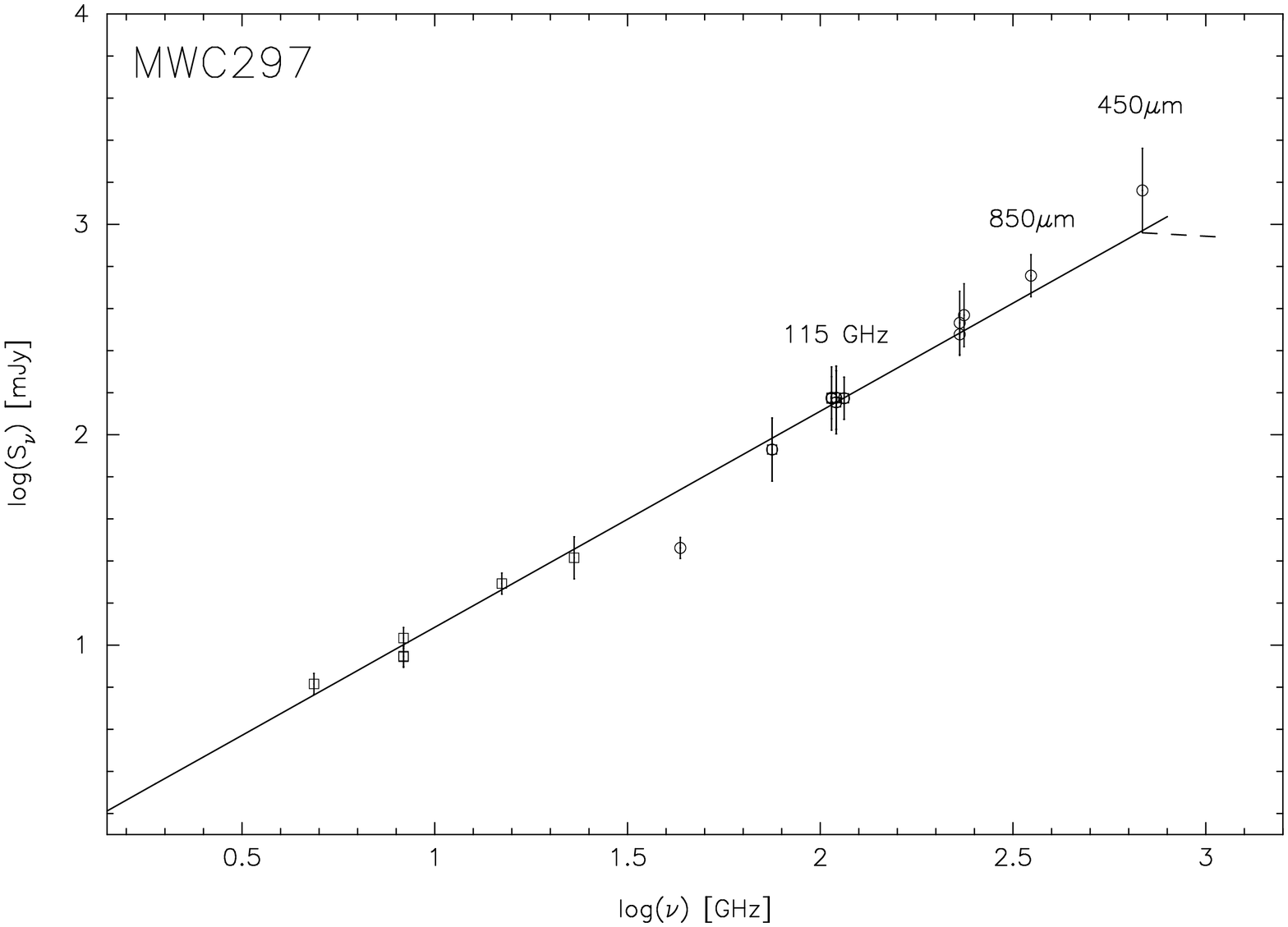}
\figcaption[]{
\label{fig-mwc297sed}
Least-squares fit to VLA and BIMA data (open squares) up to 3~mm (see
text), predicting a spectral index $\alpha$ $\sim$ 1. In this fit we
omitted the VLA data point at 43 GHz \citep{Alonso-Albi09}, which
appears anomalously low, suggesting that a significant fraction of the
emission at this wavelength has been resolved out by the interferometer.
There may be some dust excess at 850 $\mu$m and almost certainly at 450
$\mu$m.  The dashed line shows that the free-free emission will flatten
out when the whole jet becomes optically thin. }

\includegraphics[angle=-90,scale=0.65]{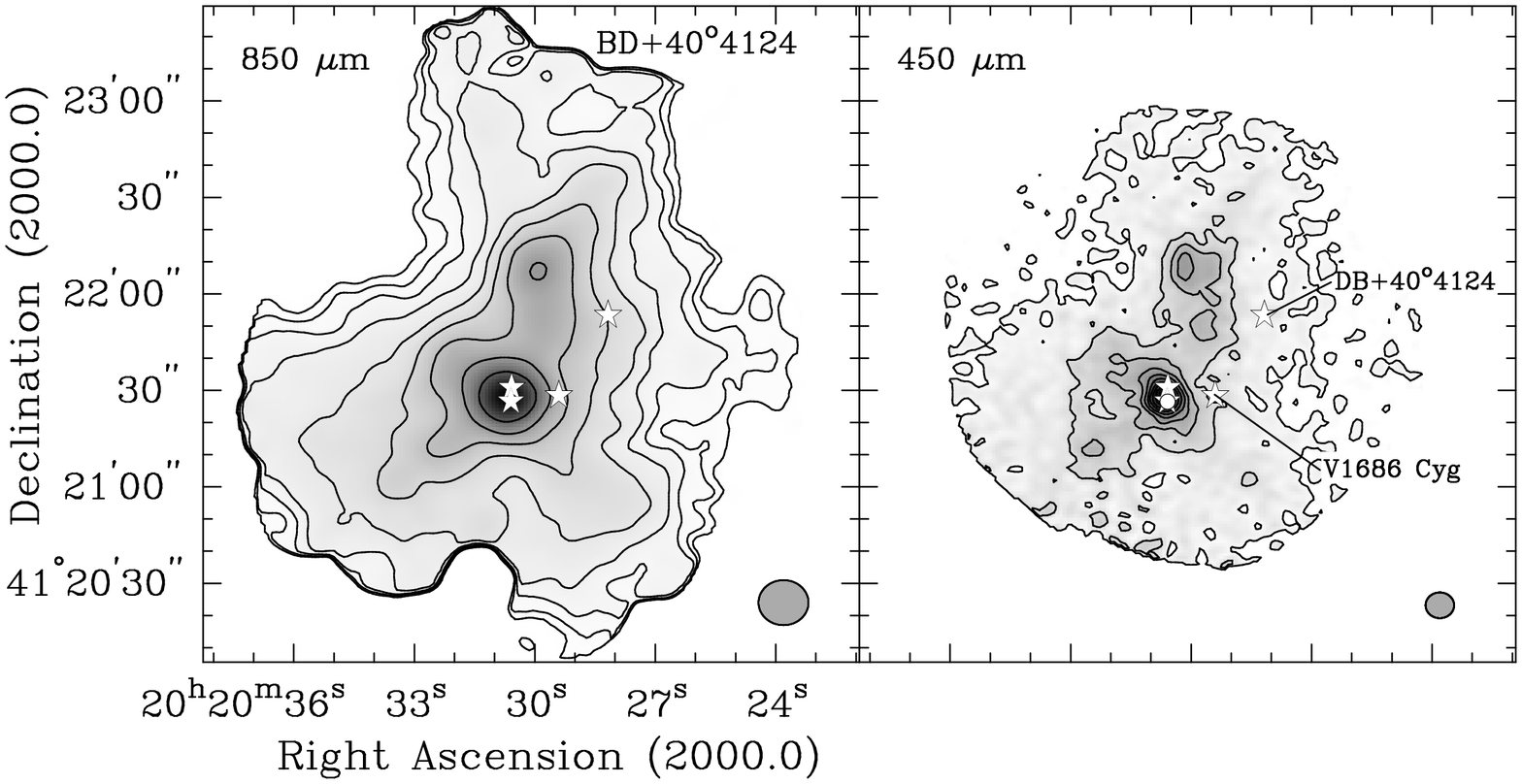}
\figcaption[]{
\label{fig-BD40}
SCUBA 850 and 450 $\mu$m images of the BD 40$^o$4124 field in grayscale
overlaid with contours. The contour levels are evenly spaced with 10 contours
from the 3-$\sigma$ level to the peak flux density. At 850 $\mu$m the
lowest contour is at 40 mJy~beam$^{-1}$ and the highest at 2.45
Jy~beam$^{-1}$. At 450 $\mu$m the lowest contour is at 300
mJy~beam$^{-1}$ with the highest contour at 8.9 Jy~beam$^{-1}$. The
stars,  BD 40$^o$4124, V\,1686~Cyg and V\,1318 Cyg N and S are marked by
star symbols and the first two are labeled on the 450 $\mu$m image.
V\,1318~Cyg N and S coincide with a strong sub-millimeter source,
although the emission is likely completely dominated by an invisible
Class 0 source situated about one arcsecond northeast of V\,1318~Cyg S
(see text). The HPBWs are shown in the bottom right corner of each
image.}

\includegraphics[angle=-90,scale=0.65]{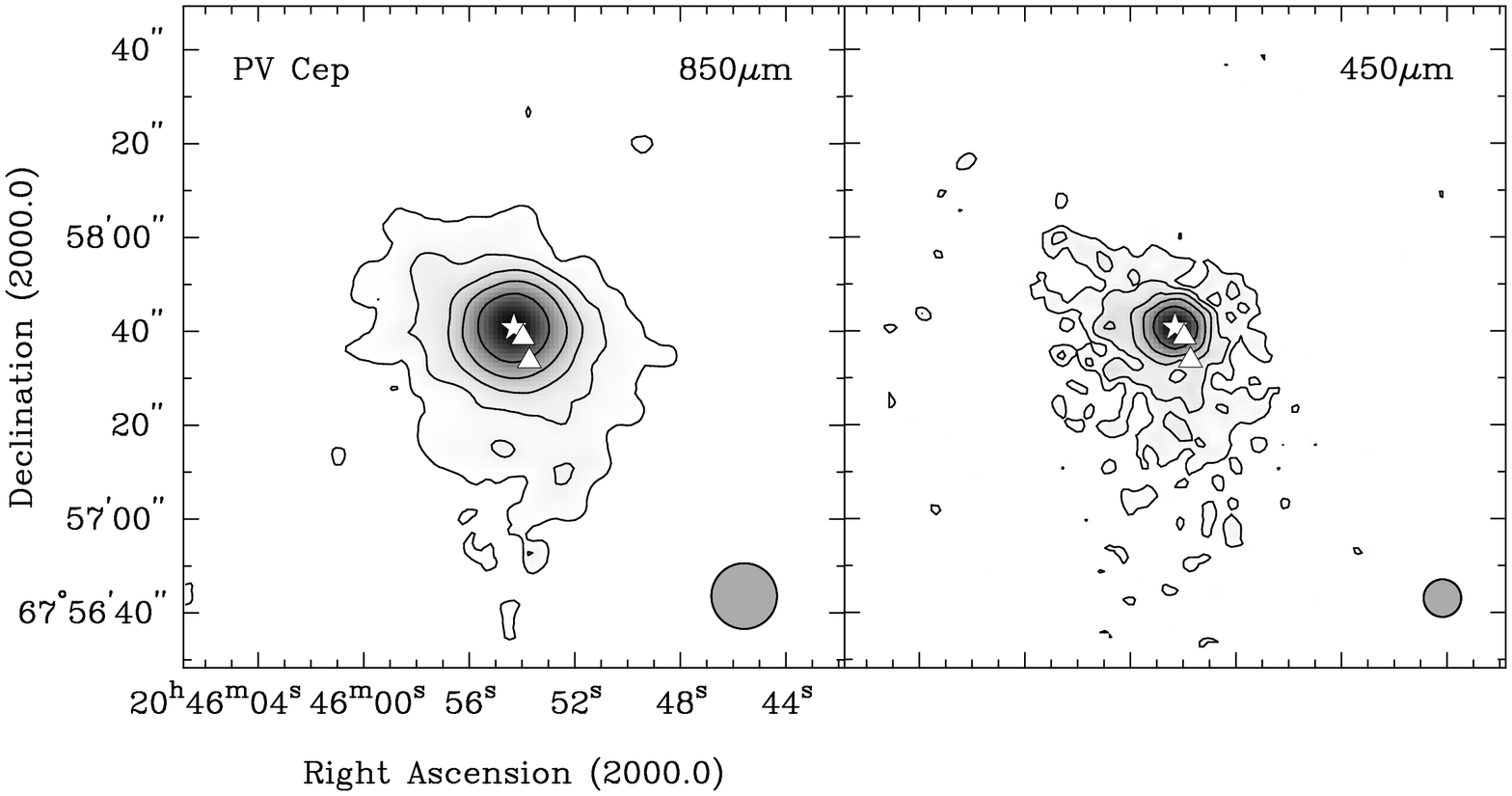}
\figcaption[]{
\label{fig-pvcep}
SCUBA 850 and 450 $\mu$m images of PV~ Cep in grayscale overlaid with
contours. The contours are logarithmic with the lowest contour at the
3-$\sigma$ level and the highest at the peak flux density.  At 850
$\mu$m the lowest contour is at 50 mJy~beam$^{-1}$ with the peak flux
density at 1 Jy~beam$^{-1}$. At 450  $\mu$m the corresponding values for
the contour levels are 300 mJy~beam$^{-1}$ and 5.5 Jy~beam$^{-1}$. PV
Cep is marked by a star symbol and H$_2$O masers with triangles. The
circumstellar disk/envelope is well resolved in these SCUBA images with
a size of $\sim$3\farcs6. We also see faint emission from the
surrounding dust cloud in which PV Cep is embedded. The HPBWs are shown in the bottom right corner of each
image.}

\includegraphics[angle=-90,scale=0.65]{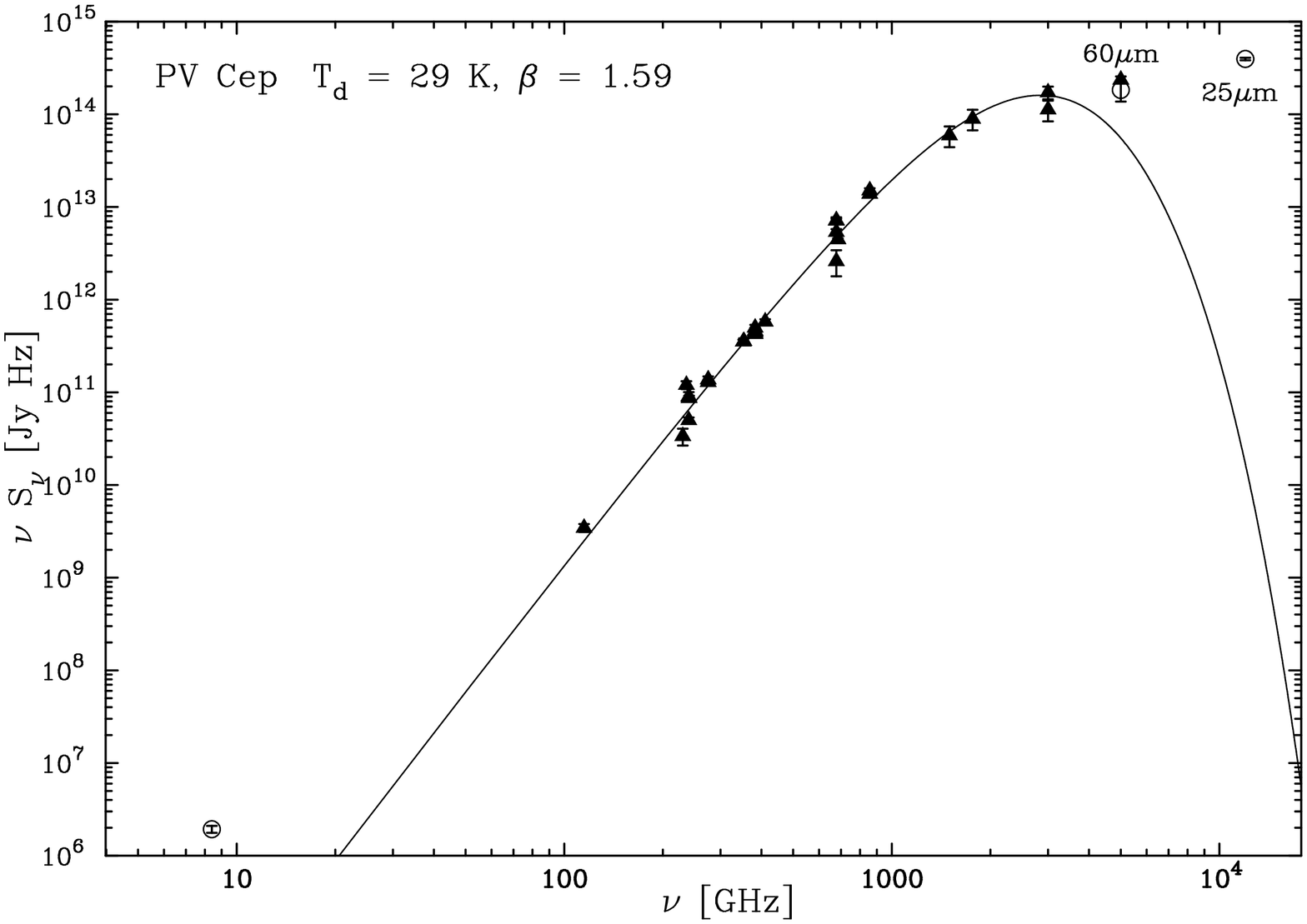}
\figcaption[]{
\label{fig-pvcep_sedfit}
An isothermal graybody fit to the SED for PV Cep.}

\includegraphics[angle=-90,scale=0.65]{f25.ps}
\figcaption[]{
\label{fig-v645cyg}
SCUBA 850 $\mu$m  image of V\,645~Cyg in grayscale overlaid with
contours. The contours are logarithmic with eight contours between 40
mJy~beam$^{-1}$ and 1.78 Jy~beam$^{-1}$. The HPBW is shown in the bottom right corner. }
\newpage

\includegraphics[angle=-90,scale=0.65]{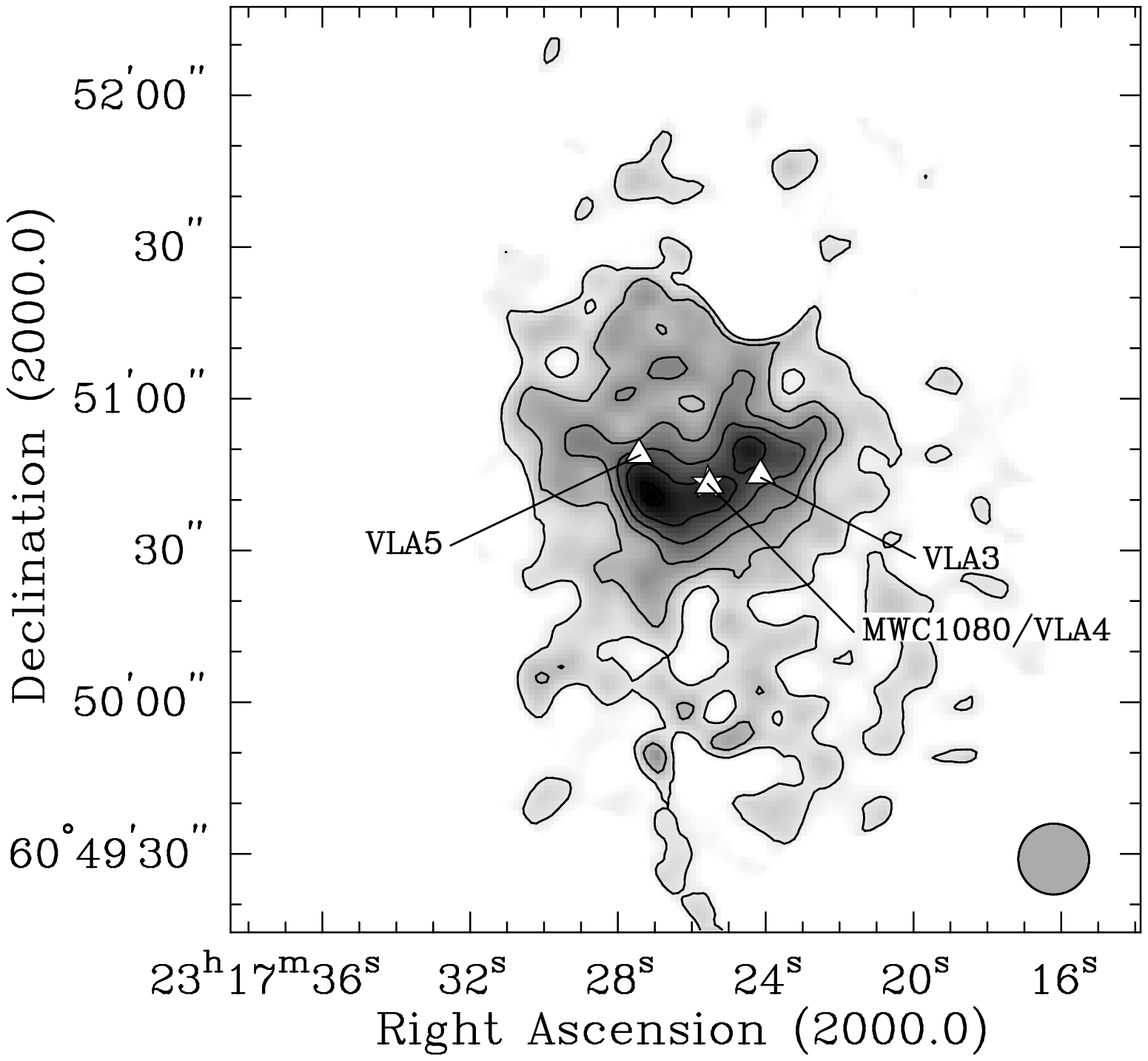}
\figcaption[]{
\label{fig-mwc1080}
SCUBA 850 $\mu$m  image of the MWC\,1080 field in grayscale overlaid
with contours. The contours are logarithmic  with six contours between
150 mJy~beam$^{-1}$ and 700 mJy~beam$^{-1}$. The VLA sources 3, 4 and 5
\citep{Girart02} are marked with triangles. VLA\,4 coincides with the B0
star MWC\,1080\,A, also marked with a star symbol. The HPBW is shown in the bottom right corner. }

\includegraphics[angle=-90,scale=0.65]{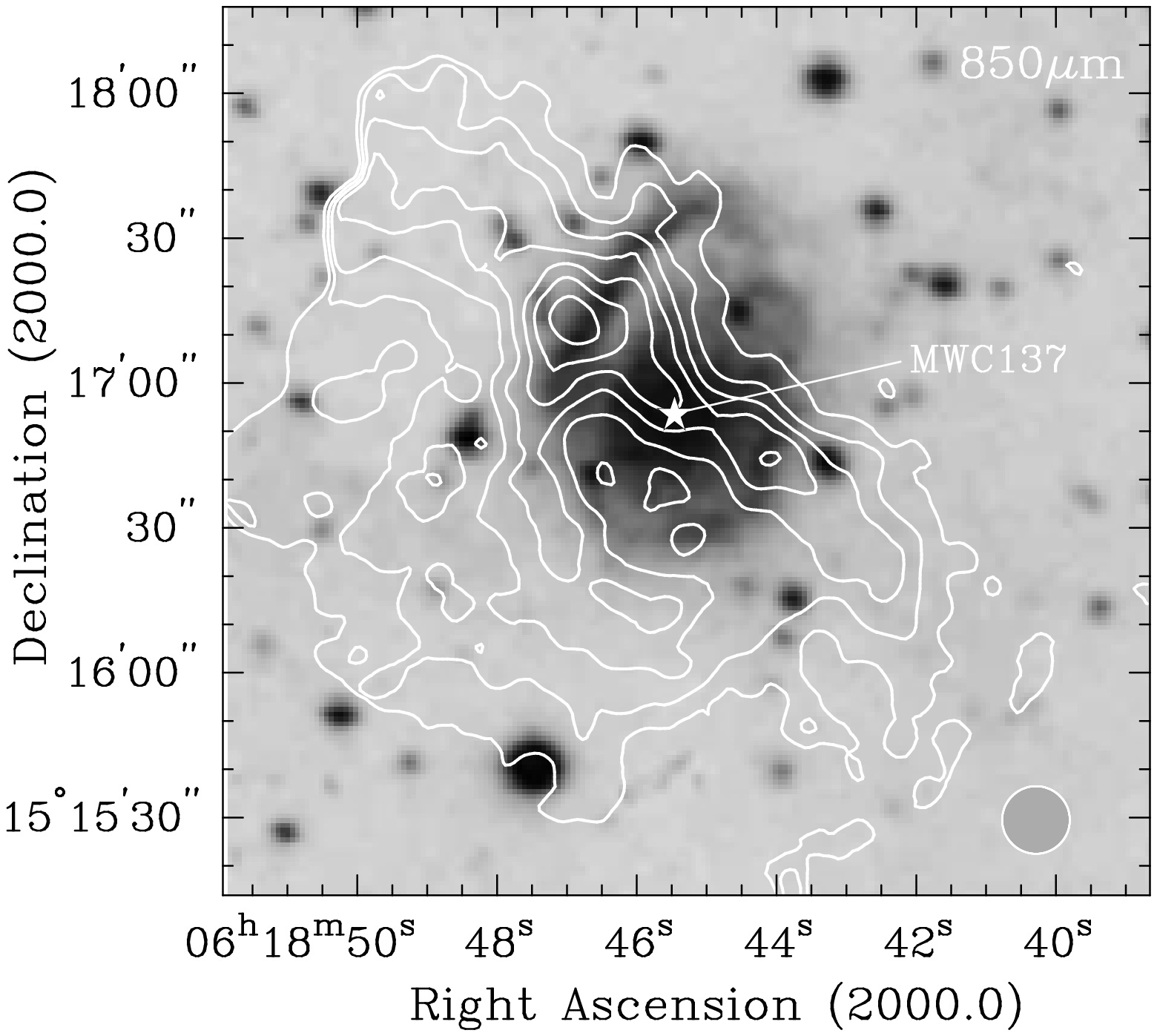}
\figcaption[]{
\label{fig-mwc137}
An 850 $\mu$m contour map of  MWC\,137 overlaid on an optical image in
grayscale. The contour levels are linear starting at 50 mJy~beam$^{-1}$
with a step of 50 mJy~beam$^{-1}$.  MWC\,137 is marked by a star symbol.
The HPBW is shown in the bottom right corner.
}
\clearpage

\includegraphics[angle=-90,scale=0.65]{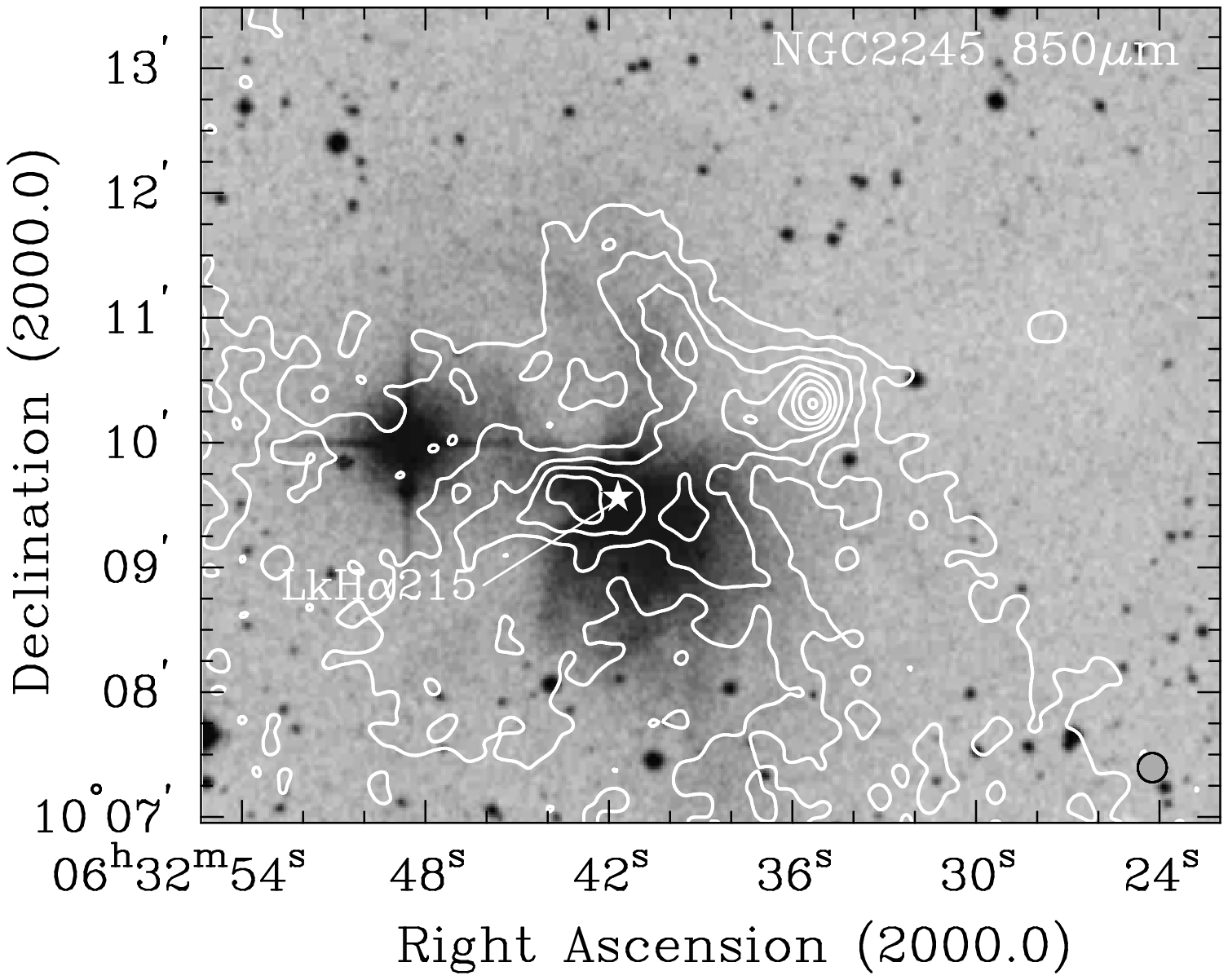}
\figcaption[]{
\label{fig-NGC2245}
An 850 $\mu$m contour map of the reflection nebula NGC\,2245 overlaid on
an optical image in grayscale. The contour levels are linear starting at
60 mJy beam$^{-1}$ with a step of 60 mJy beam$^{-1}$. LkH$\alpha$\,215
is not detected  at 850 $\mu$m. Northwest of the star and outside the
reflection nebula there is a compact sub-millimeter source with an 850
$\mu$m flux density of  150 mJy, which has neither an optical nor a
near-infrared counterpart. The HPBW is shown in the bottom right corner.}

\clearpage

\includegraphics[angle=-90,scale=0.65]{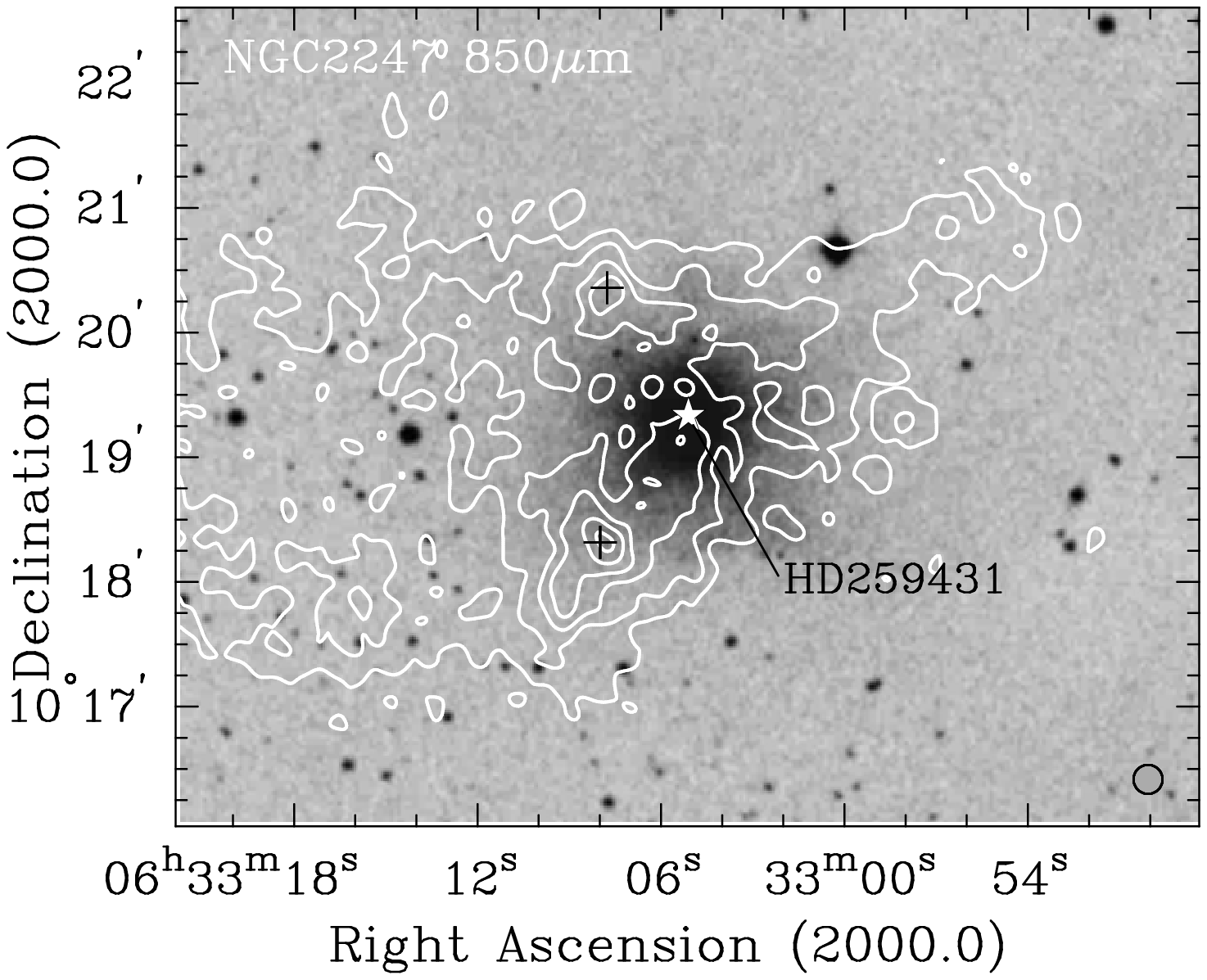}
\figcaption[]{
\label{fig-NGC2247}
An 850 $\mu$m contour map of the reflection nebula NGC\,2247 overlaid on
an optical image in grayscale. The contour levels are linear, starting
at 50 mJy~beam$^{-1}$ with a step of 50 mJy~beam$^{-1}$.  The nebula is
illuminated by HD\, 259431, which is close to a minimum in the dust
emission. Two faint sub-millimeter sources, marked with black plus
signs, are seen outside the nebula. The HPBW is shown in the bottom right corner.
% the one to the sw has a flux density of 380 mJy, the northern one, about 160 mJy
}

\includegraphics[angle=-90,scale=1.0]{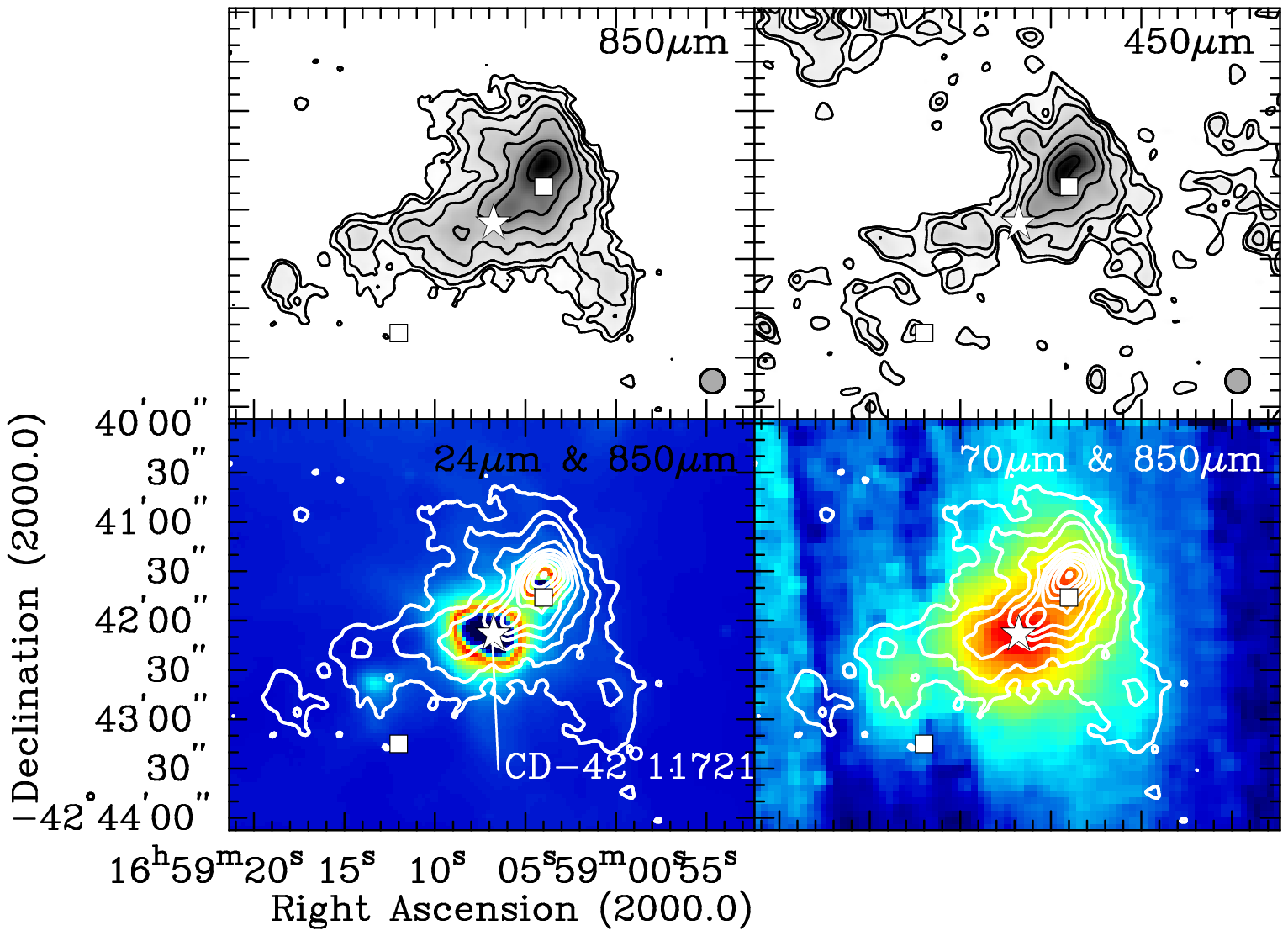}
\figcaption[]{
\label{fig-CoD}
{\bf Upper panels:} 850 and 450 $\mu$m SCUBA maps of the CD$-$42$^\circ$
1172 field. The position of CD$-$42$^\circ$ 1172 is marked by a star
symbol, and the two 13~cm radio sources \citep{Thompson04} are marked by
filled squares. The contour levels are logarithmic. At 850 $\mu$m there
are seven contour levels between 0.1 Jy beam$^{-1}$ and 1.58 Jy
beam$^{-1}$, at 450 $\mu$ they go from 1 Jy beam$^{-1}$ to 15.8 Jy
beam$^{-1}$ The SCUBA HPBWs are shown in the bottom right corner of each
panel.  {\bf Lower panels:}  False color images of 24 and 70 $\mu$m MIPS
images overlaid with the same 850 $\mu$m contours as shown in the left
upper panel. Note that the 24 $\mu$m image is saturated  on
CD$-$42$^\circ$ 1172 and the nebulosity surrounding it. The source to
the northwest of the star is also saturated. }

\includegraphics[angle=-90,scale=0.7]{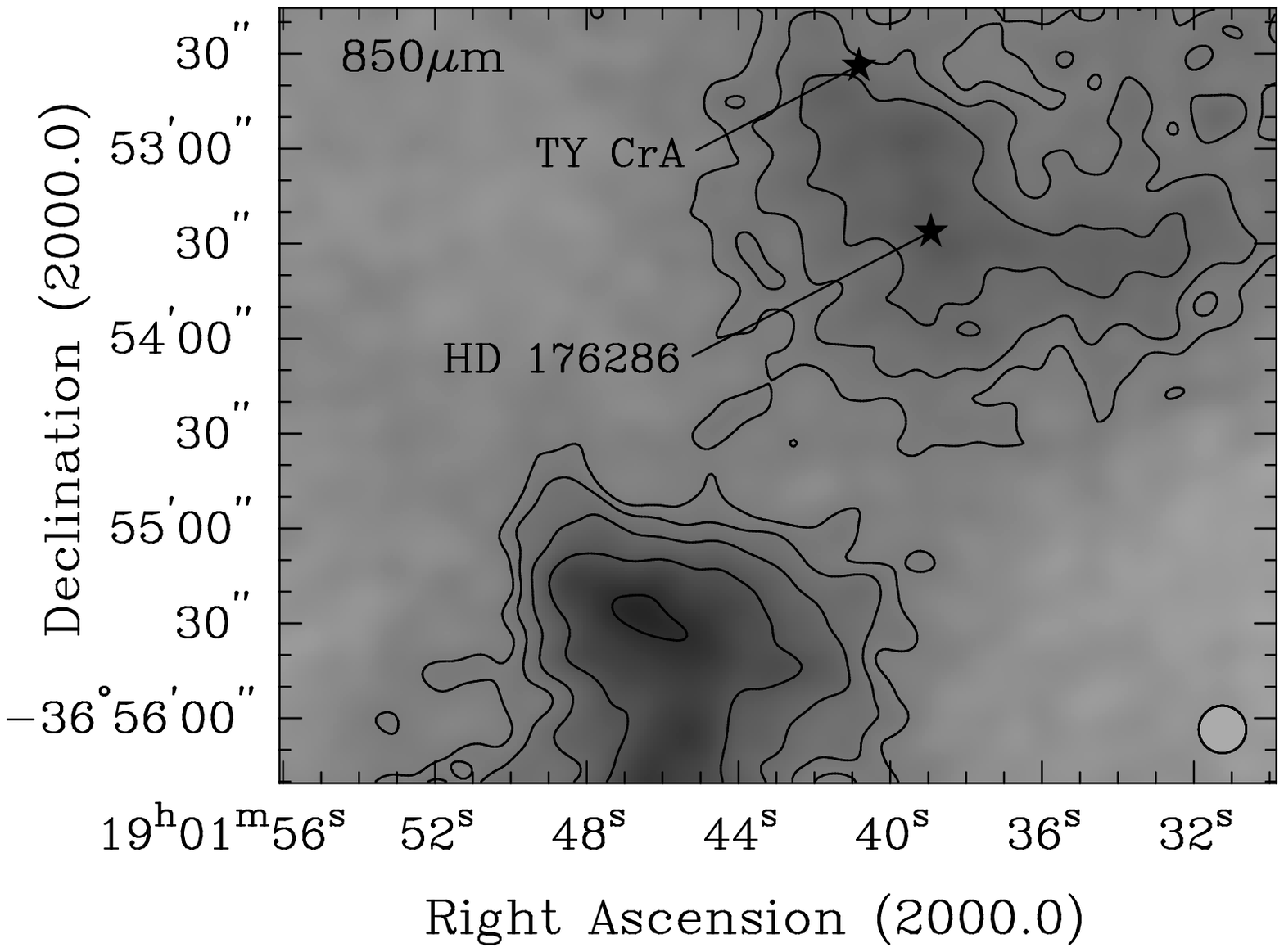}
\figcaption[]{
\label{fig-tycra}
Sub-image of large SCUBA 850 $\mu$m scan map showing HD\,176386 and
TY~CrA. Although both stars are definitely embedded in a dense dust
cloud, neither of them are associated with any sub-millimeter continuum
emission.  The HPBW is shown in the bottom right corner.}

\newpage

\includegraphics[angle=0,scale=1.0]{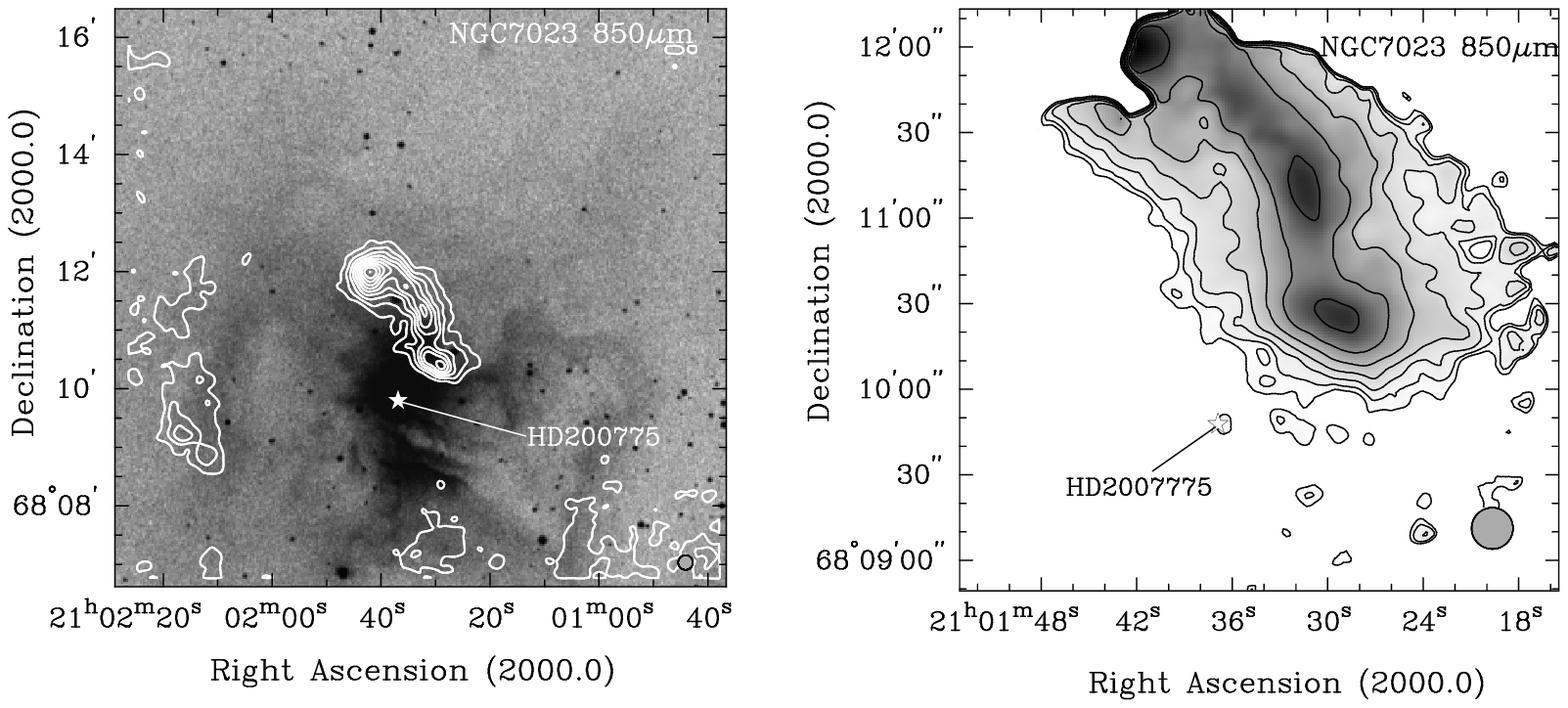}
\figcaption[]{
\label{fig-NGC7023}
{\bf Left panel:} Large 850 $\mu$m scan map of the reflection nebula
NGC\,7023 plotted with contours and overlaid on an optical image in gray
scale. We applied a logarithmic stretch to the optical image to enhance
the faint nebulosity. The contour levels are linear, starting at 40  mJy
beam$^{-1}$ with ten contours up to the peak flux density of  712  mJy
beam$^{-1}$. The HPBW is shown in the bottom right corner. {\bf Right panel:} A much deeper 850 $\mu$m jiggle map
with a 3-$\sigma$ rms of 25  mJy beam$^{-1}$. The lowest contour in this map is at 25  mJy beam$^{-1}$.  The rest of
the contours are plotted  logarithmically to the peak flux density, 530  mJy beam$^{-1}$. Even though HD\,200775 falls inside a 3-$\sigma$ contour, we consider it a non-detection, because it is not distinguishable from other noise features in the map. However, our observations are consistent with \citet{Okamoto09}, who detected HD\,200775 with the SMA at 350 GHz (856  $\mu$m)  with a flux density of 35 $\pm$  5 mJy, The SCUBA HPBW is shown in the bottom right corner of the image.
 }

\newpage

\includegraphics[angle=-90,scale=0.68]{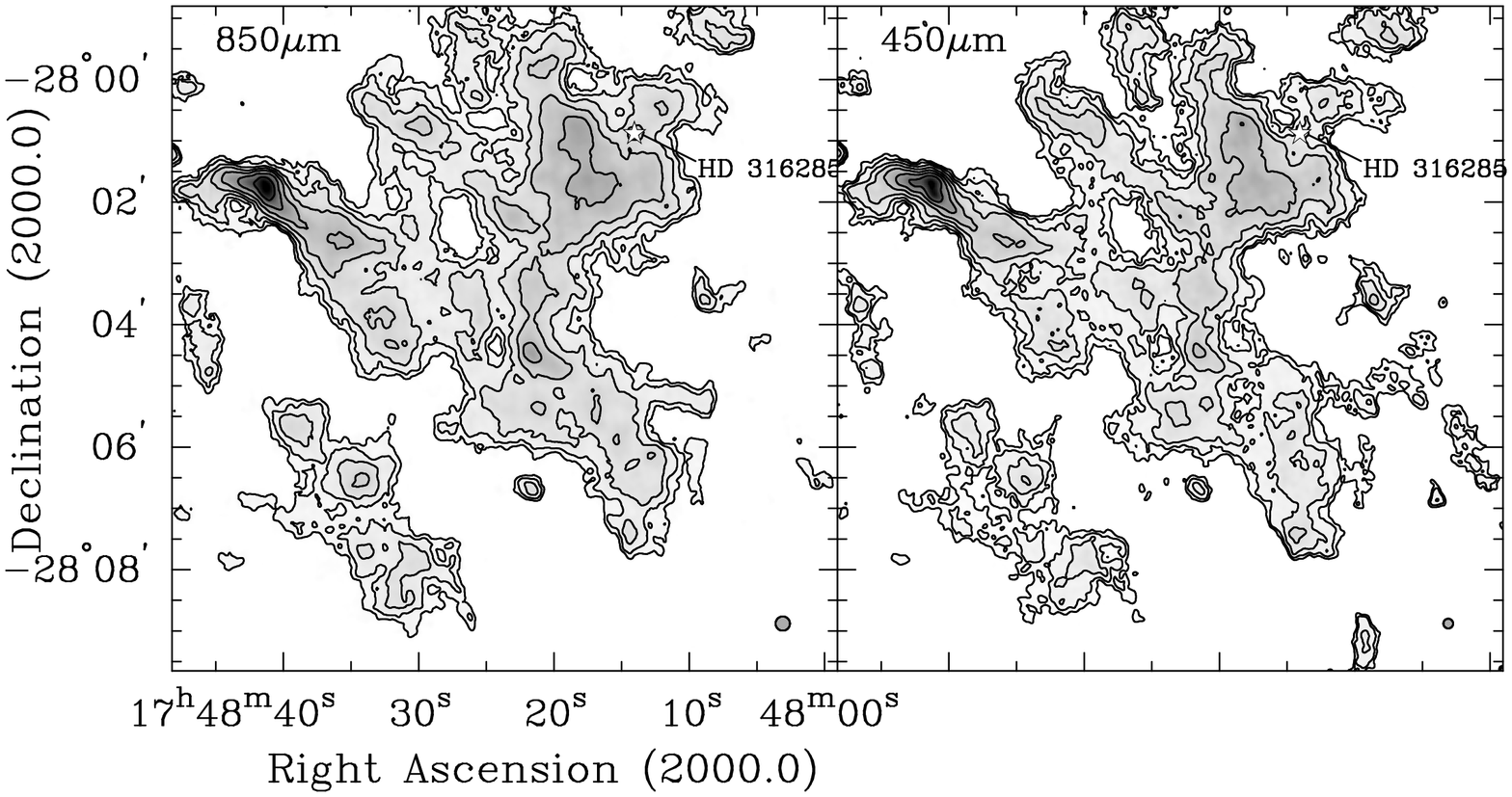}
\figcaption[]{
\label{fig-hd316285}
Large SCUBA scan maps at 850 and 450 $\mu$m in the direction of
HD\,316285. There is a faint peak at the position of the star at 850
$\mu$m (hidden by the star symbol), but not at 450 $\mu$m. The contours
are logarithmic for both images. For the 850 $\mu$m image, we plot seven
contour levels between the lowest contour, 200 mJy~beam$^{-1}$, and the
peak flux density, 3.16 Jy~ beam$^{-1}$. For 450 $\mu$m  we plot eight
contours between 1 Jy~beam$^{-1}$ and 25.1  Jy~beam$^{-1}$. 
The HPBWs are shown in the bottom right corner of each image.}
\newpage

\includegraphics[angle=-90,scale=1.0]{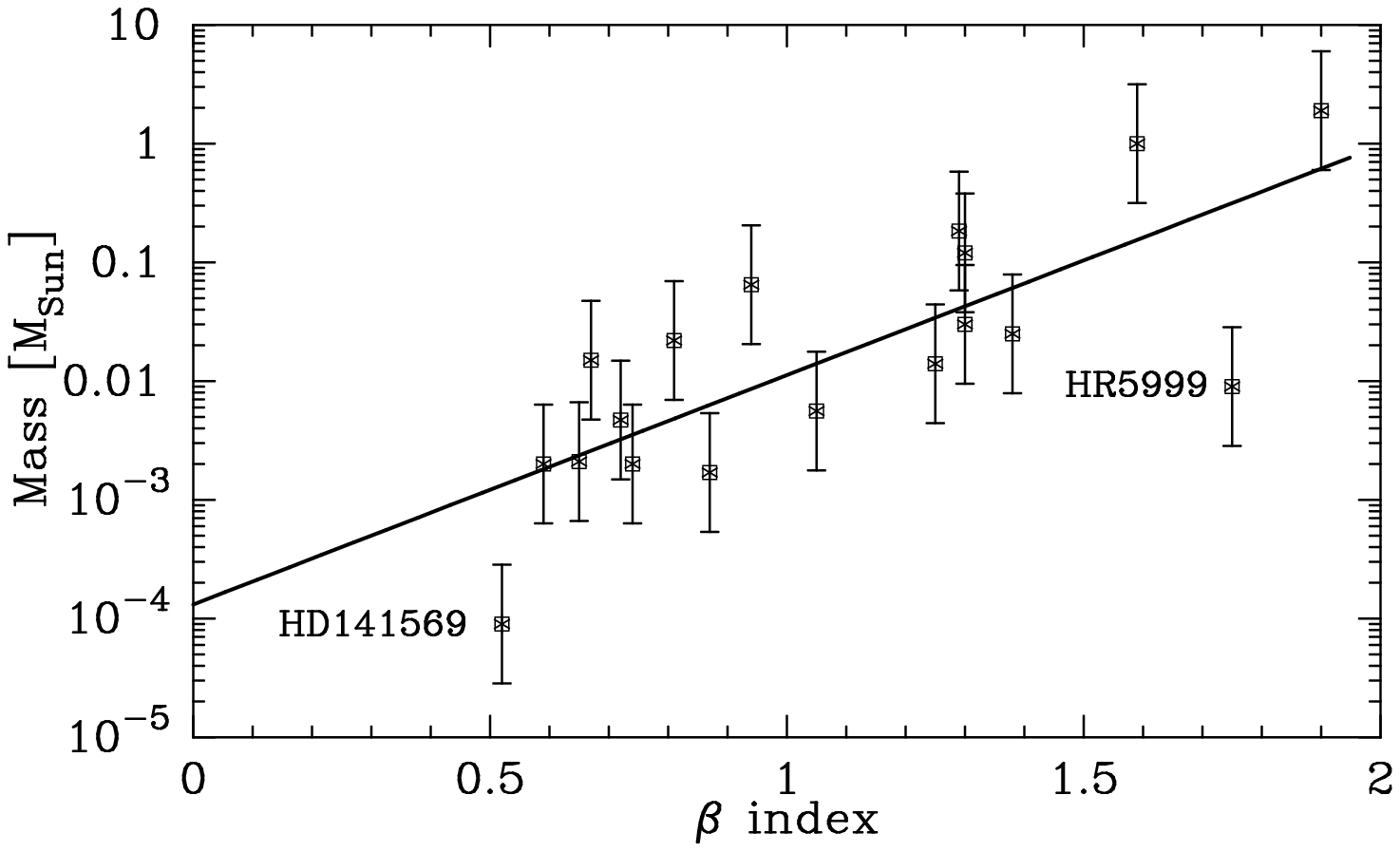}
\figcaption[]{
\label{fig-mass_beta}
Disk mass as a function of dust emissivity, $\beta$. The error bars show a worst case uncertainty of a factor of two for disk mass. The errors for $\beta$ are of the order of 0.1 -- 0.2. The two extreme
outliers, HD\,141569 and HR\,5999 are labeled in the Figure.
}
\end{document}